\documentclass[letterpaper,11pt]{article}
\usepackage[margin=1in]{geometry}
\usepackage{palatino}

\newcommand{\old}[1]{{}}
\usepackage[utf8]{inputenc}
\usepackage[T1]{fontenc}
\usepackage[noadjust,compress]{cite}
\usepackage{a4wide}
\usepackage{hyperref}
\hypersetup{colorlinks=true,citecolor=blue,linkcolor=purple,urlcolor=blue}
\usepackage{amsmath, mathtools}
\usepackage[displaymath,mathlines,running]{lineno}
\usepackage{amssymb}
\usepackage{amsthm}   
\usepackage{enumerate}
\usepackage{thm-restate}
\usepackage{caption}
\usepackage{subcaption}
\usepackage{xspace}
\usepackage{cleveref}
\usepackage{comment}
\usepackage{stmaryrd}
\usepackage{mdframed}
\usepackage{bm}
\renewcommand{\cref}{\Cref}

\usepackage{soul}

\usepackage{tikz}
\usetikzlibrary{fit,positioning, intersections, shapes.geometric,shapes.symbols,arrows,decorations.markings,decorations.pathreplacing,calc}

\definecolor{MyPurple}{cmyk}{0.45,0.86,0,0}
\renewcommand{\emph}[1]{{\color{MyPurple}{\em #1}}}

\usepackage{algorithm}
\usepackage[noend]{algpseudocode}


\newtheoremstyle{mystyle}
  {}
  {}
  {\itshape}
  {}
  {\bfseries}
  {.}
  { }
  {\thmname{#1}\thmnumber{ #2}\thmnote{ (#3)}}
\theoremstyle{mystyle}
\newtheorem{theorem}{Theorem}[section]
\newtheorem{corollary}[theorem]{Corollary}
\newtheorem{lemma}[theorem]{Lemma}
\newtheorem{remark}[theorem]{Remark}
\newtheorem{observation}[theorem]{Observation}

\newtheorem{claim}{Claim}[theorem]

\crefname{theorem}{Theorem}{Theorems}
\Crefname{theorem}{Theorem}{Theorems}
\crefname{claim}{Claim}{Claims}

\Crefmultiformat{lemma}{Lemmas~#2#1#3}%
{ and~#2#1#3}{, #2#1#3}{, and~#2#1#3}

\theoremstyle{definition}
\newtheorem{definition}[theorem]{Definition}

\newenvironment{claimproof}[1][\unskip]{\noindent {\emph{Proof of Claim #1.\space}}}{\hfill$\triangleleft$ \smallskip}

\newcommand{\defproblem}[3]{
  \par\addvspace{\baselineskip}%
  \begin{center}\fbox{
  \begin{minipage}{0.96\textwidth}
  \begin{tabular*}{\textwidth}{@{\extracolsep{\fill}}lr} #1 \\ \end{tabular*}
  {\bf{Input:}} #2  \\
  {\bf{Question:}} #3
  \end{minipage}
  }%
\end{center}
}
\newcommand{\defoptproblem}[4]{%
  \par\addvspace{\baselineskip}%
  \begin{center}\fbox{%
    \begin{minipage}{0.96\textwidth}
      \begin{tabular*}{\textwidth}{@{\extracolsep{\fill}}lr} #1 \\ \end{tabular*}
      {\bf{Input:}} #2 \\
      {\bf{Find:}} #3 \\
      {\bf{Goal:}} #4
    \end{minipage}%
}%
\end{center}
}

\newcommand{\kSAT}[1]{#1\textsc{-SAT}}
\newcommand{\gtiling}{\textsc{Grid Tiling}}
\newcommand{\ngtiling}{\textsc{Narrow Grid Tiling}}
\newcommand{\mongtiling}{\textsc{Monotone Narrow Grid Tiling}}
\newcommand{\stsquare}{\textsc{Steiner Tree of Unit Squares}\xspace}
\newcommand{\stgeo}{\textsc{Geometric Steiner Tree}\xspace}
\newcommand{\stplanar}{\textsc{Planar Object Steiner Tree}\xspace}
\newcommand{\stplanarx}{\textsc{Extended Planar Object Steiner Tree}\xspace}

  \mdfdefinestyle{highlight}{hidealllines,linewidth=0,frametitlebackgroundcolor=gray!40,backgroundcolor=gray!20}
  \newcommand{\assbox}[1]{\begin{mdframed}[style=highlight]#1\end{mdframed}}

\newcommand{\assgeox}[1]{(A#1)}
\newcommand{\asspx}[1]{(B#1)}
\newcommand{\assplx}[1]{(C#1)}

\newcommand{\assgeo}{\assgeox{$\alpha$}}
\newcommand{\assp}{\asspx{$\alpha$}\xspace}
\newcommand{\asspl}{\assplx{$\alpha$}\xspace}

\newcommand{\xmax}[1]{x_{\text{max}}(#1)}
\newcommand{\xmin}[1]{x_{\text{min}}(#1)}
\newcommand{\ymin}[1]{y_{\text{min}}(#1)}
\newcommand{\ymax}[1]{y_{\text{max}}(#1)}

\newtheorem*{assumptiona}{Assumption \assgeo}
\newtheorem*{assumptionb}{Assumption \assp}
\newtheorem*{assumptionc}{Assumption \asspl}

\newcommand{\bin}{\mathrm{bin}}
\newcommand{\modulo}{\;\textbf{mod}\;}

\newcommand{\pathmincover}[1]{M(#1)}

\newcommand{\Oh}{\bigO}
\newcommand{\Ot}{\tilde{\bigO}}
\newcommand{\bigO}{O}

\newcommand{\bolda}{\mathbf{a}}
\newcommand{\boldb}{\mathbf{b}}

\newcommand{\boldm}{\mathbf{m}}

\newcommand{\bolds}{\mathbf{s}}

\newcommand{\rom}[1]{%
	\textup{\uppercase\expandafter{\romannumeral#1}}%
}

\mathchardef\hyph="2D
\newcommand{\abs}[1]{|#1|}
\newcommand{\eps}{\varepsilon}
\renewcommand{\epsilon}{\varepsilon}
\renewcommand{\phi}{\varphi}
\renewcommand{\geq}{\geqslant}
\renewcommand{\leq}{\leqslant}
\renewcommand{\ge}{\geqslant}
\renewcommand{\le}{\leqslant}
\renewcommand{\tilde}{\widetilde}
\newcommand{\from}{\colon}
\newcommand{\Vgeqthree}[1]{V_{\geq 3}(#1)}
\newcommand{\weight}{\bm{w}}
\newcommand{\touchgraph}[1]{\textup{Touch}( #1 )}
\newcommand{\objects}{\mathtt{OBJ}}
\newcommand{\vprev}[1]{\textrm{prev}\left( #1 \right)}
\newcommand{\vnext}[1]{\textrm{next}\left( #1 \right)}
\newcommand{\vlabel}[1]{\textrm{label}\left( #1 \right)}

\usepackage{todonotes}[disable]

\graphicspath{{figures/}}

\newcommand{\executeiffilenewer}[3]{%
\ifnum\pdfstrcmp{\pdffilemoddate{#1}}%
{\pdffilemoddate{#2}}>0%
{\immediate\write18{#3}}\fi%
} 
\newcommand{\Obj}{{\mathcal{D}}}

\newcommand{\Vorsepfam}{\mathcal{N}}

\newcommand{\Fam}{{\mathcal{F}}}

\newcommand{\dist}{\mathrm{dist}}

\newcommand{\yes}{\ensuremath{\text{\textsc{yes}}}}

\newcommand{\lrconn}[1]{\textrm{left-right-}#1 \textrm{-connected}}

\newcommand{\unitsquare}[1]{\textsf{UnitSquare}\left( #1 \right)}

\newcommand{\Block}[1]{\textsf{Block}\left( #1 \right)}

\newcommand{\WireGadget}[1]{\textsf{WireGadget}\left( #1 \right)}
\newcommand{\CrossingGadget}{\textsf{CrossingGadget}}

\newcommand{\TopGadget}{\textsf{TopGadget}}
\newcommand{\BottomGadget}{\textsf{BottomGadget}}
\newcommand{\StemGadget}[1]{\textsf{StemGadget}\left( #1 \right)}

\newcommand\nth{\textsuperscript{th}\xspace}

\begin{document}

\title{On Subexponential Parameterized Algorithms for Steiner Tree on Intersection Graphs of Geometric Objects}

\author{Sujoy Bhore\thanks{Department of Computer Science \& Engineering, Indian Institute of Technology Bombay, Mumbai, India.\\ Email: \href{sujoy@cse.iitb.ac.in}{sujoy@cse.iitb.ac.in}}
 \quad
 Bar{\i}\c{s} Can Esmer\thanks{CISPA Helmholtz Center for Information Security, Germany. Email: \href{baris-can.esmer@cispa.de}{baris-can.esmer@cispa.de}}
 \quad
 D\'aniel Marx\thanks{CISPA Helmholtz Center for Information Security, Germany. Email: \href{marx@cispa.de}{marx@cispa.de}}
 \quad
Karol W\k{e}grzycki\thanks{Max Planck Institute for Informatics, Saarbrücken, Germany. Supported by the Deutsche Forschungsgemeinschaft (DFG, German Research Foundation) grant number
559177164.\\ Email: \href{kwegrzyc@mpi-inf.mpg.de}{kwegrzyc@mpi-inf.mpg.de}
}}

\date{}
\begin{titlepage}
\def\thepage{}
\thispagestyle{empty}
\maketitle

\begin{abstract}
We study the Steiner Tree problem on the intersection graph of most natural families of geometric
objects, e.g., disks, squares, polygons, etc.   Given a set of $n$ objects in the plane and a subset $T$ of $t$
terminal objects, the task is to find a subset $S$ of $k$ objects such that the intersection graph of $S\cup T$ is connected. Given how typical parameterized problems behave on planar graphs and geometric intersection graphs, we would expect that exact algorithms with some form of subexponential dependence on the solution size or the number of terminals exist. Contrary to this expectation, we show that, assuming the Exponential-Time Hypothesis (ETH), there is no $2^{o(k+t)}\cdot n^{\Oh(1)}$ time algorithm even for unit disks or unit squares, that is, there is no FPT algorithm subexponential in the size of the Steiner tree. However, subexponential dependence can appear in a different form: we show that Steiner Tree can be solved in time $n^{\Oh(\sqrt{t})}$ for many natural
classes of objects, including:
\begin{itemize}
\item Disks of arbitrary size.
\item Axis-parallel squares of arbitrary size.
\item Similarly-sized fat polygons.
\end{itemize}
This in particular significantly improves and generalizes two recent results: (1) Steiner Tree on unit disks can be solved in time $n^{\Oh(\sqrt{k + t})}$ (Bhore, Carmi, Kolay, and Zehavi, Algorithmica 2023) and (2) Steiner Tree on planar graphs can be solved in time $n^{\Oh(\sqrt{t})}$  (Marx, Pilipczuk, and Pilipczuk, FOCS 2018).
We complement our algorithms with lower bounds that demonstrate that the class of objects cannot be significantly extended, even 
if we allow the running time to be $n^{o(k+t)/\log(k+t)}$.
\end{abstract}
\end{titlepage}

\newpage
\tableofcontents
\thispagestyle{empty}

\newpage
\setcounter{page}{1}
\section{Introduction}
The \emph{Square Root Phenomenon} in parameterized algorithms refers to the observation that
many natural algorithmic problems admit improved running times when restricted
to planar graphs or to 2-dimensional geometric objects, and often the best-possible running time (under standard complexity assumptions) involves a square
root in the exponent. More precisely, for fixed-parameter tractable problems in
general graphs, it is very common that $2^{O(k)}\cdot n^{O(1)}$ is the
best-possible running time, while for most natural W[1]-hard problems, $n^{O(k)}$ is the best we can achieve. On the other hand, running times of the form
$2^{\Ot(\sqrt{k})}\cdot n^{O(1)}$ appear for quite some number of planar and geometric problems (even for some problems that are W[1]-hard in general graphs!) and there are cases where the best possible running time becomes $n^{O(\sqrt{k})}$. There is a long line of research devoted to exploring this phenomenon on planar and geometric problems (see, e.g., \cite{DBLP:conf/focs/MarxPP18,ChitnisFHM20,DBLP:conf/focs/FominLMPPS16,DBLP:journals/talg/MarxP22,DBLP:conf/soda/KleinM14,DBLP:conf/icalp/KleinM12,DBLP:conf/icalp/Marx12,DBLP:conf/fsttcs/LokshtanovSW12,DBLP:journals/algorithmica/Verdiere17,DBLP:conf/stoc/Nederlof20a,cyganParameterizedAlgorithms2015,DBLP:journals/cj/DemaineH08,DBLP:journals/jacm/DemaineFHT05,DBLP:journals/siamdm/DemaineFHT04}).

\paragraph{Steiner Tree in planar graphs.} While these positive algorithmic
results may suggest that this is the normal and expected behavior of planar and
geometric problems, the curious case of Steiner Tree shows that we cannot take
this for granted. Given an edge-weighted graph $G$ and a set $T\subseteq V(G)$
of $t$ terminals, the task is to find a tree  of minimum total weight containing
every terminal. The Steiner Tree problem was investigated in many different
settings and from many different viewpoints \cite{ChitnisFHM20,eiben_et_al,DBLP:conf/stacs/DvorakFKMTV18,rajesh-andreas-pasin,khandekar,FeldmannM16,DBLP:journals/iandc/BermanBMRY13,DBLP:journals/siamdm/GuoNS11,feldman-ruhl,DBLP:journals/jal/CharikarCCDGGL99,winter1987steiner,hakimi,DBLP:journals/jacm/BateniHM11,DBLP:conf/soda/BateniCEHKM11,DBLP:conf/focs/MarxPP18,karp1972reducibility,ramanathan1996multicast,salama1997evaluation,Li1992267,Natu1997207,daniel-grid-tiling,levin}. In general graphs, a classic dynamic programming algorithm by Dreyfus and Wagner~\cite{DBLP:journals/networks/DreyfusW71} solves the problem in time $3^t\cdot n^{\Oh(1)}$, which can be improved to $2^t\cdot n^{\Oh(1)}$ using the technique of fast subset convolution~\cite{DBLP:conf/stoc/BjorklundHKK07}. Contrary to what one would expect in planar graphs, Marx, Pilipczuk, and Pilipczuk~\cite{DBLP:conf/focs/MarxPP18} showed that the running time cannot be improved to subexponential in planar graphs, assuming the Exponential-Time Hypothesis (ETH).
\begin{theorem}[\cite{DBLP:conf/focs/MarxPP18}]\label{th:planarlb}
Assuming ETH, there is no $2^{o(t)}\cdot n^{\Oh(1)}$ algorithm for Steiner Tree in planar graphs.
\end{theorem}  

 This shows that it is not evident that the search for planar/geometric
 subexponential parameterized algorithms should always be successful.

Despite this negative result, there are two ways in which subexponential running
times become relevant for the planar Steiner Tree. First, instead of parameterizing
by the number $t$ of terminals, we can parameterize by the total number of
vertices in the solution tree (including terminals), which can be much larger than the number of
terminals. Following the notation of other papers, we denote by $k$ the number of non-terminal vertices of the solution and we consider $k+t$ to be the parameter. 
The
pattern-covering technique of Fomin et
al.~\cite{DBLP:journals/siamcomp/FominLMPPS22} gives a subexponential parameterized algorithm for the problem with this parameterization.
\begin{theorem}[\cite{DBLP:journals/siamcomp/FominLMPPS22}]\label{th:planarsubexpfpt}
Steiner Tree in planar graphs can be solved in time $2^{\Ot(\sqrt{k+t})}\cdot
n^{O(1)}$, where $k$ is the number of non-terminal vertices of the solution and $t$ is the
number of terminals.
\end{theorem}
Another approach is to settle for a different form of improved running time. A
brute-force way of solving Steiner Tree is to guess at most $t-1$ branch
vertices of the solution tree in time $n^{\Oh(t)}$ and then complete them into a
solution by finding shortest paths. Of course, the $2^t\cdot n^{O(1)}$ time
algorithm is better than the $n^{O(t)}$ time brute-force algorithm. However, if
we improve the exponent to $\Oh(\sqrt{t})$, then this running time is
incomparable with $2^{t}\cdot n^{O(1)}$. Marx, Pilipczuk, and
Pilipczuk~\cite{DBLP:conf/focs/MarxPP18} showed that such a running time is indeed possible.
\begin{theorem}[\cite{DBLP:conf/focs/MarxPP18}]\label{th:planarsubexpxp}
Steiner Tree in planar graphs can be solved in time $n^{\Oh(\sqrt{t})}$, where $t$ is the number of terminals.
\end{theorem}
Note that, assuming ETH, Theorem~\ref{th:planarsubexpxp} cannot be improved to $n^{o(\sqrt{t}/\log t)}$: as $t\le n$, this would imply a $2^{o(\sqrt{n})}$ time algorithm for Steiner Tree on planar graphs, violating ETH~\cite{pilipczuk2013subexponential}.

\paragraph{Steiner Tree in geometric intersection graphs.} Parameterized
problems that become simpler on planar graphs often also become simpler on
(certain classes of) geometric intersection graphs. Perhaps the simplest and
the most widely-studied such graph class is the class of unit disk graphs, that is,
intersection graphs of unit disks in the 2-dimensional plane. Can we 
show an analog of~\cref{th:planarsubexpfpt} and show
that the Steiner Tree on the intersection graph of unit disks also admits
$2^{\Ot(\sqrt{k+t})} \cdot n^{O(1)}$ time algorithm? 
Surprisingly, we can show that the problem becomes much harder in this closely
related class.

\begin{theorem}\label{th:main-lowerbound}
Assuming ETH, there is no $2^{o(k+t)}\cdot n^{\Oh(1)}$ time algorithm solving Steiner Tree on the intersection graph of a given set of unit disks in the plane. The same holds for unit squares. 
\end{theorem}

For the proof of Theorem~\ref{th:main-lowerbound}, we first define the \ngtiling\ problem, which formalizes in a convenient way the hardness technique behind the proof of Theorem~\ref{th:planarlb}. However, due to the geometric nature of the problem, we need a lower bound for the variant \mongtiling, which requires a different proof. We believe that these clean lower bounds will be convenient starting points for other geometric hardness results.

By Theorem~\ref{th:main-lowerbound}, we cannot hope for an analog of Theorem~\ref{th:planarsubexpfpt} in unit
disk or unit square graphs. But is there an analog of
Theorem~\ref{th:planarsubexpxp}? Namely, is $n^{\Oh(\sqrt{t})}$ running time possible? Bhore et al.~\cite{DBLP:journals/algorithmica/BhoreCKZ23} presented a weaker algorithmic result, parameterized by the total number $k+t$ of objects in the solution.
\begin{theorem}[\cite{DBLP:journals/algorithmica/BhoreCKZ23}]\label{th:previousalgo}
Steiner Tree can be solved in time $n^{\Oh(\sqrt{k+t})}$ on the intersection graph
of a given set of unit disks in the plane, where $k+t$ is the size of the
solution (including the terminals).
\end{theorem}

We show that the running time can be improved to $n^{\Oh(\sqrt{t})}$, matching the running time of Theorem~\ref{th:planarsubexpxp} for planar graphs.
\begin{theorem}\label{th:unitdiskalg}
Steiner Tree can be solved in time $n^{\Oh(\sqrt{t})}$ on the intersection graph of a given set of unit disks
in the plane, where $t$ is the number of terminals.
\end{theorem}
Thus, we reach the somewhat unexpected conclusion that Theorem~\ref{th:planarsubexpfpt} does not have an analog in unit disk graphs, but Theorem~\ref{th:planarsubexpxp} has.

While Bhore et al.~\cite{DBLP:journals/algorithmica/BhoreCKZ23} use subdivision
techniques for Theorem~\ref{th:previousalgo} that inherently work only for
similarly sized objects, our algorithms rely on the much more flexible Voronoi
separator techniques of Marx and Pilipczuk~\cite{DBLP:journals/talg/MarxP22}. A
careful application of this technique allows us to handle geometric objects in a
much broader generality:
\begin{enumerate}[(1)]
  \setlength{\itemsep}{0pt}
  \setlength{\parskip}{0pt}
\item We can extend Theorem~\ref{th:unitdiskalg} to disks of arbitrary radii.
\item We can extend Theorem~\ref{th:unitdiskalg} to axis-parallel squares of arbitrary side length.
\item We can extend Theorem~\ref{th:unitdiskalg} to similarly sized fat polygons
    (i.e., there exists a constant $\alpha\ge 1$ such that each polygon has radius at most $\alpha$ and contains a unit-diameter disk).
\item We can combine these two types of objects in the following way: for some
    constant $\alpha\ge 1$, we allow a combination of:
    \begin{itemize}
      \setlength{\itemsep}{0pt}
  \setlength{\parskip}{0pt}
  \item    disks of radius at least 1 and
    \item polygons having diameter at most $\alpha$ and containing unit-diameter disk.  
    \end{itemize}
\item A similar combination is possible as the previous point, with axis-parallel squares instead of disks.
\end{enumerate}
Note that by the \emph{Koebe's Theorem}, every planar graph can be represented as the intersection graph of disks (of possibly different radii). Therefore, our main algorithmic result generalizes not only Theorem~\ref{th:previousalgo} of Bhore et al.~\cite{DBLP:journals/algorithmica/BhoreCKZ23}, but Theorem~\ref{th:planarsubexpxp} of Marx, Pilipczuk, Pilipczuk \cite{DBLP:conf/focs/MarxPP18} as well. Moreover, our algorithm is based on a novel application of a technique that was not used in either of these two previous works.

We would like to point out that the Voronoi separator framework
\cite{DBLP:journals/talg/MarxP22} is inherently about problems dealing with a \emph{small} number of \emph{disjoint} objects in the solution, such as in the case of Maximum Independent Set of Objects.
Note that our parameter for Steiner Tree is the number $t$ of terminals, and the
size of the solution may be unbounded in $t$.
One of the main technical novelties of our result is that we demonstrate that the Voronoi separator framework can be used even if the solution consists of an \emph{unbounded} number of \emph{highly overlapping} objects: we introduce the technique of representing the (potentially large and overlapping) solution by a skeleton consisting of a bounded number of disjoint paths. We consider this approach to be a new, potentially reusable technique.

Our reductions in Theorem~\ref{th:main-lowerbound} show that, assuming ETH, the exponent $\Oh(\sqrt{t})$ in our algorithms is essentially optimal. In fact, it cannot be significantly improved even if we consider the size of the solution as the parameter:
\begin{theorem}\label{th:unitdisklb}
Assuming ETH, Steiner Tree cannot be solved in time $n^{o(\sqrt{k+t}/\log (k+t))}$ on
the intersection graph of a given set of unit disks in the plane, where $k+t$ is
the size of the solution (including the terminals).
\end{theorem}
  
\paragraph{Lower bounds for other classes of objects.}
Finally, we show that the restrictions on the set of objects cannot be
significantly relaxed. The main lower bound applies for the case when terminals are points and the other objects are ``almost-squares'' of roughly the same size (in the proof, we are reusing a construction of Chan and Grant~\cite{chan2014exact}).
\begin{theorem}\label{th:almostsquarelb}
Assuming ETH, for any $\epsilon>0$, there is no $n^{o((k+t)/\log (k+t))}$ time algorithm
for Steiner Tree on the intersection graph of a given set of objects, where
every terminal is a point and every other object is an axis-parallel rectangle with side length in  $[1,1+\epsilon]$.
\end{theorem}
This lower bound limits the set of objects that can be considered in
two ways. First, by replacing each point with a sufficiently small
square, we obtain a lower bound for the case where every object is an
almost square (of arbitrary size). Thus, our result for squares of
arbitrary sizes cannot be further generalized, even by a small relaxation
of the aspect ratio.

\begin{corollary}\label{th:almostsquarelb2}
 Assuming ETH, for any $\epsilon>0$, there is no $n^{o((k+t)/\log (k+t))}$ time algorithm for Steiner Tree for  the intersection graph of a given set of axis-parallel rectangles, even if every rectangle has aspect ratio in $[1-\eps,1+\eps]$.
\end{corollary}
Corollary~\ref{th:almostsquarelb2} highlights the significance of disks and
squares in our algorithmic results: it is not only the fatness of the
objects that matters, but also the metric properties of disks and
squares. Therefore, it is unavoidable that our algorithms need to consider two different
classes of objects: similarly sized fat objects and disks/squares. It is in fact
surprising that these two classes can be handled by the same algorithm.

 A different way to look at Theorem~\ref{th:almostsquarelb} is to consider each
 terminal point to be a disk of appropriately small size and to consider the
 almost squares as similarly sized fat objects. This shows that when combining
 similarly sized fat objects and disks, it is essential to require that the size
 of disks is at least comparable to the size of the fat objects.
 
\begin{corollary}\label{th:almostsquarelb3}
 Assuming ETH, for any $\epsilon>0$, there is no $n^{o((k+t)/\log (k+t))}$ time algorithm for Steiner Tree for  the intersection graph of a given set of objects, even where each object is a disk or an axis-parallel rectangle with side length in $[1,1+\epsilon]$.
 \end{corollary}

\section{Technical Overview}
\label{sec:overview}

In this section, we formally present our main results, along with all
assumptions, outline the main stages of our algorithms, and discuss key proof ideas.

\subsection{Geometric problems}

Let us formally define the Steiner Tree problem on geometric objects. To ensure generality, we define a weighted version of the problem (and in later steps of the proof, it will be essential that we can create weighted instances).

\defoptproblem{\stgeo}{
    Set $\objects$ of connected objects in the plane with a positive weight function $w$, and a subset $T\subseteq \objects$.
}{
    A set $S \subseteq \objects$ such that the intersection graph of $S\cup T$ is connected
  }{Minimize the total weight of $S$}

  To make this problem definition precise, we need to specify how the set
  $\objects$ is represented in the input. We restrict our attention to simple polygons and disks. We consider the input size $|I|$ of the instance $(\objects,w,T)$ to be the total complexity of these objects, that is, the total number of vertices of the polygons plus the number of disks. To avoid dealing with degeneracies and numerical issues, we assume that an arrangement of $\objects$ is given in the input: a description of all intersection points, intersecting sides, and how they are connected.  Note that the total size of the arrangement is polynomial in the total complexity of the objects; such differences do not matter in our running times.

 As mentioned above, we need the requirement that the disks cannot be smaller than the similarly sized fat objects. For every $\alpha\ge 4$, we formalize this assumption as follows:

\assbox{\begin{assumptiona}
As set $\objects$ of geometric objects in the plane satisfies \assgeo\ if every object in $\objects$ is either:
\begin{itemize}
\item \textbf{(fat)} a simple polygon of diameter at most $\alpha/4$ and that contains a unit-diameter disk, or
  \item \textbf{(disk)} a disk of radius at least 1.
  \end{itemize}
\end{assumptiona}
}
\clearpage
Our main geometric result is the following:
\begin{theorem}[Main Result]\label{thm:maingeo}
  For every $\alpha\ge 4$, \stgeo with Assumption \assgeo\ can be solved in time $|I|^{\Oh_\alpha(\sqrt{|T|})}$.
\end{theorem}

In particular, Theorem~\ref{thm:maingeo} implies $|I|^{\Oh(\sqrt{|T|})}$ time algorithms for
unit squares, unit disks, and disks of arbitrary radii.

We remark that squares are in some sense very similar to disks: an axis-parallel square rotated 45 degrees can be described as a set of points at $L_1$-distance at most $r$ from a center. This can be used to show that the reduction from geometric problems to planar graph problems described in the next section also works for axis-parallel squares. Therefore, all our algorithmic results work if disks are replaced by axis-parallel squares.

\subsection{Translation to planar graphs}
\label{sec:geotoplanar}
The main technical tool we rely on is the Voronoi separator technique of Marx
and Pilipczuk \cite{DBLP:journals/talg/MarxP22}, which is formulated for planar
graphs. Therefore, at some point, we need to create a planar graph
representation of our geometric problem. To this end, we define the Steiner Tree
problem on objects that are connected subgraphs of a planar graph. There is one
important technicality: we consider two objects connected not only if they
intersect, but also if they are adjacent by an edge. The reason for this is that the
Voronoi separator technique applies to a pairwise nonintersecting collection of
connected objects. Therefore, we need to allow that pairwise disjoint objects
form a connected solution, which would be clearly impossible if connectivity
required direct intersection.

An {\em object} in a planar graph $G$ is a connected subset of vertices in $G$.
Two objects $O_1$ and $O_2$ in $G$ {\em touch} if they share a vertex or there
is an edge of $G$ between a vertex of $O_1$ and a vertex of $O_2$. The {\em
touching graph} of a set $\objects$ of objects in $G$, denoted by
$\touchgraph{\objects}$, has
$\objects$ as vertex sets and two objects are adjacent if and only if the two objects
touch in $G$. We refer to the set of objects in $\objects$ and the corresponding
vertices in their touching graph interchangeably.
A set $S$ of objects in $G$ is {\em connected} if $\touchgraph{S}$ is a connected graph.

We formulate our planar problem, to include a cardinality constraint in addition
to the optimization goal (if we want to ignore the cardinality constraint, we
may set $k=\infty$).

\defoptproblem{\stplanar}{
An edge-weighted planar graph $G$,  integer $k$,
   a set $\objects$ of connected objects in $G$ with a positive weight function $w$, and a subset $T\subseteq \objects$.
}{
    A set $S \subseteq \objects$ of cardinality
    at most $k$ such that $S\cup T$ is connected in $G$.
}{Minimize the total weight of $S$.}
\medskip

Given an instance $I=(G,k,\objects,w,T)$ of \stplanar, the size $|I|$ of the instance is the number
of vertices of the graph plus the total number of vertices of the objects in
$\objects$. Clearly, $|I|$ is polynomial in $|V(G)|$ and $|\objects|$. Formally,
the set $S\subseteq \objects$ (excluding the terminals) is the solution to the optimization problem \stplanar, but sometimes we informally refer to $S\cup T$ as the solution, as connectivity is required for this set.

 We would like to reduce an instance of \stgeo\ satisfying \assgeo\ to an
 instance of \stplanar. While we cannot solve \stplanar\ efficiently for
 arbitrary instances, we aim to exploit the fact that the original instance of \stgeo\ satisfies \assgeo. Therefore, we need some analog of \assgeo\ in the planar graph setting. Being a disk has a natural analog in planar graphs a {\em disk} (or {\em ball}) $B(v,r)$ in an edge-weighted graph $G$ is the set of vertices at distance at most $r$ from $v$. The diameter of a connected subgraph $S$ of a planar graph $G$ also has a natural interpretation: it is the largest distance between any two vertices of $S$ (note that here we allow that the shortest path between two vertices of $S$ goes outside $S$, i.e., we mean weak diameter).

 However, formalizing fatness by saying that some disk $B(v,\alpha)$ is fully contained in the object does not have the intended effect. Suppose that every object has a radius at most $\alpha$. Let some object $S$ be contained in the disk $B(v,\alpha)$. Let us attach a new pendant vertex $v'$ to $v$ and let $\alpha$ be the length of the edge $vv'$. Let us extend $S$ with this extra vertex $v'$; the new object $S'=S\cup\{v'\}$ intersects the same objects as $S$ does. Moreover, $S'$ is still in $B(v,\alpha)$ (hence has radius at most $\alpha$) and $S'$ contains the disk $B(v',\alpha)$. Repeating this for every vertex, we obtain an equivalent instance where every object has radius at most $\alpha$ and every object contains a disk of radius $\alpha$. This shows that the property of containing a disk of radius $\alpha$ does not bring any extra algorithmic advantage in the context of planar graphs.

A key insight in our algorithmic framework is to formalize fatness in planar graphs in a different, more abstract way. We want to capture an important property of similarly sized fat 
objects: while the input may contain a large clique of such objects, it is
unlikely that a solution will contain a large clique. Intuitively, if we remove the clique, the neighborhood cannot break into too many components (as there is no space for many independent objects in the neighborhood). As a result, a few objects of the clique are already sufficient to reconnect them.\footnote{Note that if $T$ has a large clique, then of course there is no way to avoid large cliques in the solution. However, we will make sure that this does not happen in our instances.} Inspired by this property, we formalize fatness by requiring that there exists an optimum solution where every ball of radius $2\alpha$ contains a bounded number of objects. Instead of imposing a constraint on the instance, we define an assumption on the optimum solutions.

Formally, for every $\alpha\ge 1$, we define the following assumption. There is a slight technicality about terminal objects overlapping or being fully contained in disks. For the precise handling of these types of issues, we make a very mild assumption about the existence of a function $\tau$ describing distinct representatives.

 \assbox{
   \begin{assumptionb}
A set $S\cup T$ of objects in an edge-weighted graph $G$ satisfies \assp\ if there is an $r\ge 4\alpha$ and an injective function $\tau: S\cup T\to V(G)$ with $\tau(O)\in O$ for every $O\in S\cup T$ such that the objects in $S\cup T$ can be partitioned into 
     \begin{itemize}
\item \textbf{(fat)} has diameter at most $\alpha$ or
\item \textbf{(disk)} is $B(v,r)$ for some $v\in V(G)$ with $\tau(B(v,r))=v$.
\end{itemize}
Moreover, this partition satisfies:
  \begin{itemize}
	\item For every $v\in V(G)$, the ball $B(v,4\alpha)$ intersects at most
		$1000\alpha^2$ fat objects in $S\cup T$.
  \end{itemize}
\end{assumptionb} 
}

We present a reduction that transforms an instance of \stgeo\ satisfying \assgeo\ into an instance of the planar problem with optimum solutions satisfying \assp. Note that we can assume by scaling that the instance has integer weights; thus, if the optimum changes by less than $1/2$, then it is possible to recover the original optimum.
\begin{restatable}{lemma}{geotoplanar}\label{lem:geotoplanar}
  For every fixed $\alpha\ge 4$, there is a polynomial-time algorithm that transforms an instance $(\objects,T)$ of \stgeo with assumption \assgeo\ to an instance $(G',\infty,\objects',T')$ of \stplanar with $|T'|\le |T|$ such that the optimum value of the two instances differ by less than $1/2$ and every optimum solution of the new instance satisfies \assp.
\end{restatable}
Note that the reduction creates an instance where the cardinality constraint is $\infty$, i.e., we are ignoring the cardinality constraint. However, in a later step of the algorithm, it will be important to consider instances where the number of objects in the solution is bounded.

  Let us formulate now our main result in the language of the \stplanar problem.
  \begin{theorem}{\label{lem:mainplanar}}
  For every $\alpha\ge 1$, if there is an optimum solution of an instance of \stplanar satisfying Assumption \assp, then an optimum solution can be found in time $|I|^{O_\alpha(\sqrt{|T|})}$.
\end{theorem}
Using Lemma~\ref{lem:geotoplanar}, Theorem~\ref{lem:mainplanar} implies Theorem~\ref{thm:maingeo}.

The natural approach to prove Lemma~\ref{lem:geotoplanar} is to let $G'$ be the graph of the arrangement of the objects in $\objects$ (with edge length defined as the geometric distance of the endpoints), and for each object $O\in \objects$, introduce into $\objects'$ the set $O'\subseteq V(G')$ of vertices that is contained in $O$. However, there are three issues here:
\begin{itemize}\itemsep0pt
\item The set $O'$ is not necessarily connected if an object $O^*$ is fully contained in $O$. However, this can be solved by additional edges that make the graph connected.

\item Let $O\in\objects$ be a disk of radius $r$ and center $v$, and let $O'=B(v,r)$ be the corresponding object in $G'$. It is not clear at all whether some vertex $v'$ of the arrangement is in $O'$, then $v'$ is in $O$ as well. The problem is that even though the distance of $v$ and $v'$ is at most $r$ in the plane, it is possible that the shortest $v-v'$ path in the arrangement is longer than $r$. Therefore, it is possible that two disks $O_1$ and $O_2$ intersect in the plane, but the corresponding objects $B(v_1,r)$ and $B(v_2,r)$ do not intersect. In order to avoid this situation, we introduce additional lines and intersections into the arrangement to make sure that if a disk in the plane intersects an object, then the corresponding disk $B(v,r)$ reaches a vertex of that object. 
\item Even if a fat object $O$ has diameter at most $\alpha$, the corresponding vertex set $O'$ may not have bounded diameter in the graph of the arrangement of the objects: it could be that the shortest path between two vertices $x,y\in O'$ goes around the perimeter of $O$. To solve this, we introduce additional edges and intersections to make sure that two vertices of $O'$ cannot be too far from each other.

\end{itemize}
  The second part of the proof of Lemma~\ref{lem:geotoplanar} is to argue that Assumption \assp\ holds. The argument is similar to the argument why no large cliques appear in the solution.
  However, here is the first point where it is essential that the disk objects are not very small: otherwise, a large clique may be needed to cover many small disk terminals. Furthermore, we need to preprocess the instance such that the terminals are independent.

  When proving Lemma~\ref{lem:geotoplanar} in Section~\ref{sec:geotoplanarprove}, we show that the reduction works even if we have axis-parallel squares instead of disks. Therefore, our algorithmic results carry over to axis-parallel squares as well.

\subsection{Reducing the number of objects in the solution}

The size of the solution $S$ can be, of course, much larger than the number $t=|T|$ of terminals: if the terminals are far from each other, then they may be connected by long chains of intersecting objects. These chains may meet and overlap each other, but our next crucial insight is that such interactions are somewhat rare: there are only $O(t)$ objects that participate in such interactions, and the remaining objects are inside chains of objects (see Figure~\ref{fig:degree3}). Therefore, the solution can be described by $O(t)$ objects and $O(t)$ independent chains connecting these objects. The following combinatorial lemma shows that this is indeed possible:
\begin{figure}
		\centering
		\includegraphics[width=0.4\textwidth]{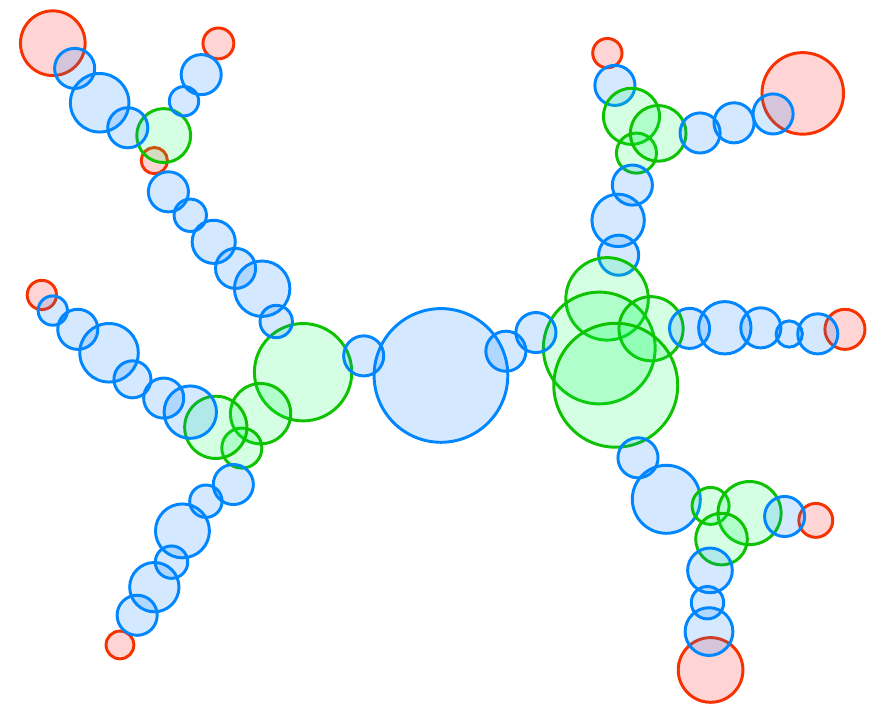}
		\caption{A minimal solution with 10 terminals and 15 objects with degree at least 3.}
		\label{fig:degree3}

\end{figure}  

\begin{restatable}{theorem}{criticalmain}\label{th:criticalmain}
Let $G$ be a connected graph and $T$ be a subset of vertices with $|T|\ge 2$.
Suppose that, for any $x\in V(G)\setminus T$, the graph $G-x$ has at least two
connected components containing vertices from $T$. Then $G$ has $O(|T|)$
vertices of degree at least three.
\end{restatable}

\begin{figure}[t]
    \centering
    \includegraphics{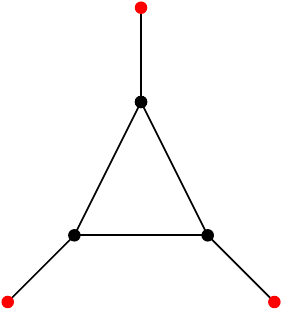}
    \caption{An example of a graph $G$ with the leaves $T$ depicted in red, such that $\abs{T} = 3$ and $\Vgeqthree{G} = 3 = 3 \cdot \left( \abs{T} - 2\right)$. It is easy to verify that $T$ is a critically connected set.}
    \label{fig:structural}
\end{figure}

Notice that $G$ in~\cref{th:criticalmain} is not necessarily a tree (see Figure~\ref{fig:structural} for an example), but thinking about $G$ as a tree is a useful intuition as it has $O(|T|)$ vertices of degree at least three.

We enrich the set $\objects$ by new objects in a way that allows us to represent each long chain with a single object. For every $x,y\in V(G)$, let us compute the cheapest connected set $P_{xy}$ of objects covering both $x$ and $y$ (this can be computed by an appropriate shortest path computation in the touching graph of $\objects$). Let us introduce an object $S_{xy}$ that is an arbitrary $x-y$ path in $\bigcup P_{xy}$. Let us set the cost of $S_{xy}$ to be the total cost of $P_{xy}$. Observe that adding such objects does not change the cost of the optimum solution: if $S_{xy}$ appears in the solution, then it can always be replaced by the set $P_{xy}$ of objects without increasing the cost. Thus, if such objects are available, then we can replace each of the $O(|T|)$ chains with a single object, ensuring that there is an optimum solution consisting of only $O(|T|)$ objects.

The introduction of these objects may ruin the property \assp\ of the solution: in particular, it may no longer be true that every non-disk object has diameter $\alpha$. However, if we perform the replacement of chains with such ``long'' objects in a careful way, then we can ensure that these new objects are completely disjoint from every other object in the solution. To take into account this third type of object, we define the following assumption.

\pagebreak
\assbox{
  \begin{assumptionc}
    A set $S\cup T$ of objects in an edge-weighted graph $G$ satisfies \asspl\ there is an $r\ge 4\alpha$ and an injective function $\tau:S\cup T\to V(G)$ with $\tau(O)\in O$ for every $O\in S\cup T$ such that the objects in $S\cup T$ can be partitioned into
     \begin{itemize}
\item \textbf{(fat)} has diameter at most $\alpha$,
\item \textbf{(disk)} is $B(v,r)$ for some $v\in V(G)$ with $\tau(B(v,r))=v$, and
\item \textbf{(long)} is an arbitrary connected subset.
\end{itemize}
Moreover, this partition satisfies:
  \begin{itemize}
  \item Every long object in $S\cup T$ is disjoint from every other object in $S\cup T$.
  \item For every $v\in V(G)$, the ball $B(v,4\alpha)$ intersects at most $1000\alpha^2$ fat objects in $S\cup T$.
  \item For every long or fat object $L\in S\cup T$, there are at most $1000\alpha^2$ fat objects in $S\cup T$ at distance at most $\alpha$ from $L$. 
  \end{itemize}
 \end{assumptionc} 
}

The last point is an additional technical requirement, which can be satisfied by making sure that if a fat object $O$ is close to a long object $L$, then $O$ is close to one of the endpoints of $L$. We ensure this requirement by carefully splitting long objects whenever necessary. The following lemma formalizes that solutions satisfying \assp\ can be transformed into solutions satisfying \asspl.

 \begin{lemma}\label{lem:longred}
   For every fixed $\alpha>0$, there is a polynomial-time algorithm that transforms
   an instance $(G,\infty,\objects,T)$ of \stplanar that has an optimum solution with assumption \assp\ into an equivalent instance $(G',k', \objects',T')$ of \stplanar that has an optimum solution with assumption \asspl\ such that $|T'|=|T|$ and  $k'=O_\alpha(|T|)$ hold. 
\end{lemma}

In the following, our goal is to solve an instance of \stplanar\ where some optimum solution satisfies
\asspl. Importantly, now the exponent of the running time can depend on the cardinality of the solution: we have created instances where this can be assumed to be linear in the number of terminals. We formalize this goal in the following theorem.

\begin{theorem}{\label{lem:smallplanaralg0}}
  For every $\alpha\ge 1$, if there is an optimum solution of an instance of \stplanar satisfying Assumption \asspl, then an optimum solution can be found in time $|I|^{O_\alpha(\sqrt{k+|T|})}$.
\end{theorem}

\subsection{Voronoi separators}
Marx and Pilipczuk~\cite{DBLP:journals/talg/MarxP22} present two versions of their main result on enumerating Voronoi separators: a simpler version (Lemma 2.1 in \cite{DBLP:journals/talg/MarxP22}) that considers disjoint sets of objects, and a more powerful version (Theorem 4.22 in \cite{DBLP:journals/talg/MarxP22}) that can handle, in particular, disks of arbitrary radii that may intersect other objects and each other. It may seem that the more powerful version is precisely the right tool for our application, as 
overlapping disks are also possible in a solution for the Steiner Tree problem. However, even this more powerful version does not handle overlapping fat objects; hence, we need to obtain a disjoint representation of the solution anyway. This is a technical detail that we will discuss shortly in~\cref{sec:disjointrep} (see Lemma~\ref{lem:createindep}). As we can obtain this disjoint representation even for overlapping disks, there is no real gain in using the more powerful version. Therefore, we use the simpler Lemma 2.1 from \cite{DBLP:journals/talg/MarxP22}, which requires considerably simpler definitions and involves much fewer technical details.

Let $G$ be an edge-weighted graph and $\Obj$ be a set of $d$ objects (connected subsets of vertices). A {\em guarded
  separator} is a pair $(Q,\Gamma)$ consisting of a set
$Q\subseteq \Obj$ of objects and a subset
$\Gamma\subseteq V(G)$ of vertices. We assume that distances are unique in $G$, that is, $\dist_G(a,b)\neq \dist_G(c,d)$ if $\{a,b\}\neq \{c,d\}$. This can be achieved by slightly perturbing the weights of the edges. 

\begin{theorem}[Lemma 2.1 in \cite{DBLP:journals/talg/MarxP22}]\label{th:guardedenum0}
  Let $G$ be an edge-weighted $n$-vertex planar graph, $\Obj$ a set of
  $d$ connected subsets of $V(G)$, and $k$ an integer. We can
  enumerate (in time polynomial in the size of the output) a set
  $\Vorsepfam$ of $d^{\Oh(\sqrt{k})}$ pairs $(Q,\Gamma)$ with
  $Q\subseteq \Obj$, $|Q|=\Oh(\sqrt{k})$, $\Gamma\subseteq V(G)$ such
  that the following holds. If $\Fam\subseteq \Obj$ is a set of $k$
  pairwise disjoint objects, then there is a pair $(Q,\Gamma)\in
  \Vorsepfam$ such that
\begin{enumerate}\itemsep0pt
\item[(a)] $Q\subseteq \Fam$,
\item[(b)] for every $v\in \Gamma$ and  $p\in \Fam\setminus Q$, we have $\dist_G(p,v)>\min_{p'\in Q}\dist_G(p',v)$.
\item[(c)] for every connected component $C$ of $G-\Gamma$, there are at most
  $\frac{2}{3}k$ objects of $\Fam$ that are fully contained in $C$.
\end{enumerate}
\end{theorem}
Theorem~\ref{th:guardedenum0} is typically used in the following way. The set $\Fam$ of disjoint objects is an unknown solution to a problem. We enumerate the collection $\Vorsepfam$ of pairs described in the Theorem, guess one of these pairs $(Q,\Gamma)$, and assume that it satisfies items (a)--(c). In particular, by item (a), we guessed that $Q$ is part of the solution. Item (b) implies that no object of $\Fam\setminus Q$ intersects $\Gamma$: otherwise, the distance of that object would be zero to a vertex $v$ of $\Gamma$, strictly larger than the distance of any object in $Q$ to $v$. Therefore, depending on the problem, it could be possible to recursively solve subproblems, where each subproblem consists of $Q$ and the objects in a component of $G-\Gamma$. Item (c) shows that the number of objects in the solution for these subproblems is a constant factor lower than in the original problem.

An algorithm using Theorem~\ref{th:guardedenum0} can be imagined as a branching algorithm that branches into $n^{O(\sqrt{k})}$ directions in each step of the recursion, where $k$ is the solution size of the current subproblem. The size of the branching tree can be bounded by the product of the number of directions in each level. Since the size $k$ of the solution decreases by a constant factor in each level of the recursion,  the exponent of the running time can be bounded by a geometric series starting with $O(\sqrt{k})$. That is, the running time is $n^{O(\sqrt{k})}$. 

We need a version of Theorem~\ref{th:guardedenum0} that is stronger in two ways:
\begin{itemize}\itemsep0pt
\item Instead of a balance condition on the components of $G-\Gamma$, we want to have a balanced bipartition of $V(G)\setminus\Gamma$. A {\em guarded separation} is a tuple $(Q,\Gamma,A,B)$ where $(\Gamma,A,B)$ is a partition of $V(G)$ such that there is no edge between $A$ and $B$. In item (c), we want to bound the number of objects fully contained in $A$ and the number of objects fully contained in $B$.
  \item Instead of giving a balance condition on the number of objects from $\Fam$, we need to satisfy the balance condition with respect to an unknown subset $\Fam_0\subseteq\Fam$.
  \end{itemize}
  We can prove these strengthening of Theorem~\ref{th:guardedenum0}, with a
  slight loss in the balance condition. To achieve the first property, we
  observe that given the components of $G-\Gamma$, we can enumerate a bounded
  number of bipartitions $(A,B)$ of these components such that one of them
  results in a balanced separator. For the second property, we slightly modify
  the graph $G$ and the set $\Obj$ of objects, and then whenever we want to have
  a separation of $\Fam$ that is balanced with respect to the objects in
  $\Fam_0$, we invoke the result for the set $\Fam'\supseteq \Fam$ that contains
  some number $c$ additional objects close to each object in $\Fam_0$. Thus
  whether a separation is balanced with respect to $\Fam'$ mainly depends on
  whether it is balanced with respect to these additional objects, hence in the
  end depends on whether it is balanced with respect to the objects in $\Fam_0$.
  
\begin{restatable}{theorem}{guardedenum}\label{th:guardedenum1}
  Let $G$ be an edge-weighted $n$-vertex planar graph, $\Obj$ a set of
  $d$ connected subsets of $V(G)$, and $k$ an integer. We can
  enumerate (in time polynomial in the size of the output) a set
  $\Vorsepfam$ of $(d+n)^{\Oh(\sqrt{k})}$ guarded separations $(Q,\Gamma,A,B)$ with
  $Q\subseteq \Obj$, $|Q|=\Oh(\sqrt{k})$ such
  that the following holds. If $\Fam\subseteq \Obj$ is a set of $k$
  pairwise disjoint objects and $\Fam_0$ is a subset of $\Fam$, then there is a tuple $(Q,\Gamma,A,B)\in
  \Vorsepfam$ such that
\begin{enumerate}
\item[(a)] $Q\subseteq \Fam$,
\item[(b)] for every $v\in \Gamma$ and  $p\in \Fam\setminus Q$, we have $\dist_G(p,v)>\min_{p'\in Q}\dist_G(p',v)$.
\item[(c)] there are at most
  $\frac{3}{4}|\Fam_0|$ objects of $\Fam_0$ that are fully contained in $A$, and
  there are at most
  $\frac{3}{4}|\Fam_0|$ objects of $\Fam_0$ that are fully contained in $B$.
\end{enumerate}
\end{restatable}

\subsection{Recursion}
\begin{figure}
		\centering
		\includegraphics[width=0.4\textwidth]{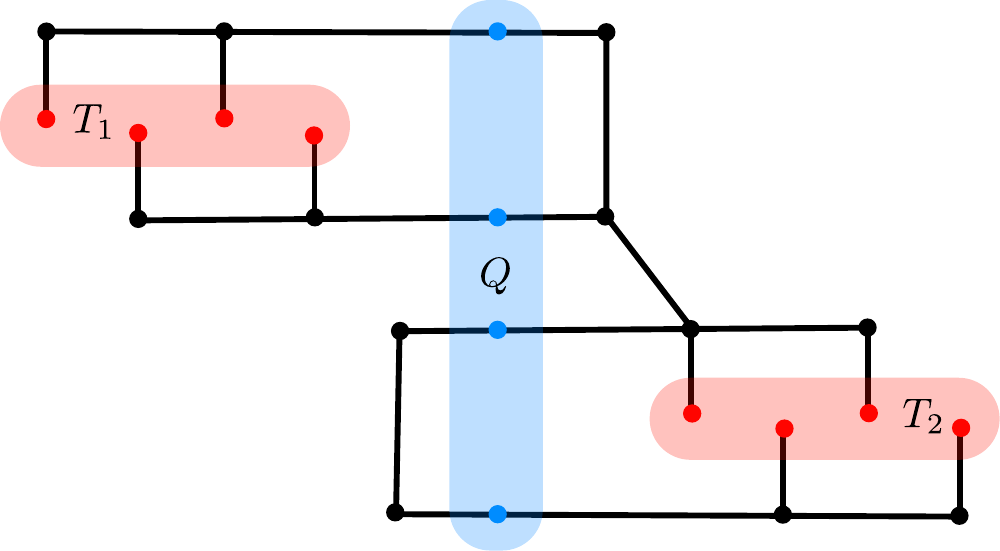}
		\caption{A solution with a separator $Q$ that breaks the solution into two disconnected parts.}
		\label{fig:connissue1}

\end{figure}

To facilitate recursions, we need to consider a slightly more general problem. In the recursion step, we guess a set $Q$ of objects in the solution that breaks the solution into two parts, one part containing terminals $T_1$, and the other part containing terminals $T_2$. It is now tempting to try to solve the problem recursively, with the first subproblem finding a tree connecting $Q\cup T_1$, and the second subproblem finding a tree connecting $Q\cup T_2$. While the union of these two solutions would certainly connect $Q\cup T_1\cup T_2$, it is not necessarily true that an optimum solution arises this way. It is possible that in the optimum solution, the first part does not provide a full connection of $Q\cup T_1$, the second part does not provide a full connection of $Q\cup T_2$, but the solution still fully connects $Q\cup T_1\cup T_2$ (see Figure~\ref{fig:connissue1}).

Intuitively, when solving a subproblem, we should assume that there is already some kind of connection on $Q$ (provided by another subproblem), and we want to achieve full connection of $Q$ and the terminals. Therefore, we define the following extension of \stplanar, where the input contains a distinguished set $X$ of terminals and a ``provided connectivity'' on $X$, described by a graph $F$. 

\defoptproblem{\stplanarx}{
An edge-weighted planar graph $G$,  integer $k$, 
    set $\objects$ of connected objects in $G$ with a positive weight function, subsets $X\subseteq T\subseteq \objects$, and a forest $F$ with vertex set $X$.
}{
    A set $S \subseteq \objects$ of cardinality
    at most $k$ such that the union of $\touchgraph{S\cup T}$ and $F$ is connected in $G$.
}{Minimize the total weight of $S$.}

To prove~\cref{lem:smallplanaralg}, we will focus on establishing the following
theorem.
\begin{theorem}{\label{lem:smallplanaralg}}
  For every $\alpha\ge 1$, if there is an optimum solution of an instance of \stplanarx satisfying Assumption \asspl, then an optimum solution can be found in time $|X|^{|X|}|I|^{O_\alpha(\sqrt{k+|T|})}$.
\end{theorem}

We formalize now the properties of a triple $(T_1,T_2,Q)$ that allows us to execute the recursion.

\begin{definition}\label{def:triplenew}
  Consider an instance $(G,k,\objects,T,X,F)$ of \stplanarx and let $(T_1,T_2,Q)$ be a triple where $Q\subseteq \objects$ and $(T_1,T_2)$ is a partition of $T\setminus Q$. We say that $(T_1,T_2,Q)$ is a {\em $\beta$-balanced triple} of a solution $S$ if $Q\subseteq S\cup T$ and there is a bipartition $(A,B)$ of $(S\cup T)\setminus Q$ such that 
  \begin{itemize}
  \item $A\cap T=T_1$ and $B\cap T=T_2$,
  \item $|A|,|B|
\le \beta|S\cup T|$, and
    \item every object in $T$ is in the same component of the graph $H$ defined by removing every $A-B$ edge from $\touchgraph{S\cup T}$ and then adding the edges of $F$.
  \end{itemize}
\end{definition}

We show that, given access to a balanced triple, we can reduce the problem to
``smaller'' subproblems. However, there is a technicality related to Assumption
\asspl\ that needs to be addressed. Ideally, we would like to say that if every optimum solution of the original instance satisfies \asspl, then every optimum solution of every subproblem also satisfies \asspl\ as well. However, it is both cumbersome and unnecessary to prove this. In light of the way Theorem~\ref{lem:smallplanaralg} is stated, it is sufficient to say that if there is a ``nice'' solution, then we return an arbitrary solution that is at least as good. Formally, we say that an algorithm {\em weakly solves instance $I$ of \stplanarx\ under Assumption \asspl} if it returns a solution that is not worse than any solution of $I$ satisfying Assumption \asspl. The main recursion step can be formulated in the following way:

\begin{restatable}{lemma}{recursion}\label{lem:recursion0}
Let $I=(G,k,\objects,T,X,F)$ be an instance of \stplanarx. If a $\beta$-balanced triple $(T_1,T_2,Q)$ is given for a solution $S$ that has minimum cost among those satisfying Assumption \asspl, then in time $2^{O(|Q\cup X|\log |Q\cup X|)}\textup{poly}(|I|)$ we can reduce weakly solving $I$ under Assumption \asspl\ to weakly solving $k\cdot 2^{O(|Q\cup X|\log |Q\cup X|)}$ instances under Assumption \asspl, where each such instance $(G,k',\objects,T',X',F')$ satisfies $|T'|\le |T\cup Q|$, $|X'|\le |X\cup Q|$, $k'\le \beta k$.
\end{restatable}

\begin{figure}
		\centering
		\includegraphics[width=0.5\textwidth]{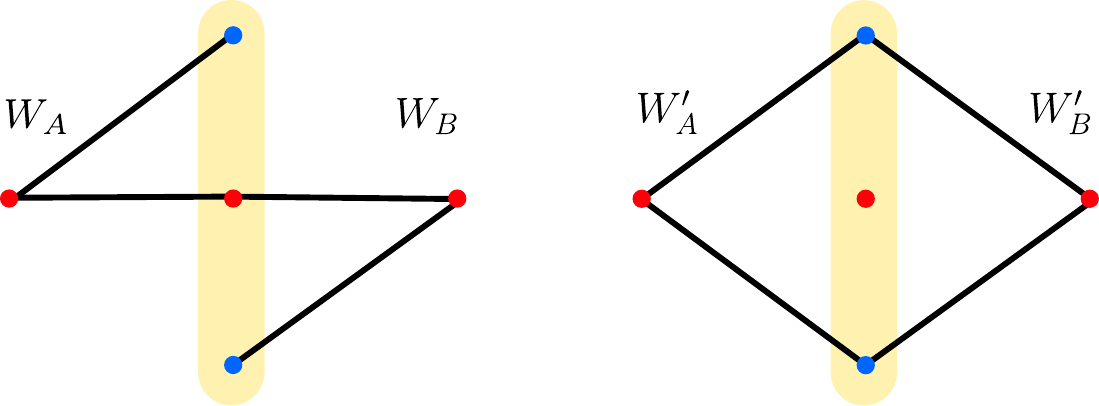}
		\caption{Replacing two parts of the solution by alternative solutions that are compatible with the original other part may not result in a solution connecting all three red terminals. }
		\label{fig:connissue2}

\end{figure}

The proof of Lemma~\ref{lem:recursion0} uses the definition of a
$\beta$-balanced triple in a fairly straightforward way. However, we have to be
careful not to follow a natural approach that does not work: Suppose that
splitting the optimum solution at the separator creates two forests, $W_A$ and
$W_B$. Let $F_A$ (resp., $F_B$) be a forest representing how the components of
$W_A$ (resp., $W_B$) connect the vertices of $Q$. The natural approach would be
to guess these forests $F_A$ and $F_B$, move $Q$ into the set $X'$ of
distinguished terminals, and solve two subproblems: in the first, we need to
connect $T_1\cup Q$ assuming connectivity $F_B$ on $Q$ is present, and in the
second, we need to connect $T_2\cup Q$, assuming connectivity $F_A$ on $Q$ is
present. However, somewhat counterintuitively, if $W'_A$ and $W'_B$ are
solutions of these two instances, respectively, then it is not necessarily true that $W'_A\cup W'_B$ really connects $Q\cup T_1\cup T_2$: see Figure~\ref{fig:connissue2} for an example.

Therefore, the proof of Lemma~\ref{lem:recursion0} proceeds in a slightly different way. First, we guess the connectivity $F_B$ provided by $W_B$. Then we solve a subproblem where 
we need to connect $T_1\cup Q$ assuming connectivity $F_B$ on $Q$ is present; let $W'_A$ be the solution obtained this way. Let $F_A$ be a forest representing how the components of $W'_A$ connect the vertices of $Q$, and let us solve the subproblem where we need to connect $T_2\cup Q$ assuming connectivity $F_A$ on $Q$ is present; let $W'_B$ be the resulting solution. Now $W'_A\cup  W'_B$ is indeed a solution and we can argue that $|W'_A|\le |W_A|$ and $|W'_B|\le |W_B|$, hence $W'_A\cup  W'_B$ is an optimum solution.

The last ingredient in our algorithmic framework is showing that we can list $|I|^{O_\alpha(\sqrt{k+|T|})}$ candidates for the balanced triple $(T_1,T_2,Q)$.
For technical reasons, the following lemma has two very mild conditions.  We say that a set $T$ of objects is {\em irredundant} if  for every $O\in T$, the vertex set $O\setminus \bigcup_{O'\in T, O\neq O'} O'$ is nonempty. Note that if $O\subseteq \bigcup_{O'\in T, O\neq O'} O'$ for some $O\in T$, then removing $O$ from $T$ does not change the problem at all (besides decreasing the cost by the cost of $O$). Indeed, removing $O$ cannot disconnect the solution of the original instance, and adding $O$ to the solution of the new instance maintains connectivity. We say that a solution $S$ is {\em inclusionwise minimal} if no proper subset of $S$ is a solution; clearly, it is sufficient to restrict our attention to inclusionwise minimal solutions.

\begin{restatable}{lemma}{listtriples}\label{lem:listtriples}
  Given an instance $(G,k,\objects,T,X,F)$ where $T$ is irredundant, in time
  $|I|^{O_\alpha(\sqrt{k+|T|})}$ we can list a set of
  $|I|^{O_\alpha(\sqrt{k+|T|})}$ triples $(T_1,T_2,Q)$ such that $|Q| =
  O_\alpha(\sqrt{k+|T|})$ and for every inclusionwise minimal solution $S$ satisfying Assumption \asspl\ at least
  one of them is a $\frac{3}{4}$-balanced triple.
\end{restatable}

In the proof of Lemma~\ref{lem:listtriples}, we first invoke Lemma~\ref{lem:createindep} to argue that there is a disjoint representation $W$ of the objects in the hypothetical solution $S\cup T$. Theorem~\ref{th:guardedenum1} lists a set of guarded separations $(Q,\Gamma)$, one of them being balanced for this disjoint representation $W$. We need to turn guarded separations $(Q,\Gamma)$ for the objects in $W$ into balanced triples $(T_1,T_2,Q')$. There is an intuitive way of doing this:
let $Q'$ contain an object of $S\cup T$ if a corresponding object of $W$ is in $Q$, and let us define $T_1$ and $T_2$ somehow based on the partition $(A,B)$. However, there are some technical challenges. Recall that in the disjoint representation $W$, multiple objects could correspond to a single object $S\cup T$. Therefore, being balanced with respect to the objects in $W$ does not necessarily mean being balanced with respect to the objects in $S\cup T$. To solve this issue, we use that Theorem~\ref{th:guardedenum1} can guarantee balance with respect to a subset of the objects: we define a subset $W^*$ containing one object of $W$ corresponding to each object of $S\cup T$, and require balance with respect to $W^*$.

An additional fundamental issue is that two objects of $W$ could be in different components of $G-\Gamma$, even if they both correspond to the same object of $S\cup T$. For disk objects in $S\cup T$, Lemma~\ref{lem:createindep} guarantees that the corresponding objects form a connected set. This means that if these objects are in two different components of $G-\Gamma$, then one of them intersects $\Gamma$. This means that the object intersecting $\Gamma$ is in $Q$, and hence the disk object was selected into the separator $Q'$. However, for a fat object $O$ of $S\cup T$ we have no such guarantee: the corresponding set $W_O$ of objects in $W$ is potentially not connected. This is the point where the properties of Assumption \asspl\ come into play. If $W_0$ contains objects from two different components of $G-\Gamma$, then the diameter property of $O$ implies that one object of $W_0$ is close to a vertex $v$ of $\Gamma$. Now item (b) of Theorem~\ref{th:guardedenum1} implies that there is an object in $Q$ that is even closer to $v$. This means that $O$ has to be an object close to some object in $Q'$. Therefore, we can avoid the issue if we extend $Q'$ with every object that is close to some object in $Q'$. The conditions of Assumption \asspl\ imply that for each object in $Q'$, we introduce only $O_\alpha(1)$ additional objects into $Q'$. 

\subsection{Disjoint representation of the solution}\label{sec:disjointrep}
\begin{figure}
		\centering
		\includegraphics[width=0.5\textwidth]{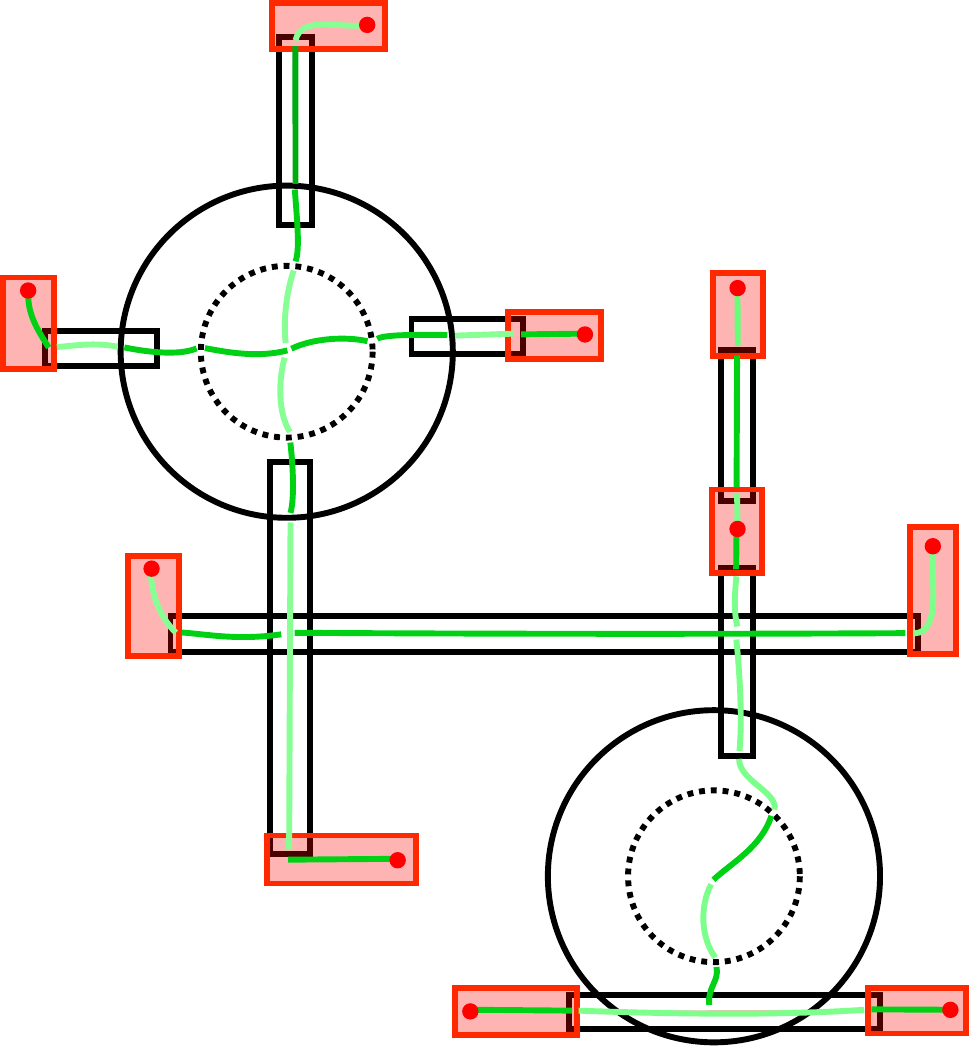}
		\caption{Representing a solution (red terminals and black non-terminals) with disjoint objects (green).}
		\label{fig:pathization}
              \end{figure}

The Voronoi separator tools that we would like to use from the work of Marx and Pilipczuk \cite{DBLP:journals/talg/MarxP22} are applicable to disjoint subsets of connected objects in graphs. However, in general, the objects of the solution in the Steiner Tree problem are not disjoint; thus, it seems that those tools, as stated, are not applicable in our case. Nevertheless, we show that, by introducing new objects (shortest paths inside objects), we can create a representation of the solution that consists of disjoint objects. This representation is then used to argue that certain balanced separators exist.
To ensure disjointness of the objects, each original object is potentially represented by multiple new objects (see Figure~\ref{fig:pathization}). 

\begin{restatable}{lemma}{createindep}\label{lem:createindep}
   Let $\objects'$ contain
  \begin{itemize}
  \item every object in $\objects$,
  \item for every object $O \in \objects$ and every $x,y\in V(G[O])$ a
      shortest path between $x$ and $y$ in $G[O]$.
  \end{itemize}
  Let $S\cup T$ be a solution of \stplanarx satisfying \asspl and let $\tau: S\cup T \to
  V(G)$ be the injective function defined in this assumption.  Then there is a subset $W\subseteq \objects'$
of size $O(|S| + |T|)$ and a mapping $\pi: W \to S\cup T$ such that
  \begin{enumerate}
    \item\label{lem:createindep:1} if $t_1,t_2\in T$ are in the same component
        of $\touchgraph{S\cup T}$, then $\tau(t_1)$ and $\tau(t_2)$ are covered by one component of $\touchgraph{W}$.
    \item\label{lem:createindep:2} objects in $W$ are pairwise disjoint.
    \item\label{lem:createindep:3} if $\pi(R')$ is long, then $R'=\pi(R')$.
    \item\label{lem:createindep:4} if $\pi(R')$ is fat, then $R'$ is a shortest
        path in $G[\pi(R')]$.
    \item\label{lem:createindep:5} if $\pi(R')$ is a disk $B(v,r)$, then $R'$ is
         either a shortest path in $G[B(v,r)\setminus B(v,r-3\alpha)]$, or
        a shortest path of length $3\alpha$ in  $G[B(v,r-3\alpha)]$.
  \item\label{lem:createindep:6} for every $R\in S\cup T$ that is a disk, the
      set $\pi^{-1}(R)$ is connected and $\{\tau(R)\}\in \pi^{-1}(R)$.
  \item\label{lem:createindep:tau} $\{\tau(R)\} \in W$ for every fat object $R \in S\cup T$ (but it is not necessarily true that $\pi(\{\tau(R)\})=R$).
  \item\label{lem:createindep:emptyfat}  if $O\in S\cup T$ is a fat object that is a subset of a disk object in $S\cup T$, then $\pi^{-1}(O)=\emptyset$.
\end{enumerate}    
\end{restatable}
Note that for each disk object, we require that its representatives form a
connected set, but we do not require this for fat objects. In fact, it would be
impossible to require that two crossing objects are represented in a way that
the representatives of each object are connected. The proof of
Lemma~\ref{lem:createindep} is an iterative process. We first identify the vertices
that need to be connected together to ensure global connectivity. Then we
iteratively add new objects from $\objects'$ to $W$, independent from those already in $W$, to ensure connectivity of these vertices. To ensure the connectivity of the objects representing disks (item \ref{lem:createindep:6} in Lemma~\ref{lem:createindep}), we need to start with the disks, and in fact always connect a vertex, if possible, to the closest disk center.
\subsection{Lower bounds}

A standard technique for proving lower bounds for planar and geometric problems is to reduce from the so-called \gtiling\ problem (see, e.g., \cite{cyganParameterizedAlgorithms2015}). 
\defproblem{$\gtiling$}{integers $k$ and $N$, a collection $\mathcal{S} = \{S_{i,j}\}_{1 \leq i, j \leq k}$ of $k^{2}$ non-empty sets where $S_{i,j} \subseteq [N] \times [N]$ for $1 \leq i,j \leq k$}{Is there $s_{i,j} \in S_{i,j}$ for each $1 \leq i, j \leq k$ such that
	\begin{itemize}
		\item If $s_{i,j} = (a,b)$ and $s_{i+1, j} = (a', b')$, then $a = a'$
		\item If $s_{i,j} = (a,b)$ and $s_{i, j+ 1} = (a', b')$, then $b = b'$.
	\end{itemize}
}

Intuitively, we can imagine the problem instance as $k\times k$ grid, where $k$ pieces of information are propagated horizontally, another $k$ pieces of information are propagated vertically, and whenever a horizontal and a vertical piece meet, there is a restriction on what combination of values they can take. Note that each piece of information is a choice of an element from $[N]$.
In the proof of Theorem~\ref{th:planarlb} (showing that there is no
subexponential FPT algorithms for Steiner Tree in planar graphs), Marx,
Pilipczuk, and Pilipczuk \cite{DBLP:conf/focs/MarxPP18} implicitly proved a
hardness result for a variant of \gtiling, where each piece of information
that propagates horizontally is a choice from $[N]$, but each piece of information
that propagates vertically is a single bit. We formally introduce this problem and state lower bounds for it, which will be used as the starting point of our lower bounds for Steiner Tree on geometric objects.

\defproblem{$\ngtiling$}{integers $x, y$, integer $N$, a collection $\mathcal{S} = \{S_{i,j}\}_{(i,j) \in [x] \times [y]}$ of $x \cdot y$ non-empty sets where $S_{i,j} \subseteq \{0,1\}  \times [N]$ for $(i,j) \in [x] \times [y]$.}{Is there $s_{i,j} \in S_{i,j}$ for each $(i,j) \in [x] \times [y]$ such that
	\begin{itemize}
		\item If $s_{i,j} = (a,b)$ and $s_{i+1, j} = (a', b')$, then $a = a'$
		\item If $s_{i,j} = (a,b)$ and $s_{i, j+ 1} = (a', b')$, then $b = b'$.
	\end{itemize}
      }
Observe that there is a simple brute-force $2^{O(y)}N^{O(1)}$ time algorithm,
which guesses each of the $y$ bits of information propagated vertically. We show
that this is essentially optimal: assuming ETH, we cannot improve $2^{O(y)}$ to
$2^{o(y)}$. In fact, we prove the stronger statement that there is no algorithm
subexponential in the number $xy$ of cells: 
\begin{theorem}\label{th:ngtiling-littleo}
  Assuming ETH, there is no $2^{o(xy)}N^{O(1)}$ time algorithm for \ngtiling.
\end{theorem}        
The actual precise statement is even stronger, to allow some leeway for further reductions (see Theorem~\ref{theorem:grid_tiling_hardness}).

In reductions involving geometric objects, it is often much easier to express $\le$ constraints than $=$ constraints: for example, the horizontal position of an object has to be at least/at most some value in order for the object to intersect/not intersect some other object. This motivated the introduction of a version of \gtiling\ where $a\le a'$ and $b\le b'$ appear as requirements in the problem definition. We introduce a similar variant of \ngtiling, where the horizontally propagated information can increase monotonically, but the vertically propagated information remains the same.

\defproblem{$\mongtiling$}{integers $x, y$, integer $N$, a collection $\mathcal{S} = \{S_{i,j}\}_{(i,j) \in [x] \times [y]}$ of $x \cdot y$ non-empty sets where $S_{i,j} \subseteq \{0,1\}  \times [N]$ for $(i,j) \in [x] \times [y]$}{Is there $s_{i,j} \in S_{i,j}$ for each $(i,j) \in [x] \times [y]$ such that
	\begin{itemize}
		\item If $s_{i,j} = (a,b)$ and $s_{i+1, j} = (a', b')$, then $a = a'$
		\item If $s_{i,j} = (a,b)$ and $s_{i, j+ 1} = (a', b')$, then $b \leq b'$.
	\end{itemize}
      }
We prove a lower bound similar to Theorem~\ref{th:ngtiling-littleo} for \mongtiling\ as well. Here, it becomes important that we show a slightly stronger bound in 
Theorem~\ref{theorem:grid_tiling_hardness}, as this additional power allows us to implement extra checks to enforce that the horizontally propagated information does not change.
\begin{theorem}\label{th:ngtiling-littleomon}
Assuming ETH, there is no $2^{o(s)}N^{O(1)}$ time algorithm for \mongtiling, where $s=xy$.
\end{theorem}        

\begin{figure}[htbp]
  \centering
  \begin{subfigure}[t]{0.48\textwidth}
    \centering
    \includegraphics[page=1,width=\linewidth]{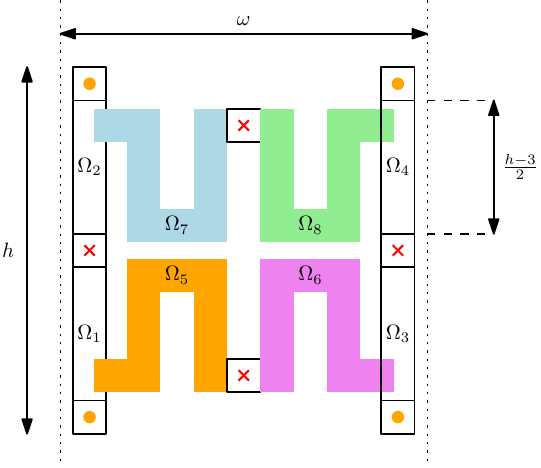}
    \caption{The gadget.}
    \label{fig:crossing_gadget_page_1}
  \end{subfigure}
  \hfill
  \begin{subfigure}[t]{0.48\textwidth}
    \centering
    \includegraphics[page=2,width=\linewidth]{crossing_gadget}
\caption{The green dashed line describes $\Delta_C^{2}$, implicitly a subgraph $H$ of $G$ of minimum size
	such that $H$ connects each terminal to an interface square. Observe
	that the interface squares $u_{NW}$ and $u_{SE}$ are the only non-basic
	squares.}
    \label{fig:crossing_gadget_page_2}
  \end{subfigure}

  \caption{A copy of the $\CrossingGadget$ for $\omega = 11$. A long rectangle is a row/column of contiguous unit squares.
	  The bottom-left corner of the gadget
	  is the point $(0,0)$. The red crosses
  denote the terminal squares, whereas the orange disks denote the interface squares.}
  \label{fig:crossing_gadget}
\end{figure}

To prove Theorem~\ref{th:main-lowerbound} (ruling out $2^{o(s)}n^{\Oh(1)}$ algorithms for Steiner Tree on unit squares), we present a reduction from \mongtiling. The reduction uses a crossing gadget (see Figure~\ref{fig:crossing_gadget}) that allows to transfer a single bit of information vertically, in a way that the solution Steiner Tree is not connected vertically. The gadget occupies an area of height roughly $h$ and width roughly $\omega$. The gadget has four interface objects $u_{NW}$, $u_{NW}$, $u_{SW}$, $u_{SE}$, one in each corner, which are the only objects that can intersect other objects outside the gadget. The gadget contains four terminal objects, and obviously, the solution needs to connect each of them to at least one of the interface objects. The construction of the gadget ensures that there are only two optimal ways of doing this: (1) every terminal is connected to either $u_{NW}$ or $u_{SE}$, or (2) every terminal is connected to either $u_{NE}$ and $u_{SW}$, and in both cases, the north and south interfaces vertices are not connected inside the gadget. Moreover, every other subset that violates these properties has a higher cost.

\begin{figure}[htbp]
\centering
\includegraphics[scale=.9]{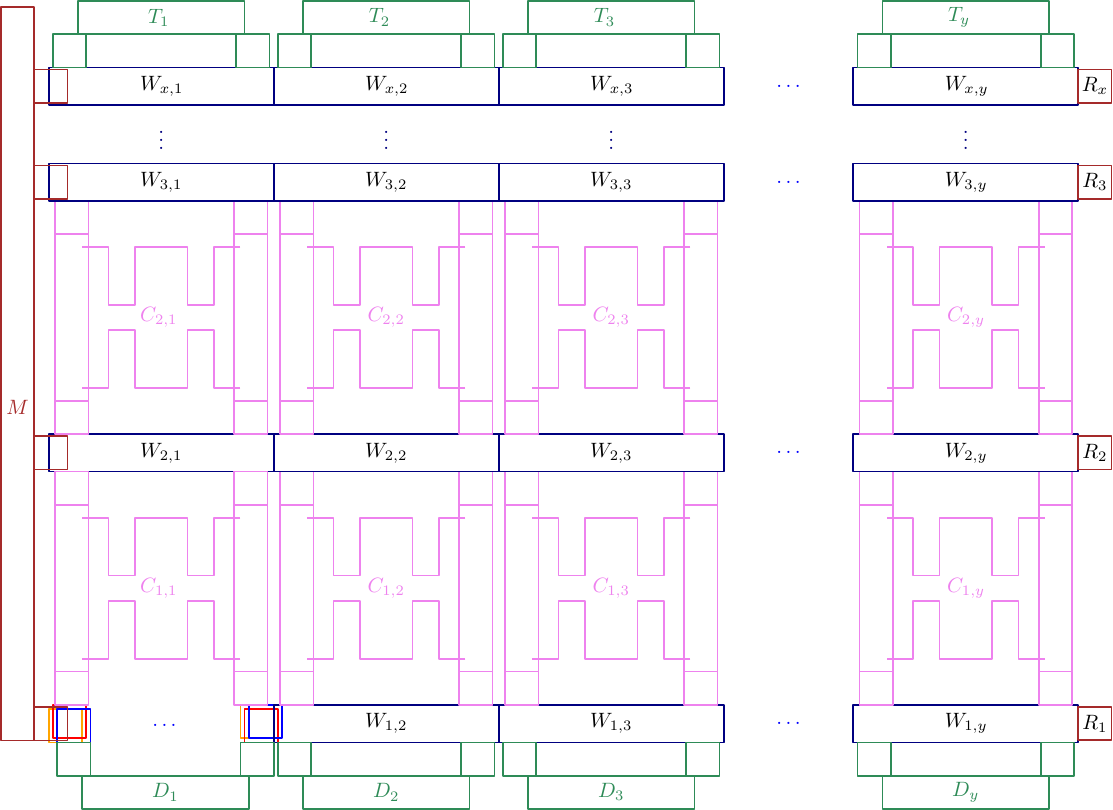}
\caption{High-level overview of the reduction. Only the first and last blocks of the gadget $W_{1,1}$ are shown. Distinct copies of crossing gadgets are pairwise disjoint, and the same holds for the top and bottom gadgets.}
\label{fig:reduction}
\end{figure}

We reduce from $\mongtiling$ using copies of this gadget. The reduction is using very crucially that we are working with intersection graphs and hence the input can contain large cliques. Each of the $x$ rows of the $\mongtiling$ instance is expressed by about $\omega\cdot y$ cliques of unit squares. In each clique, the top left corner of each square is close to the point $(p,q)$ for some integers $p,q$; more precisely, it is of the form $(p+b\epsilon,q+a\epsilon)$ for some $b\in [N]$ and $a\in\{0,1\}$. That is, each square has $N$ possible horizontal offsets and two possible vertical offsets. Each cell is represented by a block of $\omega$ such cliques next to each other, and $y$ such blocks form a line representing one row of the input $\mongtiling$ instance (see Figure~\ref{fig:reduction}). In a solution, we are supposed to select exactly one square from each clique, and the squares selected from the cliques of a line are supposed to be connected. In particular, this means that the horizontal offsets of the squares should be nonincreasing, otherwise two adjacent squares would not intersect. Therefore, we can interpret the horizontal offset in block $(i,j)$ as the second component of $s_{i,j}$ in the solution to $\mongtiling$ and we interpret the vertical offset as the first component. Then we add a $y$ copies of the crossing gadget between two rows to ensure that the vertical offset is consistent in two vertically adjacent blocks, hence the first component of $s_{i,j}$ and $s_{i+1,j}$ agrees. We need careful global arguments to ensure that every cheap solution is of the required form; in particular, no such solution creates a connection between two rows through a crossing gadget. The reduction shows that a $\mongtiling$ instance can be reduced to Steiner Tree on unit disk graphs with $O(xy)$ terminals, thus we can establish the lower bound using Theorem~\ref{th:ngtiling-littleomon}.

\section{Preliminaries and Notation}

In this section, we give an overview of the standard notation and definitions used throughout the paper and remind the reader of the definitions of concepts from the previous section.
Usually, $n \coloneqq |V(G)|$, $t \coloneqq |T|$ is the number of terminals, and $k \coloneqq |S|$ is the cardinality of the solution. 
Unless stated otherwise, costs and running times are polynomial in the size of the input, denoted $|I|$. 
We use the notations $\Ot(\cdot)$ to hide polylogarithmic factors and $O_\alpha(\cdot)$ hides constants depending only on~$\alpha$. For any $N \in \mathbb{N}$ we write $[N] \coloneqq \{1, \dots, N\}$.  All logarithms are taken to base~2.

\paragraph{Graphs and distances.}

We work with finite, undirected graphs whose edges carry non-negative real
lengths (sometimes also called weights). For a graph $G$, the shortest-path
distance between vertices $u,v\in V(G)$ is denoted by $\dist_G(u,v)$. For a
vertex set $X\subseteq V(G)$ we write $\dist_G(u,X)\coloneqq\min_{x\in
X}\dist_G(u,x)$. The (closed) ball of radius $r\ge 0$ around $v\in V(G)$ is
$B_G(v,r)\coloneqq\{\,x\in V(G)\mid \dist_G(v,x)\le r\}$, and we abbreviate
$B_G(v,r)$ as $B(v,r)$ when the graph is clear from the context. The weak
diameter of a connected vertex set $S\subseteq V(G)$ is $\max_{x,y\in
S}\dist_G(x,y)$, that is, distances are measured in the ambient graph $G$ and
paths may leave $S$. When convenient, we assume distances are made unique by an
infinitesimal perturbation of edge lengths; this yields unique shortest paths
and unique nearest objects without affecting any combinatorial argument.

\paragraph{Objects and the touching graph.}
Throughout this paper, we use $\objects$ to denote a set of objects and typically use $O$ to denote an individual object. An object is a connected subset $O\subseteq V(G)$. Two objects $O_1,O_2$ \emph{touch} if either $O_1\cap O_2\neq\emptyset$ or there exists an edge of $G$ with one endpoint in $O_1$ and the other in $O_2$. For a family $\objects$ of objects, the \emph{touching graph} $\touchgraph{\objects}$ is the graph with vertex set $\objects$ in which two vertices are adjacent exactly when the corresponding objects touch. A subfamily $\mathcal{A}\subseteq \objects$ is \emph{connected} if $\touchgraph{\mathcal{A}}$ is connected. We say that objects are \emph{disjoint} (or \emph{independent}) if the sets of their vertices are disjoint. For a subfamily $\mathcal{A}$ we use $\bigcup\mathcal{A}\subseteq V(G)$ to denote the set of vertices covered by the objects of $\mathcal{A}$, and we freely identify an object $O$ with the induced subgraph $G[O]$ when discussing paths inside $O$. We recall that in the context of Assumptions \assgeo\ and \asspl, objects are classified into the following types:
\begin{itemize}
  \setlength{\itemsep}{0pt}
  \setlength{\parskip}{0pt}
\item \textbf{(fat)} A simple polygon (in the geometric setting) or a connected subgraph (in the planar setting) of diameter at most $\Oh(\alpha)$, which also contains a unit-diameter disk (geometrically) or disk of radius at least 1 (in planar graphs).
\item \textbf{(disk)} A disk of radius at least 1 (in the geometric setting) or a ball $B(v,r)$ in an edge-weighted graph $G$ for some vertex $v$ and radius $r \geq 4\alpha$ (in the planar setting).
\item \textbf{(long)} (in Assumption \asspl\ only) An arbitrary connected subset that is disjoint from every other object in the solution.
\end{itemize}
A set $T$ of objects is \emph{irredundant} if for every $O \in T$, the vertex set $O \setminus \bigcup_{O' \in T, O' \neq O} O'$ is nonempty. A solution $S$ is \emph{inclusionwise minimal} if no proper subset of $S$ is a solution.
For a solution $S\cup T$ satisfying Assumption \asspl, a set $W \subseteq \objects'$ together with a mapping $\pi: W \to S\cup T$ is called a \emph{disjoint representation} if the objects in $W$ are pairwise disjoint and satisfy the properties stated in Lemma~\ref{lem:createindep}.

\paragraph{Hardness assumption.}
The Exponential Time Hypothesis (ETH), is a widely used assumption in fine-grained complexity. It can be formally stated as follows: There exists $\delta > 0$ such that 3-SAT with $n$ variables cannot be solved in time $\Oh(2^{\delta \cdot n})$.  
See~\cite{impagliazzoComplexityKSAT2001a,cyganParameterizedAlgorithms2015} for more details on ETH and its implications.

\section{Proofs of the algorithmic results}
In this section, we go over the proofs of the statements presented in the overview (Section~\ref{sec:overview}). The proofs are presented in the same order as the statements appeared.

\subsection{Lemma~\ref{lem:geotoplanar} (reduction to the planar problem)}
\label{sec:geotoplanarprove}
  As described in Section~\ref{sec:geotoplanar}, our first step is translating the geometric problem into a planar problem. We prove here Lemma~\ref{lem:geotoplanar}, restated for reference.

\geotoplanar*

As a first preprocessing step, we construct an instance where the terminal objects are pairwise disjoint.

\begin{lemma}\label{lem:geoindep}
  For every fixed $\alpha>0$, there is a polynomial-time algorithm that transforms an instance $(\objects,T)$ of \stgeo with assumption \assgeo\ to an instance $(\objects',T')$ of \stgeo also satisfying assumption \assgeo\ such that $|T'|\le |T|$, the optimum value of the two instances differ by less than $1/2$, and the objects in $T'$ are pairwise disjoint.
\end{lemma}
\begin{proof}
  Let $T'$ be an arbitrary maximal independent set of $T$ and let us pick an arbitrary $t\in T'$. The set $\objects'$ is the same as $\objects$, but with a weight function $w'$ that is slightly different from the original weight function $w$: we set
  \begin{itemize}
  \item $w'(O)=1/(3|T\setminus T'|)$ for every $O\in T\setminus T'$,
  \item $w'(t)=w(t)+w(T\setminus T')$, and
  \item $w'(O)=w(O)$ for every $O\in \objects \setminus (T\setminus (T'\cup \{t\}))$.
  \end{itemize}
  It is clear that \assgeo\ holds, as the set of objects is the same. Let us observe that a solution $S\cup T$ of the original instance is also a solution of the new instance, as it is connected and includes $T'$. However, the weight of the solution increases by exactly $1/3$: the original weight of the objects in $T\setminus T'$ is accounted for in $w'(t)$, and furthermore each object in $T'\setminus T$ incurs an extra weight of $1/(3|T\setminus T'|)$. Thus, the optimum of the new instance is at most $1/3$ larger than the original optimum.

  Consider now an optimum solution $S'\cup T'$ of the new instance and let $S=S'\setminus (T\setminus T)$. We claim that $S\cup T=S'\cup T' \cup (T\setminus T')$ is a feasible solution of the original instance. Indeed, the intersection graph of $S'\cup T'$ is connected and, by the maximal choice of $T'$, every object in $T\setminus T'$ intersects an object in $T'$, hence the intersection graph of $S'\cup T'\cup(T\setminus T')$ is also connected. Let us show that $w(S\cup T)\le w'(S'\cup T')$. Indeed, the only difference appears on $T\setminus T'$ and on $t$: the weight of $t$ is lower by $w(T\setminus T')$ on the left-hand side, compensating any potential increase from members of $T\setminus T'$ appearing on the left-hand side. Thus, we can conclude that the optimum of the new instance cannot be lower than the original optimum.
\end{proof}  
Next, we show a basic consequence of \assgeo: an optimum solution cannot be very dense.
\begin{lemma}\label{lem:geopacking}
Let $(\objects,T)$ be an instance of \stgeo with assumption \assgeo\ where $T$ is a set of pairwise disjoint objects and let $S\cup T$ be an optimum solution. Then every ball of radius $4\alpha$ in the plane intersects at most $1000\alpha^2$ objects in $S\cup T$.
\end{lemma}
\begin{proof}
  Let $B$ be a ball of radius $4\alpha$ centered around some point $p$.
  Let $S'\subseteq S$ be obtained in the following way: we start with $S'\subseteq S$ containing those objects that do not intersect $B$ and then  we repeatedly add objects from $S$ to $S'$ if they do not intersect any object already in $S'\cup T$. Observe that when this process stops, the objects in $S'\cup T$ intersecting the ball $B$ are pairwise independent, as no two objects in $T$ intersect by assumption.

Let us bound the number of objects in $S'\cup T$ intersecting $B$.   Let $B'$ be the ball of radius $4\alpha+1$ centered at $p$. If $O$ is a fat object intersecting $B$, then its diameter is at most $\alpha/4\ge 1$, hence $O$ is contained in $B'$, implying that the unit-diameter disk in $O$ is in ball $B'$.  If $O$ is a disk intersecting $B$, then it has radius at least 1. In particular, this means that it contains a unit-diameter disk that intersects $B$ and hence this disk is contained in $B'$. It follows that the objects in $S'\cup T$ intersecting $B$ are pairwise disjoint and each of them contains a unit-diameter disk contained in the ball $B'$ or radius $4\alpha+1$. Now, a simple area bound shows that there are at most $100\alpha^2$ such objects.

Unlike $S\cup T$, which had a connected intersection graph, the intersection graph of $S'\cup T$ may have some number $c>1$ of connected components. Each such component has to have an object that intersects the ball $B$ (otherwise, $S\setminus S'$ would not be able to reconnect $S'\cup T$). Let us select one such object from each component; this gives us a set $X$ of $c$ pairwise disjoint objects, each intersecting the ball $B$. By the previous paragraph, we have $c=|X|\le 100\alpha^2$.

  Next, we reconnect $S'\cup T$ by reintroducing some objects from $S\setminus S'$, in the following way. Let us select an arbitrary component $K$ of the intersection graph of $S'\cup T$ and let us select a minimal subset $Y$ of $S\setminus S'$ that connects this component to some other component. Let us observe that $|Y|\le 2$: by minimality, at most one object in $Y$ can intersect $K$, at most one object in $Y$ can intersect some other component, and if an object of $Y$ does not intersect any object of $S'\cup T$, then this would contradict the initial maximal choice of $S'$. Thus, by adding two objects to $S'$, we can decrease the number of components by 1. Repeating this step, we can reconnect the intersection graph by adding $2c-2<200\alpha^2$ objects. 

The optimality of the solution $S\cup T$ implies that this reconnection process reintroduces all objects in $S$ (here we use the fact that every weight is nonzero). Thus we can conclude that $S'\cup T$ had at most $100\alpha^2$ objects intersecting $B$, and $S\cup T$ can have at most $200\alpha^2$ more objects intersecting $B$. Thus $S\cup T$ has at most $300\alpha^2$ objects intersecting  $B$.
\end{proof}

We are now ready to prove Lemma~\ref{lem:geotoplanar}.
\begin{proof}[Proof of~\cref{lem:geotoplanar}]
We start from an instance $(\objects, T)$ that satisfies Assumption~\assgeo. Using Lemma~\ref{lem:geoindep}, we can assume that $T$ contains pairwise independent objects.
Our goal is to produce, in polynomial time, a planar instance $(G', \infty, \objects', T')$ whose optimal solutions satisfy Assumption~\assp. 
We achieve this through the following sequence of steps.

\paragraph{Step 1: Building the base graph.} Let $(F,D)$ be the partition of $\objects$ into fat objects and disks, as stated by Assumption~\assgeo.
We construct an intermediate geometric graph $G_T = (V_T, E_T)$ as follows.
For every fat object $f_i \in F$, we add every boundary vertex to $V_T$.
For every disk $d_i \in D$, we add its center $c(d_i)$ to $V_T$.
Next, we consider intersections between object boundaries. For any two distinct objects $O, O' \in F \cup D$ whose boundaries intersect geometrically, 
we add one intersection point between the boundaries of $O$ and $O'$ as a vertex of $V_T$.

We now define the edges $E_T$ for fat and disk objects separately.
For each fat object $f_i \in F$, let $\partial f_i$ be the set of boundary vertices of $f_i$ and let us select a vertex $s(f_i)\in \partial f_i$. We add to $E_T$: 
(1) every boundary edge of $f_i$; 
(2) for each boundary vertex $b \in \partial f_i$, a straight edge connecting $b$ to $s(f_i)$;
and (3) for every vertex $v \in V_T$ that lies strictly inside $f_i$, a straight edge \emph{fully contained in $f_i$} connecting $v$ to a boundary vertex of $f_i$ (clearly, there is a boundary vertex $b\in \partial f_i$ such that the segment $bv$ is contained in $f_i$).
Note that edges of type (1) and (3) are contained in $f_i$ by defintion, but edges of type (2) are not necessarily contained in $f_i$.
For each disk $d_i \in D$, we connect its center $c(d_i)$ by straight edges to every vertex of $V_T$ that lies within $d_i$.

\paragraph{Step 2: Planarization.}
Observe that the graph $G_T$ constructed above may still be \emph{non-planar} due to edge crossings created by the geometric connections introduced in Step~1. 
To obtain a planar representation, we construct a \emph{planar} graph $G' = (V', E')$ as follows. 
We initialize $V'$ with all vertices in $V_T$ and $E'$ with all edges in $E_T$. 
Whenever two edges in $E_T$ intersect geometrically at a point that is not already a vertex of $V_T$, 
we insert this intersection point as a new vertex in $V'$ and subdivide both edges at that location, adding the resulting subedges to $E'$. In case two edges $e_1$ and $e_2$ of $E_T$ overlap (i.e., they intersect in a segment), then we subdivide $e_1$ at the endpoint of $e_2$ contained in $e_1$ and vice versa.
After this refinement, $G'$ becomes planar. 
Each edge of $G'$ is assigned a weight equal to the Euclidean distance between its endpoints. Observe that this means that if $xy$ is an edge of $E_T$, then after the subdivisions, the graph $G_T$ contains a path whose total weight is the distance of $x$ and $y$ in the plane.

For later convenience, we perform two additional subdivision steps.
First, wherever an edge of $G'$ intersects the boundary of a disk object, we introduce a subdivision point, with the weight of the two new edges being their geometric length. Second, in a final step, every edge of $G'$ is subdivided once more, 
and the original edge weight is distributed arbitrarily between the two resulting edges. 
The newly introduced vertices in this final step are referred to as \emph{subdivision vertices}.

\paragraph{Step 3: Defining the new family of objects.}
We now define the new family of objects based on the planar graph $G' = (V', E')$. 
Let $F$ and $D$ denote the sets of fat objects and disks, respectively, in the original instance $(\objects, T)$. 
The new object family $\objects'$ consists of two corresponding subsets, $F'$ and $D'$, defined as follows.

Let $r = \max_{d_j \in D} r(d_j)$ denote the maximum disk radius in the instance.  
For each disk $d_i \in D$ with center $c(d_i)$ and radius $r(d_i)$, 
if $r(d_i) < r$, we attach a new vertex $c'(d_i)$ to $c(d_i)$ by an edge of length $r - r(d_i)$, 
and replace $c(d_i)$ by $c'(d_i)$ as the new center of the disk.
We then define $d'_i$ as the disk of radius $r$ centered at $c(d_i)$ (or $c'(d_i)$ if added), 
containing all vertices of $V'$ that lie on or within its boundary, and we add $d'_i$ to $D'$.

For each fat object $f_i \in F$, we define $f'_i$ to consist of all vertices of $V'$ that lie on edges of $E_T$ which are
(i) incident to a boundary vertex of $f_i$, and (ii) fully contained within $f_i$. 
This includes, in particular, all subdivision vertices created on such edges during planarization. 
We add $f'_i$ to $F'$. 
Observe that $f'_i$ is connected: it includes every vertex on the boundary of $f_i$
and every other vertex of $f'_i$ has a straight path to a vertex of $\partial f_i$.

Finally, we set $\objects' = F' \cup D'$, which constitutes the object family of the planar instance.

\vspace{.2cm}

\begin{claim}\label{cl:inside}
For every object $O \in \objects$ and its corresponding object $O' \in \objects'$, 
every vertex $v \in O'$ represents a point in the plane that lies within $O$.
Moreover, if $O'\in \objects'$ contains a subdivision vertex $v$, then it also
contains the two neighbors $u_1$ and $u_2$ of $v$.
\end{claim}
\begin{proof}
  If $O$ is a fat object, this follows directly from the definition, since $O'$ includes only vertices on edges of $E_T$ that are fully contained in $O$.
  If $O$ is a disk, then this follows
from the fact that in $G'$, vertex $v$ is at distance at most $r$ from the center
of $O$, which is an upper bound (by the definition of the weight of the edges)
on the geometric distance of $v$ from the center. Hence, the point corresponding to $v$ lies within the geometric disk representing $O$.

For the second statement, suppose that some $f'_i$ contains a subdivision vertex $v$. This means that $v$ was on an edge of $E_T$ whose vertices were added to $f'_i$, and these vertices include the two neighbors of $v$ (as $v$ cannot be an endpoint of an edge of $E_T$). If $v$ is in a disk $d'_i$, then $v$ is in the disk $d_i$. Moreover, if $u_1$ and $u_2$ are the two neighbors of $v$, then the $u_1u_2$ segment does not contain a boundary vertex of $d_i$: otherwise, this boundary vertex would have been introduced before the final subdivision step. 
\end{proof}
We remark that the converse of Claim~\ref{cl:inside} is not true: both for fat and disk objects, a vertex of $G'$ that is in $O$ is not necessarily in $O'$. However, it is true in the following weaker form:
\begin{claim}\label{cl:inside2}
For every object $O \in \objects$ and its corresponding object $O' \in \objects'$, 
if a $v \in V_T$ is in $O$, then $v\in O'$.
\end{claim}
\begin{proof}
If $O$ is a fat object $f_i$, then there is an edge of $E_T$ connecting $v$ to a boundary vertex $b\in \partial f_i$ such that the segment $bv$ is fully contained in $O$, and all vertices of $V_T$ on $bv$ were added to $f'_i$. If $O$ is a disk $d_i$, then $E_T$ contains a $c(d_i)v$ edge, and all vertices of $V_T$ on $bv$ were added to $d'_i$.
\end{proof}

Let us prove next that the intersection graph of $\objects$ corresponds to the touching graph of $\objects'$. 
\begin{claim}\label{cl:touch}
  Two objects $O_1, O_2 \in \objects$ intersect if and only if the corresponding connected vertex sets $O'_1, O'_2 \in \objects'$ touch, 
  that is, they share a common vertex or there exists an edge in $G'$ with one endpoint in $O'_1$ and the other in $O'_2$.
\end{claim}
\begin{proof}
\noindent \emph{Forward direction:} Suppose that $O_1$ and $O_2$ intersect geometrically. If their boundaries intersect, then there exists an intersection point $p$ between $O_1$ and $O_2$ that 
has been explicitly added as a vertex in $V_T$. Otherwise, assume without loss of generality that $O_2$ is contained in $O_1$. If $O_2$ is a fat object, then let $p$ be a boundary of a vertex of it; if $O_2$ is a disk, then let $p$ be its center. In all cases, $p\in V_T$ is contained in both $O_1$ and $O_2$. Hence Claim~\ref{cl:inside2} implies that $p\in  O'_1 \cap O'_2$.

\smallskip \noindent \emph{Reverse direction:}
Suppose that $O'_1$ and $O'_2$ touch in $G'$. 
If they share a common vertex $v$, then by Claim~\ref{cl:inside} $v$ corresponds to a point lying in both $O_1$ and $O_2$, 
and therefore $O_1$ and $O_2$ intersect. 
Suppose that $O'_1$ and $O'_2$ touch, but do not intersect. Because of the subdivision vertices, we can assume that $O'_1$ contains a subdivision vertex $v$ and $O'_2$ contains its neighbor $u$. Now Claim~\ref{cl:inside} implies that $O'_1$ contains $u$ as well, further implying that $u$ is contained in both $O_1$ and $O_2$, hence they intersect.
\end{proof}

\paragraph{Satisfaction of Assumption~\assp.}
It remains to prove that the solution of the constructed instance $(G', \infty, \objects', T')$ satisfies Assumption~\assp.
Let us verify that each object $f'_i$ satisfies the diameter condition. 
Any two boundary vertices $b,b'$ of $f'_i$ admit a path $b \to s(f_i) \to b'$ consisting of 
two straight edges of total length at most $2\,\mathrm{diam}(f_i)$ (note that these paths may leave $f'_i$). 
Moreover, every interior vertex $u \in f'_i$ is connected to some boundary vertex $b$
by a straight edge fully contained in $f_i$ of length at most $\mathrm{diam}(f_i)$. 
Thus, for any $u, w \in f'_i$, a path $u \to b_u \to s(f_i) \to b_w \to w$ 
has total length at most $4\,\mathrm{diam}(f_i) \le \alpha$, where $b_u$ and $b_w$ are the respective boundary vertices to which $u$ and $w$ is connected. 
Hence, the graph-diameter of $f'_i$ in $G'$ is at most $\alpha$.

To ensure the existence of the function $\tau$, we note that the centers of the disks are distinct. 
Moreover, for each fat object $O'$, 
we attach a new degree-$1$ vertex $v'$ to an arbitrary vertex of $O'$ by an edge of length~$0$, include $v'$ into $O'$, and set $\tau(O') = v'$ (this zero-length attachment is purely symbolic and does not affect distances or planarity).

Finally, we justify the packing condition using Lemma~\ref{lem:geoindep}. Consider an optimum solution $S'\cup T'$ and a ball $B(v,4\alpha)$. Let $X'$ be the subset of $S'\cup T'$ intersecting $B(v,4\alpha)$ and let $X$ be the geometric objects corresponding to $X$. We claim that every object in $X$ is intersected by a ball of radius $4\alpha$ in the plane, which implies by Lemma~\ref{lem:geoindep} that $|X|=|X'|\le 1000\alpha^2$. Indeed, let us consider the ball $B$ of radius $4\alpha$ in the plane, centered at the geometric point corresponding to $v$ (or its neighbor, if $v$ is a degree-1 attached vertex). If an object in $X'$ has a vertex $u$ that is at distance at most $4\alpha$ from $v$ in the graph, then (as distances in the graph are not smaller than the distances in the plane), the geometric point $u$ is at distance at most $4\alpha$ from $v$ in the plane. By Claim~\ref{cl:inside}, point $u$ is in the corresponding geometric object of $X$, hence this object is intersected by the ball of radius $4\alpha$ centered at $v$ in the plane, what we wanted to show.

Therefore, the constructed instance $(G', \infty, \objects', T')$ admits an optimal solution that satisfies Assumption~\assp. 
Each step of the transformation: graph construction, planarization, and object redefinition, introduces only a polynomial increase in size. 
Hence, the overall reduction can be implemented in polynomial time. Thus, completing the proof of the lemma.
\end{proof}

\paragraph{Extension to Rotated Squares and Fat Objects.}
The above reduction and analysis extend directly to instances where the objects are either (i) fat polygons of radius at most~$\alpha$ containing a unit-diameter disk, or (ii) rotated squares at~$45^\circ$ (i.e., $L_1$ balls). 
The construction of $G_T$ and its planarization remain identical, except that all edges of $E_T$ are assigned weights equal to their $L_1$ lengths, while their geometric embedding and subdivisions are defined as in the previous case. 
Since $L_1$ distance is additive along straight segments, every subdivided path preserves its exact total $L_1$ length, and all previous arguments on planarity, intersection preservation, and graph distances remain valid up to constant factors. 
Consequently, the resulting planar instance $(G', \infty, \objects', T')$ again satisfies Assumption~\assp\ and can be constructed in polynomial time.

\begin{remark}[Extension to Rotated Squares and Fat Objects]
There is a polynomial-time transformation that maps any instance $(\objects, T)$, where the objects are either fat polygons containing a unit-diameter disk or rotated squares (i.e., $L_1$ balls), to an instance $(G', \infty, \objects', T')$ of \stplanar, such that $|T'| = |T|$ and every optimal solution satisfies the required structural properties: object connectivity is preserved and the packing condition is met.
\end{remark}

  \subsection{\cref{lem:longred} (reducing the number of objects)}
\label{section:algorithmic_part}
In this section, we collect a few graph theoretical results and then give a
short reduction that, together, prove \cref{lem:longred}. We start with a
formal definition.

\begin{definition}\label{definition:critical_conn_set}
	Let $G$ be a connected graph and $Y \subset V(G)$ be a subset of $V(G)$
	such that $\abs{Y} \geq 2$. We say that $Y$ is critically connected if,
	for every vertex $x \in V(G) \setminus Y$, the graph $G-x$ has at least two connected components that contain a vertex of $Y$.
      \end{definition}

      Let us observe that if, for some $x\in V(G)\setminus Y$, graph $G-x$ has a component that is disjoint from $Y$, then removing a vertex $x'$ of this component does not disconnect the vertices of $Y$. Thus we can conclude:

      \begin{observation}\label{obs:critical}
If $Y$ is critically connnected in $G$, then, for every vertex $x \in V(G) \setminus Y$, graph $G-x$ has at least two connected components $C_1,
	 \ldots, C_q$ ($q \geq 2$) and each connected component contains a
	 vertex of $Y$.
\end{observation}

Given a graph $G$, we denote by $\Vgeqthree{G}$ the set of vertices $v \in
V(G)$ with $d_G(v)\geq 3$. A {\it leaf} in $G$ is a vertex of degree 1. 

We now prove \cref{th:criticalmain}, which we restate here for ease of reference.
\criticalmain*

\begin{proof}
	We prove by induction that
	\begin{equation*}
		\abs{\Vgeqthree{G}} \leq  3\bigl(\abs{T} - 2\bigr)- \abs{ T_{=2}},
	\end{equation*}
	where $T_{=2}$ denotes the vertices of $T$ that have degree $2$ in $G$.

	Observe that $T$ is
	critically connected, and the claim obviously holds when $\abs{V(G)} =
	2$, because then $G$ simply consists of an edge. Now let $\abs{V(G)}
	\geq 3$ and suppose the statement holds for all graphs with less than
	$\abs{V(G)}$ many vertices.

	First, observe that if $\abs{T}=2$ and $T$ is critically connected, then
	$G$ is an induced path whose leaves are exactly $T$. Consequently,
	\begin{equation*}
		 \abs{\Vgeqthree{G}}= 0 = 3 \cdot
	\left( \abs{T} - 2 \right) - \abs{T_{=2}}.
	\end{equation*}
	Therefore the claim holds for $G$ and
	we can assume that $\abs{T} \geq 3$.
	Moreover, we can further assume that $\Vgeqthree{G} \not\subseteq T$,
	because otherwise we would immediately get
	\begin{equation*}
		\abs{\Vgeqthree{G}} \leq \abs{T_{\leq 3}} \leq \Bigl(\abs{T} - \abs{T_{=2}}\Bigr) \leq 3 \cdot (\abs{T}-2) - \abs{T_{=2}}
	\end{equation*}
	where $T_{\leq 3}$ is the set of vertices in $T$ that have degree at least $3$ in $G$.
	
	Now let $x \in \bigl(\Vgeqthree{G} \setminus T\bigr)$
	and $C_1,\ldots,C_q$ be the connected
	components of $G \setminus x$. Since $T$ is critically connected and $x \not\in T $, by
	\cref{obs:critical} we have that $q \geq 2$ and
	$V(C_i)\cap T \neq \emptyset$ for $i \in [q]$. Then, for each $i \in [q]$,
	we set $G^{(i)} = G\Big[V(C_i)\cup \{x\}\Big]$ and $T^{(i)}=\Bigl(T \cap V(C_i)\Bigr)\cup \{x\} $. 

	\begin{claim}\label{claim:crit_conn_G_i}
		For $i \in [q]$, we have that $\abs{T_i} \geq 2$ and $T_i$ is a critically connected set in $G_i$.
	\end{claim}

	\begin{claimproof}
		Observe that $T \cap V(C_i) \neq \emptyset$ and $t_i \in T^{(i)}$,
		it holds that $\abs{T^{(i)}} \geq 2$.
		Moreover, since $C_i$ is a connected component of $G \setminus x$ and $G$ is a connected graph,
		we have that $G^{(i)}$ is connected.
		Now suppose for a contradiction that there exists $i \in [q]$ such that
		$T^{(i)}$ is not critically connected in $G^{(i)}$, i.e. $T^{(i)}$ is a proper subset of $V(G^{(i)})$
		and
		there exists $x' \in
		\Bigl( V(G^{(i)}) \setminus T^{(i)} \Bigr)$ such that
		$T^{(i)}$ is connected in $G^{(i)} \setminus x'$.
		Furthermore, by the definition of $C_i$, we have that $G \setminus V(C_i)$ is
		connected and contains $x$. Therefore, $T$ is connected in $G \setminus x'$
		for $x' \not\in T$,
		contradicting the fact that $T$ is a critically connected set in $G$.
	\end{claimproof}
	
	By the induction hypothesis and \cref{claim:crit_conn_G_i},
	it holds that
	\begin{equation*}
		\abs{\Vgeqthree{G^{(i)}} }\leq \Bigl(3(\abs{T^{(i)}}-2) - \abs{T^{(i)}_{=2}}\Bigr)
	\end{equation*}
	for $i \in [q]$, where $T^{(i)}_{=2}$ is the set of vertices in $T^{(i)}$ that
	have degree $2$ in $G^{(i)}$.
	If $q \geq 3$,
	we obtain

\begin{align*}
	\abs{\Vgeqthree{G}}&\leq \left( \sum_{i \in [q]}\abs{\Vgeqthree{G_i}} \right) +1\\
			   &\leq \left( \sum_{i \in [q]} \Bigl(3(\abs{T^{(i)}}-2) - \abs{T^{(i)}_{=2}}\Bigr)\right) +1\\
			   &= 3 \cdot\left( \sum_{i \in [q]} \bigl(\abs{T^{(i)}}-2\bigr) \right) - \left( \sum_{i \in [q]} \abs{T^{(i)}_{=2}} \right)  +1\\	
			   &\leq 3 \cdot \left( \abs{T}  - q\right) - \abs{T_{=2}}+ 1 \\
			   &=3 \cdot \abs{T}-3q -  \abs{T_{=2}}+1\\
               &\leq 3 \cdot (\abs{T}-2) - \abs{T_{=2}},
\end{align*}
where the fourth step holds because $\left(  \sum_{i \in [q]} \abs{T^{(i)}} \right) = \abs{T} + q$
and $\abs{T_{=2}} \leq \left( \sum_{i \in [q]} \abs{T^{(i)}_{=2}} \right)  $.
Similarly, the last step holds because $q \geq 3$.

On the other hand, suppose that $q=2$. Then, since $x \in \Vgeqthree{G}$, either there exists
$i \in [2]$ such that $d_{G_i}(x) \geq 3$, or $\{d_{G_1}(x), d_{G_2}(x)\}
\subseteq \{1,2\}$ and there exists $i \in [2]$ such that $d_{G_i}(x) = 2$.
Let $\Bigl[\text{there is no } i \in [2] \text{ such that } d_{G_i}(x) \geq 3\Bigr]$ be equal to
1 if the condition inside the brackets is true. Moreover, in that case, we have
\begin{equation*}
	\abs{T_{=2}} \leq  \Biggl(\Bigl( \sum_{i \in [2]} \abs{T^{(i)}_{=2}} \Bigr) - 1\Biggr)= \Biggl(\Bigl( \sum_{i \in [2]} \abs{T^{(i)}_{=2}} \Bigr) - \Bigl[\text{there is no } i \in [2] \text{ such that } d_{G_i}(x) \geq 3\Bigr]\Biggr).
\end{equation*}
Therefore,
\begin{align*}
	\abs{\Vgeqthree{G}} &\leq \left( \sum_{i \in [2]}\abs{\Vgeqthree{G_i}} \right) + \Bigl[\text{there is no } i \in [2] \text{ such that } d_{G_i}(x) \geq 3\Bigr]\\
	&\leq \left( \sum_{i \in [2]} \Bigl(3(\abs{T^{(i)}}-2) - \abs{T^{(i)}_{=2}}\Bigr) \right) + \Bigl[\text{there exists } i \in [2] \text{ such that } d_{G_i}(x) \geq 3\Bigr]\\
			   &\leq 3 \cdot\left( \sum_{i \in [2]} \bigl(\abs{T^{(i)}}-2\bigr) \right) - \left( \sum_{i \in [2]} \abs{T^{(i)}_{=2}} \right) + \Bigl[\text{there exists } i \in [2] \text{ such that } d_{G_i}(x) \geq 3\Bigr]\\		
			    &\leq 3\cdot(\abs{T}-2)-\abs{T_{=2}}.
\end{align*}

Hence the theorem holds.
\end{proof}

Next, we use \cref{th:criticalmain} to prove \cref{lem:longred}.

\begin{proof}[Proof of \cref{lem:longred}]
	We describe a polynomial-time reduction that,
	given an instance $I = \left( G, \infty, \objects, \weight,  T\right)$ of \stplanar
	satisfying \assp,
	creates an instance $I' = \left( G', k', \objects', \weight', T' \right)$ of the
	same problem with $T' = T$ and $k' = O_{\alpha}(\abs{T})$ such that:
	\begin{enumerate}
		\item $I$ has a solution $S$ of weight at most $\gamma$ if and
			only if $I'$ has a solution $S'$ of weight
			at most $\gamma$ and cardinality at most $k'$,
		\item $I'$ satisfies the assumption \asspl.
	\end{enumerate}

	\paragraph{Construction of the long objects:}  For each path $P$ in $G$, let $\pathmincover{P}$ denote
	the minimum weight of objects in $\objects$ that cover $P$, i.e.

	\begin{equation*}
		\pathmincover{P} = \min \Biggl\{ \weight\left( \mathcal{S} \right)  \Bigm| \mathcal{S} \subseteq \objects, \, V(P) \subseteq \biggl( \bigcup_{S \in \mathcal{S}} V(S) \biggr) \Biggr\}.
	\end{equation*}

	For each distinct $x,y \in V(G)$, we pick a path $P_{x,y} = (x, \ldots, y)$
	such that $P_{x,y}$ minimizes $\pathmincover{P}$ among all paths
	$P$ connecting $x$ and $y$.
	We also let $N_{x,y}$ to be a set of objects that cover $P_{x,y}$ such that
	$\weight\left( N_{x,y} \right)  = \pathmincover{P_{x,y}}$.
	Moreover, we let
	\begin{equation}\label{eq:omega_xi_yi_defn}
		\begin{aligned}
			\Omega_{x,y} \coloneqq \Bigl\{\bigl(Q,N, \weight(N)\bigr) \bigm| &Q \text{ is a subpath of } P_{x,y}, \\&N \text{ is the smallest subset of $N_{x,y}$ such that $N$ covers $Q$.}  \Bigr\},			
		\end{aligned}
	\end{equation}
	in other words we define $\Omega_{x,y}$ to be the set of all subpaths $P_{x,y}$,
	with the corresponding set of covering objects and weights.
	These paths will correspond to the long
	objects in the constructed instance $I'$. To construct the set $\Omega_{x,y}$,
	we follow these steps:

	\begin{enumerate}
		\item \textbf{Find the shortest path of objects:} Let
			$A_x=\{O\in\objects \mid x\in V(O)\}$ and
			$A_y=\{O\in\objects \mid y\in V(O)\}$ be the sets of
			objects containing $x$
			and $y$, respectively.     
			To find the shortest
			path from any object in $A_x$ to any object in $A_y$, add
			two dummy vertices $s$ and $t$ of weight~$0$, adjacent
			to every object in $A_x$ and $A_y$, respectively, and
			run Dijkstra’s algorithm on the resulting
			vertex-weighted graph. Let $\psi$ denote the
			resulting shortest $s$–$t$ path, and define
			\begin{align*}
				V_\psi \coloneqq \left( \bigcup_{O \in \psi \setminus \{s,t\} } V(O) \right)
			\end{align*}

		\item \textbf{Construction $\Omega_{x,y}$:} Let $P_{x,y}$ denote an arbitrary path
			in $G[V_{\psi}]$
			that connects $x$ to $y$. Note that by definition $G[V_{\psi}]$ 
			is connected and $\{x,y\}
			\subseteq V_{\psi}$, hence $P_{x,y}$ is well-defined.
			To construct $\Omega_{x,y}$, we simply iterate over all
			subpaths of $P_{x,y}$.
	\end{enumerate}

	\paragraph{Equivalence of the instances:} We let $G' = G$, $T' = T$,
	$k' = 8 \cdot \abs{T'}=  8 \cdot \abs{T}$.
	Moreover, we define
	\begin{equation*}
		\objects^{\prime} = \objects \cup \left(  \bigcup_{\substack{x,y \in V(G)\\ x \neq y}}  \{Q \mid  (Q,N,\alpha) \in \Omega_{x,y}\}  \right),
	\end{equation*}
	where the weight of each long object $Q$ is set to be $\weight'(Q) = \alpha = \weight\left( N \right)$ for $(Q,N,\alpha) \in \Omega_{x,y}$. For $S \in \objects$ we have $\weight'(S) = \weight(S)$.
	The new instance $I'$ becomes
	\begin{equation*}
		I' = \left( G', k', \objects', \weight', T' \right) = \left( G, 8 \cdot \abs{T}, \objects^{\prime}, \weight', T \right).
	\end{equation*}

	Next, we observe that for $\left( L,N,\weight\left( N \right)  \right) \in \Omega_{x,y}$ for some
	$x,y \in V(G)$, one can replace $L$ with a set of long objects covered
	by the objects $N$ without increasing the weight of the solution.

	\begin{observation}\label{observation:partition_long_object}
		Let $x,y \in V(G)$ and $\left( L,N,\weight\left( N \right)  \right) \in \Omega_{x,y}$.
		Let $\{ E_1, \ldots, E_s \} $ be a partition of $N$ and
		$\{Q_1, \ldots, Q_s\} \in \Omega_{x,y}$ be a set of long
		objects such that $Q_i$ is covered by $E_i$ for each $i \in [s]$.
		Then, it holds that
		\begin{equation*}
			\sum_{i \in [s]} \weight'\left( Q_i \right) \leq \sum_{i \in [s]} \weight\left( E_i \right) =\weight\left( N \right) = \weight'\left( L \right).
		\end{equation*}
	\end{observation}
	In the following, we say that an object $O \in \objects$ is $\alpha$-close to another object in $O' \in \objects$ if there are $x \in O$ and $x' \in O'$ such that $x'$ is at distance at most $\alpha$ from $x$.
	\begin{claim}\label{claim:bounded_objects_to_fat_object}
		Let $S$ be a solution of the instance $I$ and let $Z$ be a fat object
		in $S$. Then, there are at most $1000\alpha^{2}$ many fat
		non-terminal objects in $S$ which are $4\alpha$-close to $Z$.
	\end{claim}

	\begin{claimproof}
		Suppose for a contradiction that there are strictly more than $1000\alpha^{2}$ many
		fat
		non-terminal objects in $S$ which are $4\alpha$-close to $Z$.
		Let $K_Z$ denote these objects.
		Then, for an arbitrary vertex $v \in Z$, the ball
		$B(v,4\alpha)$ intersects $K_Z$,
		however, this contradicts
		assumption \assp.
	\end{claimproof}

	\begin{claim}\label{claim:assp_B_to_C_i}
		Suppose that $I$ has a solution $S$ with weight at most $\gamma$.
		Then, $I'$ has a solution $S'$ with cardinality at most $8 \cdot \abs{T'}$ and weight at most $\gamma$.
		Moreover, it holds that
  			\begin{enumerate}
  				\item Long objects are disjoint from every other object in $S' \cup T'$,
  				\item For every $v\in V(G')$, the ball $B(v,4\alpha)$ intersects at most $1000\alpha^2$ fat objects in $S'$.
				\item For every long or fat object $L\in S'$, there are at most $1000\alpha^2$ fat objects in $S$ at distance at most $\alpha$ from $L$.
  			\end{enumerate}
	\end{claim}

	\begin{claimproof}
		Let $S$ be a solution of $I$ with weight at most $\gamma$
		and $\mathcal{T}$ denote $\touchgraph{S \cup T}$.
		We will construct a solution $S'$ for $I'$ in two rounds,
		and start with describing the first one.

		\textbf{Description of the first round}:
		First, observe that the terminal objects are critically connected
		in $\mathcal{T}$, because otherwise there exists another solution
		$S'$ of weight at most $\gamma$ such that $\touchgraph{S' \cup T}$
		is still connected.
		Let $D$ denote the set of objects of
		degree at least $3$ in $\mathcal{T}$, respectively.
		By the above discussion and \cref{th:criticalmain}, we have that
		\begin{equation}\label{eq:D_size_bound}
			\abs{D} = O\left( \abs{T} \right).
		\end{equation}

		Let us define $F \coloneqq T \cup D$ as the set of special
		objects, 
		and let $\ell$ be the number of connected components in $\mathcal{T}[F]$.
		Observe that $\ell \leq \abs{F}$, and each connected component
		in $\mathcal{T} \setminus F$ (which is a path by definition) connects exactly two objects in $\mathcal{T}[F]$.
		Therefore, without loss of generality, we can assume that
                there are $\ell - 1$
		many paths in $\mathcal{T} \setminus F$,
		because otherwise we can remove the extra paths from $\mathcal{T}$
		such that $\mathcal{T}$ is still connected, and the weight
		of the solution we obtain is at most the original one.

		Enumerate the paths in $\mathcal{T} \setminus F$ as
		$\mathcal{X} = \{X_1, \ldots, X_{\ell-1}\}$. For each $1 \leq i
		\leq \ell - 1$, let $A_i$ and $B_i$ be the two endpoint objects
		of $X_i$, and let $A^{\prime}_i$ and $B^{\prime}_i$ be their
		neighbours outside $X_i$, respectively. We pick $x_i \in \Bigl(
		N_G\bigl(V(A^{\prime}_i)\bigr) \cap V(A_i) \Bigr) $, $y_i \in
		\Bigl( N_G\bigl(V(B^{\prime}_i)\bigr) \cap V(B_i) \Bigr)$.

		Next, we inductively show that one can replace the paths of objects
		$X_1, \ldots, X_\ell$ in $S$ by long objects $Q_1, \ldots, Q_{\ell}$ such that
		\begin{enumerate}
			\item the long objects are disjoint from every other object, and
			\item the weight of the solution does not increase after replacing $X_i$ with $Q_i$.
		\end{enumerate}
		We let $S_0 \coloneqq S$ and let $S_i$ denote the solution set $S$
		after replacing each $X_j$ with $Q_j$ for $j \in [i]$.
		Also define $\mathcal{T}_0 \coloneqq \mathcal{T} = \touchgraph{S \cup T}$
		and $\mathcal{T}_i = \touchgraph{S_i \cup T}$.
		Let $i \in [\ell - 1]$, and suppose that we have replaced each $X_j$ with a long object $Q_j$
		that is disjoint from all other objects,
		for $j \in [i-1]$, such that $\mathcal{T}_j$ is connected.
		Moreover, suppose that $\weight'\left( S_j \right) \leq \weight(S)$ for $j \in [i-1]$.
        
		Observe that $X_i$ divides $\mathcal{T}_{i-1}$ into exactly two
		connected components, which we call $\mathcal{C}^{i}_1$ and
		$\mathcal{C}^{i}_2$. Let $P_{x_i,y_i}=(x_i,\dots,y_i)$. For
		every internal vertex $v\in
		V(P_{x_i,y_i})\setminus\{x_i,y_i\}$, denote its predecessor and
		successor on $P_{x_i,y_i}$ by $\vprev{v}$ and $\vnext{v}$,
		respectively. Then, for each $x \in V\left( P_{x_i, y_i}
		\right)\setminus \{x_i, y_i\}  $,  define
		\begin{equation*}
			\vlabel{x} \coloneqq \begin{cases}
				1, &\text{if } x \in \mathcal{C}^{i}_1\\
				2, &\text{if } x \in \mathcal{C}^{i}_2\\
				0, &\text{otherwise},
			\end{cases}
		\end{equation*}
		and let $\vlabel{x_i} = \vlabel{y_i} = 0$.
		Observe that $\vlabel{x}$ is well-defined, because if $x \in \Bigl(\mathcal{C}^{i}_1 \cap \mathcal{C}^{i}_2\Bigr)$,
		then $\mathcal{C}^{i}_1$ and $\mathcal{C}^{i}_2$ are connected,
		which is a contradiction.
		Let us write
		\begin{equation*}
			P_{x_i, y_i} = \left( x_i, \ldots, y_i \right)  = \left( v_1, \ldots, v_{s_i} \right).
		\end{equation*}

		Define the indices
		\begin{align*}
			j_{i,1} &\coloneqq \max \biggl( \Bigl\{ j \mid 1 < j \leq s_i,\, \vlabel{v_{j-1}} = 1,\, \vlabel{v_j} = 0\Bigr\}  \cup \{1\} \biggr),\\
			j_{i,2} &\coloneqq \min \biggl( \Bigl\{ j \mid j_{i,1} \leq j < s_i,  \vlabel{v_{j+1}} = 2, \vlabel{v_{j}} = 0 \Bigr\} \cup \{s_i\}  \biggr).
		\end{align*}
		Note that $j_{i,1}$ and $j_{i,2}$
		are well-defined, and by definition $1 \leq j_{i,1} \leq j_{i,2} \leq s_i$. Moreover,
		\begin{equation}\label{eq:vlabel_zero}
			\vlabel{x} = 0 \quad \text{for } x \in [j_{i,1},\, j_{i,2}].
		\end{equation}

		Let us define the subpath $Q_i \coloneqq \left( v_{j_{i,1} ,\, \ldots,\, v_{j_{i,2}}} \right)$ of $P_{x_i, y_i}$
		where $\left( Q_i, N_i, \weight(N_i) \right)  \in \Omega_{x_i, y_i}$.
		Moreover, \eqref{eq:vlabel_zero} implies that $Q_i$ is disjoint from
		other objects in $\mathcal{T}_{i-1}$. Finally,
		$Q_i$ connects $\mathcal{C}^{i}_1$ and $\mathcal{C}^{i}_2$,
		and since by our induction hypothesis it holds that $\mathcal{T}_{i-1}$ is connected,
		$\mathcal{T}_{i}$ is also connected.
		Moreover, it holds that
		\begin{equation*}
			\weight'\left( Q_i \right) \leq \weight'\left( P_{x_i, y_i} \right) = \weight\left( N_{x_i,y_i} \right) \leq \weight\left( X_i \right)
		\end{equation*}
		by \eqref{eq:omega_xi_yi_defn}.
		Therefore
		\begin{equation*}
			\weight'\left( S_i \right) = \weight'\left( S_{i-1} \right) - \weight\left(  X_i \right) + \weight'\left( Q_i \right) \leq \weight'\left( S_{i-1} \right) \leq \weight\left( S \right).
		\end{equation*}
		
		Hence, all in all, $\mathcal{T}_{\ell-1}$ is a connected graph
		of objects in $\objects^{\prime}$ which have weight at most $\weight(S) \leq \gamma$,
		and the number of objects in $\mathcal{T}_{\ell - 1}$ is bounded by
		\begin{equation}\label{eq:l_F_bound}
			\left( \ell + \abs{F} \right)  \leq 2 \cdot \abs{F} \leq 2 \cdot \left( \abs{T} + \abs{D} \right) \leq 8 \cdot \abs{T},
		\end{equation}
		where the last inequality follows from \eqref{eq:D_size_bound}.

		\textbf{Description of the second round}:
		Next, for each $i \in [\ell-1]$, we create a set of independent
		long objects $\mathcal{L}_i$ which we will replace $Q_i$ with.
		Let $\mathcal{F}_i \subseteq S$ denote the fat objects of degree at least 3
		which are $\alpha$-close to $Q_i$. Moreover, let $r_i$ denote the number of objects in $O \in N_i$
		such that there exists an $F \in \mathcal{F}_i$ that is $\alpha$-close to $O$.
		We partition $Q_i$ into at most $r_i$ many long objects $Z^{i}_1, \ldots, Z^{i}_{p_i}$ such that
		\begin{enumerate}
			\item $\left( Z^{i}_j, A^{i}_j, \weight(A^{i}_j) \right)  \in \Omega_{x_i,y_i}$ for some $A^{i}_j \subseteq N_i$, for each  $j \in [p_i]$,
			\item each $F \in \mathcal{F}_i$ is $\alpha$-close to an endpoint $Z^{i}_j$ for some $j \in [p_i]$,
			\item the long objects $Z^{i}_1, \ldots, Z^{i}_{p_i}$ are vertex-disjoint
				and form a path in the touching graph,
			\item $\{A^{i}_j\}_{j \in [p_i]}$ is a partition of $N^{i}$.
		\end{enumerate}
		By \cref{observation:partition_long_object} it holds that
		\begin{equation}\label{eq:weight_Z_ij_bounded}
			\sum_{j \in [p_i]} \weight'\left( Z^{i}_j \right) \leq \weight'\left( L^{i} \right).
		\end{equation}
		Hence, we can replace each $Q_i$ with the long objects
		$\mathcal{L}_i = \{Z^{i}_j\}_{j \in [p_i]}$ where $r_i \leq
		p_i$. Let $\tilde{S}$ denote the set of objects $S_{\ell-1}$
		after replacing each $Q_i$ with $\mathcal{L}_i$, and observe
		that $\weight\left( \tilde{S} \right) \leq \weight\left(
		S_{\ell- 1} \right) $. Moreover, observe that each long object
		introduced in the second round is disjoint from all other
		objects. Finally, we bound the number of long objects
		introduced in the second round. Consider $\sum_{i \in [\ell-1]}
		r_i$, which is an upper bound on the number of long objects
		added to $\tilde{S}$. By
		\cref{claim:bounded_objects_to_fat_object}, each fat object in
		$\mathcal{F}_i$ can touch at most $1000 \alpha^{2}$ objects in $N_i$, in other words, each $\mathcal{F}_i$ contributes at most $1000
		\alpha^{2}$ to $\sum_{i \in [\ell-1]} r_i$. Moreover, since $\abs{\mathcal{F}_i} \leq \abs{D}
		= O\left(\abs{T}\right)$,  it holds that
		\begin{equation*}
			\left( \sum_{i \in [s]} r_i  \right) = O\left( \abs{T}\right).
		\end{equation*}
		Hence, we have $\abs{\tilde{S}} = O\left(\abs{T}\right)$.
		This concludes round 2.

		Now, let us show that the properties in the statement of the claim hold.
		The first property can be easily verified by analyzing the construction,
		whereas the second property follows from $I$ satisfying \assp.
		Observe that at the end of round 2, if a fat object of degree at
		least three is $\alpha$-close to a long object, then it is
		$\alpha$-close to an endpoint of $L$. This fact, together with the
		second property implies that there are at most $1000\cdot \alpha^{2}$
		objects close to a long object. Similarly, since each fat object
		has diameter at most $\alpha$, and by the second property,
		there are at most $1000\cdot \alpha^{2}$
		objects close to a fat object.
		This proves the claim.
	\end{claimproof}

	Next, we prove the other direction of the lemma, i.e. show that if $I'$
	has a solution, then $I$ also has a solution with a smaller or  equal
	weight.
	\begin{claim}\label{claim:assp_B_to_C_ii}
		Suppose $I'$ has a solution $S'$ with cardinality at most $8
		\cdot \abs{T'}$ and weight at most $\gamma$. Then $I$ has a
		solution $S$ with weight at most $\gamma$.
	\end{claim}

	\begin{claimproof}
		We construct $S$ by going over all long objects $Q$ in $S'$.
		We first group the long objects into $\mathcal{L}_1, \ldots, \mathcal{L}_s$
		where for each $i \in [s]$ there exists $x_i, y_i \in V(G)$
		such that $\mathcal{L}_i \subseteq \Omega_{x_i, y_i}$.
		Then, we replace all the objects in $\mathcal{L}_i$
		with $N_i$ where $\biggl( P_{x_i, y_i}, N_i, \weight\left( N_i \right)  \biggr) \in \Omega_{x_i, y_i}$.
		Observe that the weight of $\mathcal{L}_i$ is
		at most $\weight\left( P_{x_i,y_i} \right)  = \weight\left( N_i \right)$, therefore
		the weight of $S$ is at most the weight of $S'$. Moreover, the
		connectivity properties are preserved since the objects in
		$N_i$ cover the long objects in $\mathcal{L}_i$.
		Therefore $S$ is a solution of $I$ with
		weight at most $\gamma$.
	\end{claimproof}

	By \cref{claim:assp_B_to_C_i,claim:assp_B_to_C_ii}, we conclude that
	$I$ and $I'$ are equivalent instances. Observe that $I'$ also satisfies
	the assumption \asspl. This is because $I'$ is constructed from $I$ by
	adding a set of long objects, ensuring that the conditions regarding
	object types and terminals are immediately satisfied for $I'$. Finally,
	if $I'$ is a yes-instance, then, by
	\cref{claim:assp_B_to_C_i,claim:assp_B_to_C_ii}, $I'$ has a solution
	$S'$ that meets the last two conditions in \asspl. Therefore, $I'$
	satisfies the assumption \asspl.

	\paragraph{Running time of the reduction:} The construction of the
	instance $I'$ boils down to the construction of the sets $\Omega_{x,y}$.
	The algorithm goes over all $x,y \in V(G)$, constructs $A_x$, $A_y$ and
	$\Omega_{x,y}$ by running
	Dijkstra's algorithm on $\touchgraph{\objects}$.
	All these operations take polynomial time in the size of the instance.
	Therefore, all in all, the whole procedure
	takes polynomial time.
\end{proof}


\newcommand{\CurrObj}{K}
\newcommand{\SpanningTree}{\mathcal{T}}
\newcommand{\Scur}{\textnormal{\texttt{Curr}}}
\newcommand{\Spath}{\textnormal{{\texttt{Inner}}}}
\newcommand{\Sshort}{\textnormal{{\texttt{Outer}}}}
\newcommand{\Sdisk}{\textnormal{{\texttt{Disk}}}}
\newcommand{\Slong}{\textnormal{{\texttt{Long}}}}
\newcommand{\Sfat}{\textnormal{{\texttt{Fat}}}}
\newcommand{\Curr}{\Scur}
\newcommand{\ord}{\sigma}

\subsection{\cref{lem:createindep} (disjoint representation)}\label{sec:proof-createindep}
The goal of this section is to prove \cref{lem:createindep}, restated here for reference:

\createindep*

Let $\Slong \subseteq S$ be the set of long objects. Let $\Sdisk \subseteq S$ be
the set of disks of $S$. For every $B(v,r) \in \Sdisk$, let $\Spath_D$ be the
set of shortest paths contained inside $B(v,r - 3\alpha)$, and let $\Sshort_D$
be the set of shortest paths of length at most $3\alpha$ contained strictly
inside $B(v,r) \setminus B(v,r - 3\alpha)$. Let $\Spath := \bigcup_{D \in \Sdisk} \Spath_D$ and $\Sshort :=
\bigcup_{D \in \Sdisk} \Sshort_D$.  Finally, let $\Sfat = S \setminus (\Sdisk
\cup \Slong)$ be the set of fat objects of $S$ (i.e., non-terminals that are fat
but are neither long nor disk objects), and let $\Sfat'$ be the set of shortest
paths inside every fat object. For every $O \in \objects'$, we define $\pi(O)$ to be
the object that gave rise to $O$ in the construction defined
in~\cref{lem:createindep}.

For the purposes of this proof, we can treat every connected component
$\touchgraph{S\cup T}$ separately.

\begin{definition}\label{def:spanning-tree-obj}
    We say that a tree $\SpanningTree = (V_T, E_T)$, where $V_T \subseteq V$ and
    $E_T \subseteq V_T \times V_T$, is a \emph{spanning tree of objects}
    $\Obj \subseteq \objects$ if:
    \begin{enumerate}[(i)]
        \item\label{def:spanning-tree-p0} For every $O \in \Obj$, $\tau(O) \in V_T$.
        \item\label{def:spanning-tree-p2} For every $uv \in E_T$ there exists $O \in \Obj$ such that $u$ and $v$ both touch $O$ or $uv \in E[G]$.
    \end{enumerate}
    An edge of the spanning tree is called \emph{important} if there is no
    object in $\Sdisk$ containing both of its endpoints. A vertex $v \in V_T$ is important if it is an endpoint of
    some important edge or there exists $O \in \Obj$ such that $v = \tau(O)$.
\end{definition}
Observe that if there exists a spanning tree of objects $\Obj$, then $\Obj$ is a solution with the
corresponding forest and set of terminals. We prove that the converse is also
true.

\begin{claim}\label{claim:size-of-spanning-tree}
    If $\Obj$ is connected set of objects, then there exists a spanning tree of $\Obj$ with $\Oh(|\Obj|)$ edges.
\end{claim}

\begin{proof}
    Let $\Psi := \{\tau(O) \; \mid O \in \Obj\}$.  Let $r \in V[\Obj]$ be an arbitrarily selected vertex of
    $\Obj$ that will serve as the root of $\SpanningTree_\Obj$.
    We consider a set of currently selected objects
    $\CurrObj \subseteq \Obj$ to be the set of objects $O \in
    \Obj$ for which there exists $v \in V[\SpanningTree_\Obj]$
    that touches $O$. Notice that $\CurrObj$ initially contains at least one
    object that touches the root $r$.  

    Next, we define a set $P_v \subseteq
    \Obj$ of \emph{pending objects} for $v \in V$ as the set of objects
    $O \in \Obj \setminus \CurrObj$ that are touched by both $\CurrObj$
    and $v$.  For each set $P_v$, let $C_v \subseteq V$ be minimal set of
    \emph{current connectors}, i.e., vertices in $G$ such that every object in $P_v$ is
    touched by some vertex in $C_v$. 

    Observe that $|C_v| \le |P_v|$.
    Next, we create a set $C'_v$ which is set $C_v$ plus the vertices of $\Psi$
    that are touched by at least one object of $P_v$ and are not part of
    $V[\SpanningTree_\Obj]$ already. This way we will guarantee that every vertex
    in the image of $\tau$ is eventually added to the spanning tree.

    We build the spanning tree $\SpanningTree_\Obj$ in a depth-first
    search manner. When considering a vertex $v \in
    V[\SpanningTree_\Obj]$, we compute the set $C_v'$ of current
    connectors augmented with the vertex of $\Psi$. For every $u \in C_v'$, we add an edge $uv$ to
    $\SpanningTree_\Obj$ and recurse into $u$ in the depth-first search manner.
    We terminate when $\Obj = \CurrObj$.

    For the size of the set spanning tree, notice that each time we add an edge
    to $\SpanningTree_\Obj$, the cardinality of $\Obj
    \setminus \CurrObj$ decreases. We stop when this set is empty, so the
    procedure terminates and $\SpanningTree_\Obj$ has size
    $\Oh(|\Obj| + |\Psi|) = \Oh(|\Obj|)$.

    Property~(\ref{def:spanning-tree-p0}) from \cref{def:spanning-tree-obj} is satisfied because every vertex of
    $\Psi$ is
    touched by some object in $\objects$ which means it will be eventually added to
    the set $C_v'$ and $\CurrObj = \Obj$ at the end.
    For Property~(\ref{def:spanning-tree-p2}) from \cref{def:spanning-tree-obj},
    consider edge $uv$ and observe that during the
    procedure there existed an object in $\Obj$ that touched both $u$ and
    $v$. Hence, $\SpanningTree_\Obj$ is indeed a spanning tree of
    $\Obj$.
\end{proof}

Let $\mathcal{T} = (V_T, E_T)$ be the spanning tree constructed in
\cref{claim:size-of-spanning-tree} for objects $S \cup T$ (for one connected
component at the time) and let important vertices be the important vertices of
$\mathcal{T}$. The procedure to construct $W$ consists of
three steps, in which we iteratively construct the set $\Curr$ of current
objects.

\subsubsection{Step 1: Connect disks}\label{sec:step1}

We begin by connecting disks and important vertices that arose from
$\mathcal{T}$. The connections realized by the fat objects will be handled later.

First, we define a function $\eta : V \to \Sdisk$. We let $\eta(v)$ be the
closest center of the disk in $\Sdisk$ to vertex $v \in V$. This function is
well-defined because, by perturbing the distances, we can assume that all
shortest paths in $G$ have distinct lengths.  Based on the centers given by
$\eta$, we define an order function $\ord : V \to \mathbb{R}$ by setting
$\ord(v) = \dist_G(v, \eta(v))$ for every $v \in V$.

Let $\Curr$ be initially the empty set. We construct this set iteratively by
adding objects from $\objects'$ using the following procedure: We consider
vertices of $V_T$ in increasing order according to the $\ord(\cdot)$ function,
until we encounter a vertex $v \in V_T$ with $\ord(v) > r$ whose every neighbor $u \in G$ has $\ord(u) > r$.

Let $v \in V_T$ be the currently considered vertex, and let $\rho$ be the
shortest path between $v$ and $\eta(v)$. Let $\rho'$ be the maximal prefix of
$\rho$ that does not contain any vertices from $V[\Curr]$. Observe that $\rho'$
can be decomposed as $\rho' = \rho_1 \cup \rho_2$ such that $\rho_2 \subseteq
B(\eta(v), r - 3\alpha)$, $\rho_1 \subseteq B(\eta(v), r) \setminus
B(\eta(v), r - 3\alpha)$, and note that both $\rho_1$ and $\rho_2$ are in
$\objects'$. We add $\rho_1$ and $\rho_2$ to $\Curr$ and define $\pi(\rho_1) =
\pi(\rho_2) = \eta(v)$. We add an extra requirement: if a path $\rho$ added in this point contained a vertex
$\tau(F)$ for some $F \in \Sfat$, then we split it into $\rho_1 \cup \{\tau(F)\}
\cup \rho_2$ and add these paths instead. Notice that these paths exist in the
disks as a subpath of a shortest path is a shortest path. Moreover, we let
$\pi(\rho_1) = \pi(\{\tau(F)\}) = \pi(\rho_2)$ be the disk that contains them
so that the connectivity of the objects of the disk is preserved.

We repeat this process until every neighbor $u$ of the remaining vertex $v \in V_T$
has $\ord(u) > r$ and $\ord(v) > r$. This concludes the construction of the set $\Curr$.

Observe that for every $D \in \Sdisk$, the set $\pi^{-1}(D)$ is connected, and
every important point within distance $r$ of a disk is connected to its center.
However, it may happen that components of $\Curr$ that were initially connected
only through disks are now disconnected. We handle these connections in the next
step.

\subsubsection{Step 2: Merge connected components with disk objects}

In the second step of the construction, we consider the connected components of $\Curr$.

\begin{claim}\label{clm:step2}
    If there exists a disk $D \in \Sdisk$ that touches two different components
    of $\Curr$, then there exist $\rho_1, \rho_1', \rho_2, \rho_2' \in \objects'$
    such that:
    \begin{itemize}
        \item The union of paths $\rho_1 \cup \rho_1' \cup \rho_2 \cup \rho_2'$
            connects two different components of $\Curr$,
        \item $\pi(\rho_1) = \pi(\rho_1')$ and $\pi(\rho_2) = \pi(\rho_2')$, 
        \item Objects in $\pi^{-1}(\pi(\rho_1)) \cap (\Curr \cup \{\rho_1, \rho_1'\})$ are connected, 
        \item Objects in $\pi^{-1}(\pi(\rho_2)) \cap (\Curr \cup \{\rho_2, \rho_2'\})$ are connected.
    \end{itemize}
\end{claim}
\begin{proof}
    Assume that there exists a disk that touches two different components of
    $\Curr$, and let $\sigma$ be the shortest path between them. By minimality,
    we can write $\sigma = \sigma_1 \cup \sigma_2$, such that there exist two distinct
    disk centers $c_1$ and $c_2$ for which, for every $i \in \{1,2\}$ and every
    $v \in V[\sigma_i]$, it holds that $\eta(v) = c_i$. Hence, adding
    $\sigma_1$ and $\sigma_2$ to $\Curr$ preserves the connectivity of
    $p^{-1}(\sigma_i)$ and touches two different components of $\Curr$.

    However, it is not necessarily true that $\sigma_i \in \objects'$ for
    $i \in \{1,2\}$. But since $\sigma_i \subseteq B(c_i, r)$, there exist paths
    $\rho_i, \rho_i' \in \objects'$ such that $\rho_i \cdot \rho_i' = \sigma_i$ by
    taking $\rho_i \subseteq B(c_i, r - 3\alpha)$. Observe that the paths
    $\rho_1, \rho_1', \rho_2, \rho_2'$ satisfy the properties of~\cref{clm:step2}.
\end{proof}

The remainder of the construction is straightforward. While there exists a disk
that touches two different components of $\Curr$, we find four paths
$\rho_1,\rho_1', \rho_2,\rho_2' \in \objects'$ that satisfy the properties of \cref{clm:step2} and
add them to $\Curr$. This concludes the second step of the procedure. Note that
the number of added objects is at most $4|\Curr|$, because for each connected
component of $\Curr$ we add at most four paths.

To summarize, at the end of the second step, we have the following property:

\begin{claim}\label{clm:sum-step2}
    If a pair of vertices $v_1, v_2 \in \SpanningTree$ is connected in
    $V[\bigcup\Sdisk]$, then $v_1$ and $v_2$ are
    connected in $V[\bigcup\Curr]$.
\end{claim}

\subsubsection{Step 3: Merge components with fat objects}

Finally, we connect components of $\Curr$ that were initially connected by
fat objects.
\begin{claim}\label{clm:step3}
    If there exists a fat object $F \in \objects$ that touches at least two
    different connected components of $\Curr$, then there exists
    $\rho \in \Sfat'$ that touches two different components of $\Curr$.
\end{claim}
\begin{proof}
    Let $F \in \objects$ be a fat object that touches at least two
    different components of $\Curr$, and let $\rho' \in \Sfat'$ be the shortest
    path inside $F$ touching two such vertices.

    For each vertex of $\rho'$, assign an identifier corresponding to
    the arbitrarily selected connected component of $\Curr$ it touches, or $0$
    if it does not touch any. Since the endpoints of $\rho'$ lie in different components, $\rho'$
    contains a path $\rho$ with (i) different labeled endpoints, and (ii) every
    vertex in between endpoints of $\rho$ has label $0$.
    Notice, that such a path $\rho$ touches two different components of $\Curr$.
\end{proof}

Using~\cref{clm:step3}, the remainder of the construction is as follows. First,
for every fat object $F \in \Sfat$ we add shortest path $\{\tau(R)\}$ to the set
of objects $W$ with $\pi(\{\tau(R)\})$ if it was not already added.
While there exists a fat object that touches at least two different
components of $\Curr$, we add a shortest path $\rho \in \objects'$ such that
$\pi(\rho)$ is a fat object. At the end, we let add long objects $L \in \Slong$ to the set with
$\pi(L) = L$ and return $W = \Curr \cup \Slong$. This concludes the
description of the procedure.

To bound the size of $W$, observe that for every vertex of the spanning tree
$\SpanningTree$, we add at most $\Oh(1)$ objects. Moreover,
by~\cref{claim:size-of-spanning-tree}, the number of vertices in the spanning
tree $\SpanningTree$ is bounded by $\Oh(|S| + |T|)$. By construction,
Properties~\ref{lem:createindep:2},\ref{lem:createindep:3},\ref{lem:createindep:4},\ref{lem:createindep:5}
are satisfied. \cref{clm:step2} guarantees Property~\ref{lem:createindep:6}.
Property~\ref{lem:createindep:tau} holds because we added object
$\{\tau(F)\}$ for every $F \in \Sfat$ in Step 2 and Step 3.

Property~\ref{lem:createindep:emptyfat} holds because if $F \in \Sfat$ was
contained in a disk then every important vertex touching $F$ is also touching a
disk. It remains to notice that in Step 3 we only add objects that are not touched
by disks. 

To complete the proof of Property~\ref{lem:createindep:1}
of~\ref{lem:createindep}, it remains to show the following claim:

\begin{claim}
    If there does not exist a fat object $F \in \objects$ that connects two
    different connected components of $\Curr \cup \Slong$, then $\Curr$ is connected.
\end{claim}
\begin{proof}
    For the sake of contradiction, assume that the spanning tree contains an edge
    $(p_1, p_2)$ touching two different components of $\Curr$.
    If $p_1p_2$ is an edge of $G$ then these two components are connected with
    themselves. Therefore, there exists an object $O \in S$ that touches both $p_1$
    and $p_2$. Notice that $O \in S$ cannot be a long object (as every long object
    is included) or a disk (as otherwise this would contradict~\cref{clm:sum-step2}). Therefore, 
    every $O \in S$ that touches both $p_1$ and $p_2$ is a fat object which 
    contradicts the assumption that such an object does not exist.
\end{proof}


\subsection{Theorem~\ref{th:guardedenum1} (Voronoi separators)}

In the first step, we show that a simple guess is sufficient to move from bounding the balance of the components to presenting a balanced separation.
\begin{theorem}\label{th:guardedenum2}
  Let $G$ be an edge-weighted $n$-vertex planar graph, $\Obj$ a set of
  $d$ connected subsets of $V(G)$, and $k$ an integer. We can
  enumerate (in time polynomial in the size of the output and using polynomial working space) a set
  $\Vorsepfam$ of $(d+n)^{\Oh(\sqrt{k})}$ guarded separations $(Q,\Gamma,A,B)$ with
  $Q\subseteq \Obj$, $|Q|\le \lambda \sqrt{k}$ (for some universal constant $\lambda$) such
  that the following holds. If $\Fam\subseteq \Obj$ is a set of $k$
  pairwise disjoint objects, then there is a pair $(Q,\Gamma)\in
  \Vorsepfam$ such that
\begin{enumerate}
\item[(a)] $Q\subseteq \Fam$,
\item[(b)] for every $v\in \Gamma$ and  $p\in \Fam\setminus Q$, we have $\dist_G(p,v)>\min_{p'\in Q}\dist_G(p',v)$.
\item[(c)] there are at most
  $\frac{2}{3}k$ objects of $\Fam$ that are fully contained in $A$, and
  there are at most
  $\frac{2}{3}k$ objects of $\Fam$ that are fully contained in $B$.
\end{enumerate}
\end{theorem}
\begin{proof}
  First, let us invoke the algorithm of Theorem~\ref{th:guardedenum0}; let $\Vorsepfam$ be the returned family of guarded separators.
  For every $(Q,\Gamma)\in \Vorsepfam$, we proceed as follows. Let $K_1$, $\dots$, $K_t$ be an ordering of the vertex sets of the  components of $G-\Gamma$. We output $2t-1$ guarded separations $(Q,\Gamma,A,B)$, with $(A,B)$ going over the following bipartitions of $V(G)\setminus \Gamma$:
  \begin{itemize}
  \item for every $1\le i \le t$, the bipartition $(K_i,\bigcup_{j\in[t],j\neq i}K_j)$, and
  \item for every $1\le i \le t-1$, the bipartition $(K_1\cup \dots \cup K_i,K_{i+1}\cup \dots \cup K_t)$.
  \end{itemize}
Note that we output at most $(2n-1)d^{\Oh(\sqrt{k})}$ guarded separations.
  
We claim that for every normal subfamily $\Fam\subseteq \Obj$ of cardinality exactly $k$, the required guarded separation exists. By Theorem~\ref{th:guardedenum0}, there is a $(Q,\Gamma)\in \Vorsepfam$ satisfying properties (a)--(c) of Theorem~\ref{th:guardedenum0}; in particular, every component of $G-\Gamma$ fully contains at most $\frac{2}{3}k$ objects from $\Fam$.  Let $K_1$, $\dots$, $K_t$ be an ordering of the vertex sets of the components of $G-\Gamma$. Any of the $2t-1$ bipartitions we have output for $(Q,\Gamma)$ satisfies properties (a) and (b) of the theorem being proved; we only have to show that the balance condition (c) holds for at least one of these bipartitions.

  We consider two cases. Suppose first that some component $K_i$ contains at least $k/3$ objects from $\Fam$. By our assumption on $(Q,\Gamma)$,  we also know that $K_i$ contains at most $\frac{2}{3}
  k$ objects of $\Fam$. Thus the bipartition $(A,B)=(K_i,\bigcup_{j\in[t],j\neq i}K_j)$ satisfies property (c).
  
  Assume now that every  $K_i$ contains less than $k/3$ objects from $\Fam$. Let $i$ be the largest value such that $K_1\cup \dots \cup K_i$ contains less than $\frac{2}{3}k$ objects from $\Fam$ (note that $i\ge 1$). We claim that the bipartition $(A,B)=(K_1\cup \dots \cup K_{i},K_{i+1}\cup \dots \cup K_t)$ satisfies property (c). The choice of $i$ ensures that $A$ fully contains strictly less than $\frac{2}{3}k$ objects of $\Fam$. If $B$ fully contains strictly more than  $\frac{2}{3}k$ such objects, then we have that $A$ contains strictly less than $k/3$ objects from $\Fam$. By the assumption that $K_{i+1}$ contains less than $k/3$ objects, this contradicts the maximum choice of $i$.  
\end{proof}  

Next, we prove Theorem~\ref{th:guardedenum1} by showing that the objects of
$\Fam_0$ can be given a higher weight by attaching additional objects next to them.

\guardedenum*
\begin{proof}
  We extend the graph $G$ to a graph $G^*$ by introducing, for every vertex $v\in V(G)$, a set of $4k$ degree-1 neighbors $v^{1}$, $\dots$, $v^{4k}$. The weight of the edge connecting $v$ and $v^i$ can be arbitrary (as long as we ensure that distances are distinct). 
  
  We extend $\Obj$ to a family $\Obj^*$ in the following way. For every object $O\in \Obj$, we pick an arbitrary $\lambda(O)\in O$. For every $1\le i \le 4k$, we introduce an object $O^i=\{\lambda(O)^i\}$ into $\Obj^*$. Note that $\Obj^*$ is multiset (since if $\lambda(O_1)=\lambda(O_2)$, then $O^i_1=O^i_2$) and has size $4kd$.
 
  We iterate over $k'\in [k]$, creating an output that prepares us for the case when $|\Fam_0|=k'$. 
  Let $\lambda$ be the universal constant from Theorem~\ref{th:guardedenum2}; we can assume that $\lambda\ge 1$.
We also assume that $k\ge \lambda$: otherwise, we can obtain a trivial collection $\Vorsepfam$ of size $|\Obj|^k=|\Obj|^{O(1)}$ by having a tuple $(Q,\emptyset,V(G)\setminus Q,\emptyset)$ for every possible set $Q\subseteq \Obj$ of size $k$.
  Let us define $\theta:=\lceil (608\lambda)k/k' \rceil \le (1216\lambda)(k/k') \le 1216\lambda k$ and $k^*=k+\theta k'\le (1217\lambda)k$.
 For a given $k'\in[k]$, let us invoke Theorem~\ref{th:guardedenum2} for $G^*$,  $\Obj^*$, and $k^*$. Let $\Vorsepfam^*$ be the returned set of Voronoi separations.
 
 Based on $\Vorsepfam^{*}$, we define the output $\Vorsepfam$ of our algorithm in the following way. Let us consider every separation $(Q^*,\Gamma^*,A^*,B^*)\in \Vorsepfam^{*}$. For each such tuple, let us introduce $(Q^*\cap \Obj,\Gamma^*\cap V(G),A^*\cap V(G),B^*\cap V(G))$ into $\Vorsepfam$. Observe that the fact that there is no edge of $G^*$ between $A^*$ and $B^*$ implies that there is no edge of $G$ between $A^*\cap V(G)$ and $B^*\cap V(G)$.

   Let $\Fam\subseteq \Obj$ be a subfamily of cardinality exactly $k$ and let $\Fam_0\subseteq \Fam$ be of cardinality exactly $k'$.
   Let us define $\Fam^*\supseteq \Fam$ the following way: for every $O\in \Fam_0$, let us extend $\Fam$ with $O^1$, $\dots$, $O^\theta$. Thus $|\Fam^*|$ is exactly $k^*=k+\theta k'$. Theorem~\ref{th:guardedenum2} guarantees that there exists a guarded separation $(Q^*,\Gamma^*,A^*,B^*)\in \Vorsepfam^{*}$ satisfying properties (a)--(c) of Theorem~\ref{th:guardedenum2}.
   There is a corresponding guarded separation  $(Q,\Gamma,A,B)=(Q^*\cap \Obj,\Gamma^*\cap V(G),A^*\cap V(G),B^*\cap V(G))\in \Vorsepfam$. We want to show that  $(Q,\Gamma,A,B)$  satisfies properties (a)--(c) of the theorem being proved. Property (a) is clear. To verify property (b), let us first observe that $\Gamma\subseteq \Gamma^*$ (so we are imposing fewer constraints) and that the distance between two vertices $u,v\in V(G)$ is the same in $G$ and in $G^*$. However, there is a corner case that needs to be checked: it is possible that for some $v\in \Gamma^*\cap V(G)$, we have that $\min_{p'\in Q^*}\dist(p',v)$ is minimized by some $p'=O^i\not\in \Obj$, hence $O^i$ is not in $Q$. In this case, as $O^i\in \Fam^*$, object $O$ is in $\Fam^*$ as well, and (as $v\in V(G)$) object $O$ is closer to $v$ than $O^i$, contradicting that $O^i$ minimizes the distance.
   
   The key part of the proof is proving property (c). Suppose that $O\in \Fam_0$ is fully contained in $A$. We claim that $O^i$ for $i\in [\theta]$ is either contained in $A^*$ is in $Q^*$. As $O$ and $O^i$ is connected by the edge $\lambda(O)\lambda(O)^i$ of $G^*$, object $O^i$ is in $A^*$, unless $\lambda(O)^i\in \Gamma^*$. However, in this case $O^i$ has to be in $Q^*$, otherwise
$\min_{p'\in Q^*}\dist(p',\lambda(O)^i)$ would be greater than $\dist(O^i,\lambda(O)^i)=0$.

   Suppose for a contradiction that the number $c$ of objects from $\Fam_0$
   fully contained in $A$ is greater than $\frac{3}{4}k'$. The argument in the
   previous paragraph implies that the number of objects of $\Fam^*$ fully
   contained in $A^*$ is at least
   \begin{align*}
     c(\theta+1)-|Q^*|&> c\theta-\lambda\sqrt{k^*}
     >
     \frac{3}{4}k'\theta-\lambda\sqrt{k^*}=\frac{2}{3}\theta k'+\frac{1}{12}\theta k' -\lambda\sqrt{1217\lambda k}\\
&\ge \frac{2}{3}\theta k'+\frac{1}{12}(8+600\lambda)(k/k')k' -\lambda\sqrt{1217k^2}\\
&\ge \frac{2}{3}\theta k'+\frac{2}{3}k + 50\lambda k  -\sqrt{1217}\lambda k
     \ge 
    \frac{2}{3}\theta k'+\frac{2}{3}k=\frac{2}{3}k^*,
  \end{align*}
contradicting the assumption that $(Q^*,\Gamma^*,A^*,B^*)$ satisfies property (c) of Theorem~\ref{th:guardedenum2}.
A similar argument shows that $B$ contains at most $\frac{3}{4}k'$ objects of $\Fam_0$.
\end{proof}

  \subsection{Lemma~\ref{lem:recursion0} (executing recursion)}

  First we prove Lemma~\ref{lem:recursion0} under the simplifying assumption
  that $Q\subseteq T$ (i.e., every object in the separator is a terminal) and
  $X=Q$ (i.e., the distinguished terminals are precisely the separator
  vertices). These assumptions make the proof notationally much simpler. Then we
  show that the instance and the balanced triple can be modified to satisfy
  these assumptions simply by making every vertex of $Q$ a terminal and moving
  $X$ into $Q$. Recall that the way weakly solving $I$ under Assumption \asspl\ is defined, the returned solution does not have to satisfy \asspl.

  \begin{lemma}\label{lem:recursion}
    Let $I=(G,k,\objects,T,X,F)$ be an instance of \stplanarx\ with $X=Q\subseteq T$. If a $\beta$-balanced triple $(T_1,T_2,Q)$ is given for a solution $S$ that has minimum cost among those satisfying Assumption \asspl, then in time $2^{O(|Q|\log |Q|)}\textup{poly}(|I|)$ we can reduce weakly solving $I$ under Assumption \asspl\ to
weakly solving multiple instances $(G,k',\objects,T',X',F')$ under Assumption \asspl, where each such instance satisfies $k'\le \beta k$, $|E(F')|\ge |E(F)|$. Moreover, there are at most $\Oh(k|X|^{2c})$ such instances where $|E(F')|\le |E(F)|+c$. The running time of the reduction is polynomial in the instance size and the number of created instances.
\end{lemma}

\begin{proof}
  For $k'=0,\dots, \beta k$ and for every forest $F'\supseteq F$ on $X$, let us consider the instance $(G,k',\objects, T_1\cup Q,X,F')$. Let us try to weakly solve this instance under Assumption \asspl; if it is successful, let $S^1_{k',F'}$ be the returned solution.
  Similarly, let us consider the instance $(G,k',\objects, T_2\cup Q,X,F')$. Let us try to weakly solve this instance under Assumption \asspl; if it is successful, let $S^2_{k',F'}$ be the returned solution. Next, we consider the union $S'=S^1_{k^1,F^1}\cup S^2_{k^2,F^2}$ for every $k^1$, $k^2$ with $k^1+k^2\le k$ and for every $F^1$, $F^2$. The algorithm checks for each such $S'$ if it is a feasible solution of $I=(G,k,\objects,T,X,F)$ and returns the best feasible solution found this way. We claim that this algorithm weakly solves instance $I$ under Assumption \asspl.

Recall that $(T_1,T_2,Q)$ is a $\beta$-balanced triple for some $S$, where $S$ is solution of $(G,k,\objects,T,X,F)$ having minimum cost among those satisfying \asspl. Let $(A,B)$ be the corresponding bipartition of $(S\cup T)\setminus Q$. Let $A^-=A\setminus T_1$ and $B^-=B\setminus T_2$; observe that $A^-\cup B^-=S$. Let $F_B$ be a forest on $Q$ where two vertices of $Q$ are in the same component if and only if they are in the same component in $\touchgraph{B\cup Q}$. Let $F'_B\supseteq F$ be a spanning forest of $F\cup F_B$. Let $k_A=|A^-|$ and $k_B=|B^-|$. Note that $k_A\le |A|\le \beta k$, hence the pair $(k',F')=(k_A,F_B)$ was considered during the algorithm.
\begin{claim}\label{cl:leftfeasible}
$A^-$ is a feasible solution of $(G,k_A,\objects, T_1\cup Q,X,F'_B)$ satisfying Assumption \asspl. 
\end{claim}  
\begin{proof}
  Let $H$ be the graph defined  by removing every $A-B$ edge from $\touchgraph{S\cup T}$ and then adding the edges of $F$. By Definition~\ref{def:triplenew}, $H$ is connected. Observe that the neighborhood of $B$ is in $Q$. By the definition of $F_B$, this means that every terminal in $T_1\cup Q$ is in the same component of $\touchgraph{A\cup Q}\cup F'_B$: any subpath involving a vertex of $B$ can be replaced by an edge of $F_B$ (and hence a path of $F'_B$) between two vertices of $Q$. Thus $A^-$ is indeed a feasible solution and it satisfies \asspl, since $A^-\cup(T_1\cup Q)\subseteq S\cup T$.
\end{proof}
Claim~\ref{cl:leftfeasible} implies that our algorithm returns a solution $S^1_{k_A,F_B}$ for  $(G,k_A,\objects, T_1\cup Q,X,F'_B)$ whose weight is not larger than the weight of $A^-$. 
Let $F_A$ be a forest on $Q$ where two vertices are in the same component of $F_A$ if and only if they are in the same component of $\touchgraph{S^1_{k_A,F_B}\cup T_1\cup Q}$. It is crucial to observe that $S^{-}{k_A,F_B}$ and $A^-$ could be very different solutions, hence the components of $\touchgraph{S^1_{k_A,F_B}\cup T_1\cup Q}$ and $\touchgraph{A^{-}\cup T_1\cup Q}$ may partition $Q$ differently. We stress that forest $F_A$ is defined based on $S^1_{k_A,F_B}$. Let $F'_A\supseteq F$ be a spanning forest of $F\cup F_A$.

\begin{claim}\label{cl:rightfeasible}
$B^-$ is a feasible solution of $(G,k_B,\objects, T_2\cup Q,X,F'_A)$ satisfying Assumption \asspl.
\end{claim}  
\begin{proof}
   Since $S^1_{k_A,F_B}$ is a feasible solution of $(G,k_A,\objects, T_1\cup Q,X,F'_B)$,
   the graph $\touchgraph{S^1_{k_A,F_B}\cup T_1\cup Q}\cup F'_B$ is connected.
   Any connection involving $S^1_{k_A,F_B}\cup T_1$ can be replaced by an edge of $F_A$ between two vertices of $Q$, hence $F\cup F_A \cup F_B$ is also connected. By definition of $F_B$, every edge of $F_B$ can be expressed by a path going through $B\cup Q$, hence the set $Q$ appears in a single component of $\touchgraph{B\cup Q}\cup F'_A$. Moreover, $\touchgraph{B\cup Q}$ connects every vertex of $T_2$ to $Q$. Thus, $T_2\cup Q$ appears in a single component of
   $\touchgraph{B\cup Q}\cup F'_A=\touchgraph{B^-\cup (T_2\cup Q)}\cup F'_A$, implying that $B^-$ is indeed a feasible solution of $(G,k_B,\objects, T_2\cup Q,X,F'_A)$. Furthermore,
    it satisfies \asspl, since $B^-\cup(T_2\cup Q)\subseteq S\cup T$.
\end{proof}
Claim~\ref{cl:rightfeasible} implies that our algorithm obtains a solution $S^2_{k_B,F_A}$ for  $(G,k_B,\objects, T_2\cup Q,X,F'_A)$ whose weight is not larger than the weight of $B^-$.

\begin{claim}\label{cl:replace2}
$S^1_{k_A,F_B}\cup S^2_{k_B,F_A}$ is a feasible solution of  $(G,k,\objects,T,X,F)$.
\end{claim}  
\begin{proof}
  By the way $S^2_{k_B,F_A}$ was obtained, every vertex of $Q\cup T_2$ is in the same component of $\touchgraph{S^2_{k_B,F_A}\cup T_2\cup Q}\cup F'_A$. By the definition of $F_A$,  every vertex of $Q\cup T_2$ is in the same component of $\touchgraph{S^1_{k_A,F_B}\cup S^2_{k_B,F_A}\cup T}\cup F$: connections using edges of $F_A$ can be replaced by paths using objects of $S^1_{k_A,F_B}$. Furthermore, the fact that $S^1_{k_A,F_B}$ is a feasible solution of $(G,k_A,\objects, T_1\cup Q,X,F'_B)$ implies that every terminal in $T_1$ is connected to some vertex of $Q$ in $\touchgraph{S^1_{k_A,F_B}\cup T_1\cup Q}\cup F$. Thus every terminal in $T$ is in the same component of $\touchgraph{S^1_{k_A,F_B}\cup S^2_{k_B,F_A} \cup T}\cup F$ as $Q$, hence $S^1_{k_A,F_B}\cup S^2_{k_B,F_A}$ is a feasible solution of  $(G,k,\objects,T,X,F)$.
\end{proof}
Claim~\ref{cl:replace2} shows that $S^1_{k_A,F_B}\cup S^2_{k_B,F_A}$ is one of the solutions considered by the algorithm. Claim~\ref{cl:leftfeasible} implies  that the weight of $S^1_{k_A,F_B}$ is at most the weight of $A^-$, while Claim~\ref{cl:rightfeasible} implies that the weight of $S^2_{k_B,F_A}$ is at most the weight of $B^-$. This shows that the weight of $S^1_{k_A,F_B}\cup S^2_{k_B,F_A}$ is at most the weight of $A^-\cup B^-=S$, showing that the solution returned by the algorithm is not worse than $S$.

Let us prove the stated bound on the number of created instances. Observe that in each created instance $(G,k',\objects, T_1\cup Q,X,F')$ or  $(G,k',\objects, T_2\cup Q,X,F')$, the forest $F'$ is a supergraph of the forest $F$, hence $|E(F')|\ge |E(F)|$. Moreover, let us observe that if $F'$ (a forest on $X$) is created from $F$ by adding at most $c$ edges to it, then there are at most $|X|^{2c}$ ways of doing this. There are at most $k+1$ possibilities for $k'$ and there are only two choices for the terminal set ($T_1\cup Q$ or $T_2\cup Q$), hence the stated bound on the number of instances follows.  
\end{proof}  

Now we complete the proof of \cref{lem:recursion0} by removing the simplifying assumption $X=Q\subseteq T$ of \cref{lem:recursion}.

\recursion*
\begin{proof}
   The fact that $(T_1,T_2,Q)$ is a $\beta$-balanced triple for $S$ implies $Q\subseteq S\cup T$; more precisely, $Q\setminus T\subseteq S$ and $Q\cap T\subseteq T$.
   Let us construct the instance $I'=(G,k,\objects,T',X',F)$ with $T':=T\cup (Q\setminus T)$ and $X':=Q\cup X \subseteq T'$. We claim that $S'=S\setminus (Q\setminus T)$ is a feasible solution of $I'$: indeed, this clearly follows from $S'\cup T'=S\cup T$. Moreover, as $S\cup T$ satisfies \asspl, so does $S'\cup T'$ as well. Conversely, if $S'$ is a solution of $I'$ satisfying \asspl, then $S'\cup (Q\setminus T)$ is a solution of $I$, also satisfying \asspl. This means that weakly solving $I$ under assumption \asspl\ can be reduced to weakly solving $I'$ under assumption \asspl: we simply need to add the set $Q\setminus T$ to the solution obtained for $I'$.

   Let $(A,B)$ be the partition showing that $(T_1,T_2,Q)$ is a $\beta$-balanced triple for solution $S$ of $I$.
   Let $T'_1=T_1\setminus X$, $T'_2=T_2\setminus X$, and $Q'=Q\cup X$. 
   We claim that the bipartition $(A\setminus X,B\setminus X)$ shows that $(T'_1,T'_2,Q')$ is a $\beta$-balanced triple for $S'$ in $I'$.  As $A$, $B$, and $Q$ are disjoint, we have $A'\cap T'=(A\setminus X)\cap (T\cup Q)=(A\cap T)\setminus X=T_1\setminus X=T'_1$ and, similarly, $B'\cap T'=T'_2$.
   Furthermore, $|A'|\le |A|\le \beta |S\cup T|=
   \beta|S'\cup T'|$ and, similarly, $|B'|\le |S'\cup T'|$.
For the last item of Definition~\ref{def:triplenew}, observe that $\touchgraph{S'\cup T'}=\touchgraph{S\cup T}$ and $A'\subseteq A$, $B'\subseteq B$ imply that we are making fewer removals if we remove the $A'-B'$ edges instead of the $A-B$ edges. This means that the fact that $(A,B)$ satisfied this condition for $S\cup T$ implies that $(A',B')$ also satisfies this condition for $S'\cup T'$. Thus we have shown that $(T'_1,T'_2,Q')$ is a $\beta$-balanced triple for $S'$ in $I'$. 
  
Observe that instance $I'$, solution $S'$, and triple $(T'_1,T'_2,Q')$ satisfy the requirements of Lemma~\ref{lem:recursion}: we have  $X'=Q'$ and $Q'\subseteq T'$. Therefore, we can weakly solve instance $I'$ under Assumption \asspl. As we have observed before, this is sufficient to weakly solve instance $I$ under Assumption \asspl.

The bound on the number of constructed instances follows from the bound given by Lemma~\ref{lem:recursion} for the instance $I'$.
\end{proof}  

\subsection{Lemma~\ref{lem:listtriples} (listing balanced triples)}

In this section, we prove \cref{lem:listtriples}. As the algorithm needs to
return a list of triples, we will be proving an algorithmic
statement. However, note that the algorithm does not have access to
solutions and these appear in the proof only in the analysis. That is, in the proof that the list contains a required
balanced triple, we imagine a hypothetical solution $S$, for which there is a
hypothetical disjoint representation $W$ by Lemma~\ref{lem:createindep}, for which there is a suitable Voronoi separator in a list given by Theorem~\ref{th:guardedenum1}. 

\listtriples*
\begin{proof}
  Consider the set $\objects'$, as defined in Lemma~\ref{lem:createindep}. As $T$ is irredundant, there is an injective function $\tau:T\to V(G)$ such that $\tau(O)\in O$ for every $T$. As $S$ is an inclusionwise minimal solution, every $O\in S$ should have a vertex not contained in any other object of $S\cup T$: otherwise, removing $S$ would not disconnect the solution. Therefore, we can extend $\tau$ to an injective function $\tau: S\cup T\to V(G)$ with $\tau(O)\in O$ for every $O\in S\cup T$.   
Let $k^*=c_1(k+|T|)$, where $c_1$ is the constant in Lemma~\ref{lem:createindep}. Let us invoke the algorithm of Theorem~\ref{th:guardedenum1} with the set of objects $\objects'$ and $k^*$; let $\Vorsepfam'$ be the returned set of guarded separations.
For each guarded separator $(Q',\Gamma',A',B')\in \Vorsepfam'$, we proceed as follows.
Let $c_2$ be a constant depending on $\alpha$ that we will set later in Claim~\ref{cl:Qsizebound}.
We guess a set $Q$ of $c_2|Q'|=\Oh_\alpha(\sqrt{k^*})$ objects from $\Obj$ that includes every terminal $O\in T$ with $\tau(O)\in \Gamma'$ (if there are more than $c_2|Q'|$ such terminals, then obviously there is no such set $Q$). We let $T_1$ (resp., $T_2$) contain every terminal $O\in T$ with $\tau(O)\in A'$ (resp., $\tau(O)\in B'$).  We output $(T_1,T_2,Q)$  as one of the possible triples. This way, we output at most $|\Vorsepfam'|\cdot |\objects'|^{O(\sqrt{k^*})}=|I|^{O(\sqrt{k+|T|})}$ triples.

We have to show that one of the triples in the output is a $\beta$-balanced triple. Let $S$ be an inclusionwise minimal solution satisfying Assumption \asspl.
Lemma~\ref{lem:createindep} for the set $S$ and the function $\tau$ gives a set $W$ and a mapping $\pi:W\to S\cup T$.

 Let $\Fam=W$. For every $O\in S\cup T$, let $\Fam_0$ contain the objects $\{\tau(O)\}$, which is in $\Fam$ by Lemma~\ref{lem:createindep}~\ref{lem:createindep:tau}.   By the statement of
 Theorem~\ref{th:guardedenum1}, there is a guarded separation
 $(Q',\Gamma',A',B')\in \Vorsepfam'$ in the output of the enumeration algorithm satisfying
 properties (a)--(c). Let $Q_0\subseteq \objects$ contain an object $O$ if $\pi^{-1}(O)\cap Q' \neq \emptyset$ or $\{\tau(O)\}\in Q'$. Let us define $Q$ to be the superset of $Q_0$ that contains every fat object $O$ of $\objects$ that has $\pi^{-1}(O)\neq \emptyset$ and is at distance at most $\alpha$ from $O'$ for some $O'\in Q_0$. 

\begin{claim}\label{cl:Qsizebound}
  $|Q|\le c_2\sqrt{k^*}$ for some constant $c_2$ depending on $\alpha$.
\end{claim}
\begin{proof}
The size of $Q'$ is $O(\sqrt{k^*})$ by Theorem~\ref{th:guardedenum1} and we have $|Q_0|\le 2|Q'|=O(\sqrt{k^*)}$. For every
$O'\in Q_0$, if $\pi(O')$ is a fat or long object, then the
conditions of Assumption \asspl\ imply that there are at most $O_\alpha(1)$ fat
objects  at distance $\alpha$ from $O'$. If $\pi(O')$ is a disk $B(v,r)$, then
$O'$ is either fully contained in the disk $B(v,r-3\alpha)$ or $O'$ has length
at most $3\alpha$. In the former case, every fat object $O^*$ at distance at
most $\alpha$ from $O'$ is fully inside the disk $B(v,r)$, in which case
property~\ref{lem:createindep:emptyfat} in Lemma~\ref{lem:createindep} implies that $\pi^{-1}(O^*)=\emptyset$, which means that $O^*$ was not added to $Q$.
In the latter case, every fat object at distance at most $\alpha$ from $O'$ intersects a ball of radius $4\alpha$, hence Assumption \asspl\ implies that there are at most $O_\alpha(1)$ such objects. Therefore, we can conclude that $|Q\setminus Q_0|$ is $O_\alpha(|Q_0|)$, hence $|Q|=O_\alpha(\sqrt{k^*})$.
\end{proof}
This means that the triple $(T_1,T_2,Q)$ was part of the output of our enumeration, where  $T_1$ (resp., $T_2$) contains every terminal $O\in T$ with $\tau(O)\in A'$ (resp., $\tau(O)\in B'$). We show that this tuple is a $\frac{3}{4}$-balanced triple. 
To this end, we show first that the guarded separation $(Q',\Gamma',A',B')$ can be used to define a bipartition of the objects in $\objects\setminus Q$.

\begin{claim}\label{cl:bipartition}
  For every $O\in \objects\setminus Q$, if $\tau(O)\in A'$ (resp., $\tau(O)\in B'$), then all the sets in $\pi^{-1}(O)$ are fully contained in $A'$ (resp., $B'$).
\end{claim}  
\begin{proof}
  If an object $O'\in \pi^{-1}(O)$ intersects $\Gamma$, then $O'\in Q$ and hence
  $O$ is in $Q_0\subseteq Q$. As there is no edge between $A'$ and $B'$, this
  means that each object $O'\in \pi^{-1}(O)$ is fully contained either in $A'$ or
  in $B'$. If $\tau(O)\in \Gamma'$, then $O\in Q_0\subseteq Q$, thus we can assume that $\tau(O)$ is in either $A'$ or $B'$.

  If $O$ is a long object, then $\pi^{-1}(O)=\{O\}$ and $\tau(O)\in O$, hence there is nothing to prove.

  If $O$ is a disk object, then $\pi^{-1}(O)$ is a connected set of objects
  (i.e., the touching graph of $\pi^{-1}(O)$ is connected). This means that
  either all of them are in $A'$, all of them are in $B'$, or there are two
  touching objects $O_1,O_2\in \pi^{-1}(O)$ such that $O_1\subseteq A'$ and
  $O_2\subseteq B'$. However, this last situation would contradict the
  assumption that no edge of $G$ connects $A'$ and $B'$. Moreover, by property~\ref{lem:createindep:6} in Lemma~\ref{lem:createindep}, $\{\tau(O)\}\in \pi^{-1}(O)$, thus $\tau(O)$ is in the same part as $\pi^{-1}(O)$.

 Assume now that $O$ is a fat object and, without loss of generality, $\tau(O)\in A'$, but there is an $O'\in  \pi^{-1}(O)$ with
 $O' \subseteq B'$; in particular, this means that $\pi^{-1}(O)\neq\emptyset$. As every fat object has a diameter at most $\alpha$, there is a path $P$ of length at most
 $\alpha$ between $\tau(O)$ and $O'$. As there is no edge between $A'$ and $B'$,
 path $P$ has to go through a vertex $v\in \Gamma'$. Vertex $v$ is at distance at most $\alpha/2$ either from $\tau(O)$ or $O'$. This means
 that, by property (b) of Theorem~\ref{th:guardedenum1}, there is a $p\in Q'$
 that has distance at most $\alpha/2$ from $v$, and hence distance at most
 $\alpha$ either from $\tau(O)\in O$ or from $O'\subseteq O$. In both cases, it follows that $\pi(p)\in Q_0$ is at distance at most
 $\alpha$ from $O$, hence $O\in Q$  by construction (as $\pi^{-1}(O)\neq \emptyset$), a contradiction.
\end{proof}

Based on Claim~\ref{cl:bipartition}, we define a bipartition $(A,B)$ of $(S\cup
T)\setminus Q$ by letting $O\in A$ (resp., $O\in B$) if all the sets in $\pi^{-1}(O)$ are in $A'$ (resp., in $B'$). In particular, for some $O\in T$, this means that if $\tau(O)\in A'$ (resp., $\tau(O)\in B'$), then $O\in A$ (resp., $O\in B$). Thus $T_1=T\cap A$ and $T_2=T\cap B$.
\begin{claim}\label{cl:balance}
We have $|A|,|B|\le \frac{3}{4}(k+|T|)$.
\end{claim}  
\begin{proof}
If $O\in \objects$ is in $A$, then by Claim~\ref{cl:bipartition} the unique
object $O'\in \pi^{-1}(O)\cap \Fam_0$ is in $A'$. By property (c) of Theorem~\ref{th:guardedenum1}, at most $\frac{3}{4}|\Fam_0|=\frac{3}{4}(k+|T|)$ of the objects in $\Fam_0$ can be in $A'$, implying $|A|\le \frac{3}{4}(k+|T|)$. A similar argument bounds $|B|$.
\end{proof}  
Finally, we show that the optimum solution set $S$ and the bipartition $(A,B)$ satisfy the requirements of Definition~\ref{def:triplenew}, showing that  $(T_1,T_2,Q)$ is indeed a $\frac{3}{4}$-balanced triple. We have seen that $T_1=T\cap A$ and $T_2=T\cap B$, and the property  $|A|,|B|\le \frac{3}{4}|S\cup T|$ was shown in Claim~\ref{cl:balance}.

Let $H$ be the graph defined by removing every $A-B$ edge from
$\touchgraph{S\cup T}$ and then adding $F$. Suppose that there are two terminals $t_1,t_2\in T$ that are in different components of $H$. Let $P$ be an $t_1-t_2$ path in $\touchgraph{S\cup T}\cup F$; let us choose $t_1$, $t_2$, $P$ such that $P$ is shortest possible. If path $P$ goes through an edge $x_1x_2$ of $F$, then $x_1,x_2\in X\subseteq T$ and the edge $x_1x_2$ is in $H$. Thus, either $x_1$ and $t_1$ are in different components of $H$, or $x_2$ and $t_2$ are in different components of $H$, contradicting the minimal choice of $P$. Therefore, we can assume that $P$ does not use edges of $F$, hence $t_1$ and $t_2$ are in the same component of $\touchgraph{S\cup T}$.
As $W$ was provided by Lemma~\ref{lem:createindep}, property \ref{lem:createindep:1} implies that there are $t'_1,t'_2\in W$ (not necessarily terminals) with $\tau(t_i)\in t'_i$, $i=1,2$, and $t'_1$, $t'_2$ are in the same component of $\touchgraph{W}$. Now $\tau(t_i)\in t'_i$ implies that $t_i$ and $\pi(t'_i)$ touch, thus they are in the same component of $H$ (it is not possible that $t_i\pi(t'_i)$ is an $A-B$ edge, as they intersect). Therefore, $t'_1$ and $t'_2$ are in the same component of $\touchgraph{W}$, but $\pi(t'_1)$ and $\pi(t'_2)$ are in different components of $H$.

The previous paragraph shows that we can select two objects $O'_1,O'_2\in W$ that are in the same component of $\touchgraph{W}$, but $\pi(O'_1)$ and $\pi(O'_2)$ are in different components of $H$. This means that we can select objects $O'_1,O'_2\in W$ (with $\pi(O'_i)$ not necessarily being a terminal) such that they are adjacent in $\touchgraph{W}$, but $\pi(O'_1)$ and $\pi(O'_2)$ are not adjacent in  $H$. In other words, $\pi(O'_1)\in A$, $\pi(O'_2)\in B$, and $O'_1$ and $O'_2$ touch each other.
However, $\pi(O'_1)\in A$ (resp., $\pi(O'_2)\in B$) implies by definition $O'_1\in A'$ (resp., $O'_2\in B'$), contradicting that $O'_1$ and $O'_2$ touch. Thus we have proved that every terminal of $T$ is in the same component of $H$, hence $(T_1,T_2,Q)$ is indeed a $\frac{3}{4}$-balanced triple.
\end{proof}

\subsection{Analyzing the recursion}

Here we conclude the proof of~\cref{lem:smallplanaralg}: weakly solving an instance of \stplanarx\ under assumption \asspl. 
The algorithm is a branching procedure described as follows. Assume we are given an instance
$(G,k,\objects,T,X,F)$ of \stplanarx.

In each step, we need to first handle a technicality. If the set $T$ of terminals is not irredundant, then we can replace it with a subset $T_0\subseteq T$ of irredundant terminals that has the same union as $T$. This decreases the cost of the optimum solution exactly by the cost of $T\setminus T_0$: removing $T\setminus T_0$ does not disconnect any solution of the original instance, and adding $T\setminus T_0$ to any solution of the instance maintains connectivity. Moreover, if a solution of the original instance satisfies Assumption \asspl, then this is true even after removing $T\setminus T_0$ (the conditions about the bound on the number of certain objects still hold). Therefore, weakly solving the original instance under Assumption \asspl\ can be reduced to weakly solving the modified instance under Assumption \asspl.

In the following, we assume that $T$ is irredundant.
Therefore, we can use~\cref{lem:listtriples} to list
$|I|^{O_\alpha(\sqrt{k+|T|})}$ triples and guess a $\frac{3}{4}$-balanced
triple $(T_1,T_2,Q)$ for the instance $(G,k,\objects,T,X,F)$. Then
we invoke~\cref{lem:recursion0} and branch into
instances $(G,k',\objects,T',X',F')$ that
satisfy  $|X'| \le |X| + |Q| \le
|T| + |Q|$, $|E(F')| \ge |E(F)|$. Moreover, as stated by Lemma~\ref{lem:recursion0}, for every integer $c$,  we branch into
$O(k|X\cup Q|^{2c}) = |I|^{2c}$ instances with $|E(F')| \le |E(F)| +c$.
Because $(T_1,T_2 ,Q)$ is a
$\frac{3}{4}$-balanced triple we have that $k'+|T'| \le \frac{3}{4}(k+|T|) + |Q|$.
As $|Q|=O_\alpha(\sqrt{k+|T|})$, we can assume that $k+|T|$ decreases by a constant factor in each step. The base case of the algorithm occurs when $k' = O(1)$, which can be solved naively in
$|I|^{O(1)}$ time.  Note that $|X'|\le |X|+|Q|$ and since $|Q|=O_\alpha(\sqrt{k+|T|})$ decreases by a constant factor in each step, the size of $|X|$ is always $O_\alpha(\sqrt{k+|T|})$ for the original $k$ and $|T|$.

The correctness of the algorithm follows from the correctness of
\cref{lem:listtriples,lem:recursion0} (here we implicitly use that a solution has the minimum cost among those satisfying Assumption \asspl\ can be assumed to be an inclusionwise minimal solution). For the running time,
observe that the degree of the branching procedure is
$\sum_{c=|F|}^{|X|} |I|^{2c} \cdot |I|^{O_\alpha(\sqrt{k+|T|})}$.
More precisely, in each step, each new constructed instance can be described by a triple $(i,c,j)$: we selected the $i$-th triple $(T_1,T_2,Q)$ for the list returned by \cref{lem:listtriples}, the constructed instance satisfies $|E(F')|=|E(F)|+c$ for some $c$, and this is the $j$-th such instance. Each leaf of the recursion tree can be described by a sequence of such triples. To bound the number of leaves of the recursion tree, we bound the number of possible such sequences. 

First, in each step, there are $|I|^{O_\alpha(\sqrt{k+|T|})}$ possibilities for
$i$. Since $k+|T|$ decreases by a constant factor in each step, the product of
these possibilities has a decreasing geometric series in the exponent, hence the
total number of possibilities for the values of $i$ in this sequence is also
$|I|^{O_\alpha(\sqrt{k+|T|})}$. The sequence of $c$'s is a sequence of
non-negative integers that add up at most to the final size of $|X|$, which is
$O_\alpha(\sqrt{k+|T|})$. It is known that there are
$2^{O_\alpha(\sqrt{k+|T|})}$ such sequences. Finally, in each triple $(i,c,j)$,
after fixing $i$ and $c$, the value of $j$ can take $|I|^{O(c)}$ possibilities.
Thus, if the $c$'s are fixed in a sequence, then the number of possibilities for
the $j$'s can be bounded by $|I|$ to a power that is the sum of the $c$'s, that
is, $|I|^{O_\alpha(\sqrt{k+|T|})}$. Therefore, we can conclude that there are
$|I|^{O_\alpha(\sqrt{k+|T|})}$ such possible sequences and hence the  branching
tree has at most that many leaves. As the total work to be done at all nodes of the branching tree is polynomial in $|I|^{O_\alpha(\sqrt{k+|T|})}$, this bounds the total running time.


\section{Lower Bounds for Narrow Grid Tiling Problems}
In this section we present an ETH based lower bound for the $\ngtiling$ problem.
Let $\mathcal{I} = (x,y,N,\mathcal{S})$ be an instance of $\ngtiling$ ($\mongtiling$).
We say that a set $T = \{s_{i,j}\} _{(i,j) \in [x] \times [y]}$ is consistent with $\mathcal{I}$
if $s_{i,j} \in S_{i,j}$ for $(i,j) \in [x] \times [y]$ and $T$ satisfies the conditions
stated in the definition of $\ngtiling$ (respectively, $\mongtiling$). Therefore $\mathcal{I}$ is a yes
instance if and only if there exists a set consistent with $\mathcal{I}$.

\begin{theorem}\label{theorem:grid_tiling_hardness}
	For every $b > 0$, there exists $\eps > 0$ such that for every $C > 0$ there
    is no algorithm that solves every instance $(x,y,N,\mathcal{S})$ of
    $\ngtiling$ in time $C \cdot 2^{\eps \cdot (x \cdot y + x^{2} \cdot \log(N))} \cdot N^{b}$, unless ETH fails.
\end{theorem}

\begin{proof}
	Let $\mathcal{I}$ be an arbitrary instance of $\kSAT{3}$ with $n$ variables and $m$ clauses.
	Using sparsification lemma \cite{impagliazzoWhichProblemsHave2001}, we can assume that $m \leq c \cdot n$ for some integer $c > 0$.
	Assume now that there exists $b > 0$ such that for all $\eps > 0$ there exists an algorithm that solves every instance $(x,y,N,\mathcal{S})$ of $\ngtiling$ in time $2^{\eps \cdot (x \cdot y + x^{2} \cdot \log(N))} \cdot N^{b}$.
	Let $\delta > 0$ be as in ETH, i.e. there is no algorithm that solves $\kSAT{3}$ in time $2^{\delta \cdot n}$. In the following we will describe a reduction from $\kSAT{3}$ to $\ngtiling$, which will imply an algorithm for $\kSAT{3}$ that solves each instance $\mathcal{I}$ with $n$ variables and $m$ clauses in time $2^{\delta \cdot n}$, and therefore contradicts ETH.

	Define $g \coloneqq \left\lceil \frac{12 \cdot b \cdot c}{\delta} \right\rceil $ and $\eps \coloneqq  \frac{\delta}{4} \cdot \frac{1}{g \cdot (1 + 3 \cdot c)} $. Note that both $g$ and $\eps$ depend on $b,c$ and $\delta$.

	\paragraph{Construction of the instance.} We start by grouping the clauses of $\mathcal{I}$ into $g$ groups $M_1, \ldots,  M_g$, each of size $\frac{m}{g}$ (we can assume without loss of generality that $g$ divides $m$).
	We let $x \coloneqq g$, $y \coloneqq n$, $N \coloneqq 2^{\frac{3m}{g}}$.
	For $1 \leq i \leq g$, let $V_i$ be the variables that appear in at least one of the clauses in $M_i$.
	By definition, we have $\abs{V_i} \leq 3 \cdot \abs{M_i} = \frac{3m}{g}$.
	Since $N = 2^{\frac{3m}{g}}$, for each assignment $W$ to $V_i$, we can assign a unique number $\alpha(W) \in [N]$.
	For each $1 \leq i \leq x$, we construct the sets $\{S_{i,j}\}_{1 \leq j \leq y}$ as follows.

	For $1 \leq j \leq y$, if $x_j \not\in V_i$, then for each assignment $W$ to $V_i$ that satisfies $M_i$, we add $(0, \alpha(W))$ and $(1, \alpha(W))$ to $S_{i,j}$.
	On the other hand, if $x_j \in V_i$, then for each assignment $W$ that satisfies $M_i$, we add $(a, \alpha(W))$ to $S_{i,j}$, where $a$ is the value assigned to $x_j$ by $W$.

	This is the whole construction for the instance $\mathcal{I}' = (x,y,N,\{S_{i,j}\}_{(i,j) \in [x] \times [y]})$.

	\paragraph{Equivalence of the instances.} Suppose $\mathcal{I}$ is a yes instance, i.e. there is an assignment $W$ to the variables $x_1, \ldots, x_n$ of $\mathcal{I}$ such that each clause is satisfied.

	For each $1 \leq i \leq g$, let $W_i$ denote the restriction of $W$ to $V_i$.
	By definition, $W_i$ satisfies $M_i$.
	We construct a solution for $\mathcal{I}'$ as follows.
	For each $(i,j) \in [x] \times [y]$, we pick $s_{i,j} = (a_j, \alpha(W_i) )$
	where $a_j$ is the value assigned to $x_j$ by $W_i$.
	Note that $s_{i,j} \in S_{i,j}$ by our construction, because $W_i$ satisfies $M_i$ and
	the value assigned to $x_j$ by $W$ and $W_i$ is the same for all $i \in [g]$ such that
	$x_j \in V_i$.

	For each $1 \leq i \leq x$, the elements in $\{s_{i,j}\}_{1 \leq j \leq y}$
	agree in the second coordinate which is equal to $\alpha(W_i)$.
	Similarly, for each $1 \leq j \leq y$, the elements in $\{s_{i,j}\}_{1 \leq i \leq x}$
	agree in the first coordinate which is equal to $a_j$, the value assigned to $x_j$ by $W$.
	Therefore the solution is consistent with $\mathcal{I}'$ and hence $\mathcal{I}'$ is a yes instance.

	Next, suppose that $\mathcal{I}'$ is a yes instance, i.e. there exists
	$s_{i,j} \in S_{i,j}$ for each $(i,j) \in [x] \times [y]$ such that
	$\{s_{i,j}\}_{(i,j) \in [x] \times [y]}$ is consistent with $\mathcal{I}'$.
	We construct an assignment to the variables of $\mathcal{I}$ as follows.
	For each $1 \leq j \leq n$, the value assigned to $x_j$ is the
	first coordinate of $s_{1,j}$, which is also equal to the first coordinate of
	$s_{i,j}$ for $1 \leq i \leq g$. Let $A$ denote this assignment.

	Now we will prove that this assignment satisfies all clauses of $\mathcal{I}$.
	For each group of clauses $M_i$ for $i \in [g]$, let $r_i$ denote the second coordinate of $s_{i,1}$
	which is equal to the second coordinate of $s_{i,j}$ for $j \in [y]$.
	Let $W_i$ denote the assignment corresponding to $r_i$,
	i.e. $r_i = \alpha(W_i)$. By definition, $W_i$ satisfies
	all the clauses in $M_i$. Finally, to see that $A$ satisfies all
	the clauses of $\mathcal{I}$, observe that $A$ and $W_i$ agree on all variables in
	$V_i$. Hence $A$ is a solution for $\mathcal{I}$ and $\mathcal{I}$ is a yes instance.

	\paragraph{Running Time.} The construction of the instance $\mathcal{I}'$
	takes time at most
	\begin{align}
		C \cdot x \cdot y \cdot N \cdot n^{\Oh(1)} &= C \cdot n \cdot g \cdot 2^{\frac{3m}{g}} \cdot n^{\Oh(1)}\notag\\
					     &= C \cdot 2^{\frac{3m}{g}} \cdot n^{\Oh(1)}\notag\\
					     &\leq C \cdot 2^{\frac{3 \cdot c \cdot n}{g}} \cdot n^{\Oh(1)}\notag\\
					     &= C \cdot 2^{\frac{\delta}{4 \cdot b} \cdot n} \cdot n^{\Oh(1)} \notag\\
					     &\leq C \cdot 2^{\frac{\delta}{4} \cdot n} \cdot n^{\Oh(1)}\notag\\
					     &\leq C \cdot 2^{\frac{\delta}{3} \cdot n},\label{eq:gtil_red_rt}
	\end{align}
	where the last step holds for large enough $n$.

	Running the hypothetical algorithm for $\ngtiling$ takes time
	\begin{align}
		C \cdot 2^{\eps \cdot (x \cdot y + x^{2} \cdot \log(N))} \cdot N^{b} &=
        C \cdot 2^{\eps \cdot (g \cdot n + g^{2} \cdot \frac{3m}{g})} \cdot 2^{\frac{3 \cdot m \cdot b}{g}}\notag\\
									     &\leq C \cdot 2^{\eps \cdot (g \cdot n + 3 \cdot c \cdot n \cdot g)} \cdot 2^{\frac{3 \cdot c \cdot b}{g} \cdot n}\notag\\
									     &\leq C \cdot 2^{\eps \cdot g \cdot (1 + 3c) \cdot n} \cdot 2^{\frac{\delta}{4} \cdot n}\notag\\
									     &= C \cdot 2^{\frac{\delta}{4} \cdot n} \cdot 2^{\frac{\delta}{4} \cdot n}\notag\\
									     &= C \cdot 2^{\frac{\delta}{2} \cdot n}.\label{eq:gtil_algo_rt}
	\end{align}
	By \eqref{eq:gtil_red_rt} and \eqref{eq:gtil_algo_rt}, it holds that there exists an algorithm that solves $\kSAT{3}$ in time 
	\begin{align}
		C \cdot 2^{\frac{\delta}{3} \cdot n} + C \cdot 2^{\frac{\delta}{2} \cdot n}
        &\leq 2 C \cdot 2^{\frac{\delta}{2} \cdot n}\notag\\
									    &\leq 2^{\delta \cdot n}\label{eq:gtil_final_rt}.
	\end{align}
    where the last step holds for large enough $n >
    \frac{2\log_2(2C)}{\delta}$. Therefore ETH fails by \eqref{eq:gtil_final_rt}.
\end{proof}

\begin{remark}\label{remark:grid-tiling}
    By setting parameters $x = g = m$, $N = 8$, $y = n$  in the~\cref{theorem:grid_tiling_hardness} we can
    show that for every $b > 0$ and $\eps = \delta/(4c)$ and $C > 0$ no
    $C\cdot (xyN)^{\eps\sqrt{xy}/\log(xy)+b}$ time algorithm solves every instance
    $(x,y,N,\mathcal{S})$ of $\ngtiling$.
\end{remark}

Next, we will prove an intermediary result which we will use in the hardness result for $\mongtiling$.

A bitstring is a sequence of bits where each bit is either 0 or 1.
For an integer $x \geq 0$, we let $\bin(x)$ denote its binary representation.
For two bitstrings $\bolda$ and $\boldb$, we let $\bolda \circ \boldb$ denote the
concatenation of the two strings. The bitwise complement of a bitstring
$\bolda$ is the bitstring $\overline{\bolda}$ obtained by inverting each bit in
$\bolda$. Similarly, for an integer $s$, we let $\overline{s}$ denote the bitwise
complement of the number $s$ as well. For bitstrings $\bolda, \boldb$ we say
that $\bolda \leq \boldb$ if the integers they represent satisfy the
inequality. Finally, for an integer $s \geq 0$, we define the $i$'th
bit of $s$ as its $i$'th least significant bit, which is denoted by $s[i]$.

\begin{lemma}\label{lemma:complement_eq}
	Let $N \geq 1$ be a power of two and define $\ell \coloneqq \log(N)$.
	For each $1 \leq i \leq \ell$, define the sets $A_i, B_i$ as
	\begin{align*}
		A_i &\coloneqq \Big\{(s[i], s) \mid 1 \leq s \leq N\Big\} \\
		B_i &\coloneqq \Big\{\Bigl(\overline{s[i]}, s\Bigr) \mid 1 \leq s \leq N \Big\}.
	\end{align*}
	Suppose that for each $1 \leq i \leq \ell$ there exists $x_i \in \{0,1\}$,
	$1 \leq a_i, b_i \leq N$ such that $(x_i, a_i) \in A_i$, $(x_i, b_i) \in B_i$ and
	\begin{align*}
		a \coloneqq a_1 &\geq a_2 \geq \ldots \geq a_\ell,\\
		b \coloneqq b_1 &\geq b_2 \geq \ldots \geq b_\ell.
	\end{align*}
	Then, $a$ is bitwise complement of $b$.
\end{lemma}

\begin{proof}
	For each $1 \leq i \leq \ell$, $(x_i, a_i )\in A_i$ implies that $x_i = a_i[i]$.
	Similarly, $(x_i, b_i) \in B_i$ implies that $x_i = \overline{b_i[i]}$.
	Therefore, the $i$'th bits of $a_i$ and $b_i$ are complements of each other.

	In the following, we will prove by reverse induction that for each $1 \leq i \leq \ell$,
	the first $\ell - i + 1$ most significant bits of $a_i$ and $b_i$ are complements of each other.
	For $i = \ell$, this holds by the argument above. Therefore, let $i \leq \ell - 1$
	and suppose that the claim holds for $i + 1$. We can write the
	binary representation of $a_{i + 1}$ and $b_{i + 1}$ as
	\begin{align*}
		\bin\left( a_{i + 1} \right)  &= \bolds \circ \alpha\\
		\bin\left( b_{i + 1} \right)  &= \overline{\bolds} \circ \beta
	\end{align*}
	where $\bolds$ is a binary string of length $\ell - i$ and $\alpha, \beta$ are
	binary strings of length $i$.

	Similarly, we can write
	\begin{align*}
		\bin\left( a_i \right)  &= \boldm_1 \cdot t \cdot \alpha'\\
		\bin\left( b_i \right)  &= \boldm_2 \cdot \overline{t} \cdot \beta'
	\end{align*}
	where $\boldm_1, \boldm_2$ are bitstrings of length $\ell - i$, $t \in \{0,1\}$ and
	$\alpha', \beta'$ are bitstrings of length $i - 1$.
	Observe that the $i$'th bit of $a_i$ and $b_i$, i.e. $t$ and $\overline{t}$,
	are complements as discussed above.

	Since $a_i \geq a_{i + 1}$ and $b_i \geq b_{i + 1}$, we have
	\begin{equation*}
		\boldm_1 \geq \bolds, \quad \boldm_2 \geq \overline{\bolds}.
	\end{equation*}
	Since $\bolds$ is a bitstring of length $i - 1$, we also have
	\begin{equation}\label{eq:m_12_lb}
		\boldm_1 + \boldm_2 \geq \bolds + \overline{\bolds}  = 2^{i} - 1.
	\end{equation}
	Note that \eqref{eq:m_12_lb} holds only if $\boldm_1$ and $\boldm_2$ are bitwise complements
	as both of them are bitstrings of length $i$. Therefore the claim holds.

	Finally, by setting $i = 1$, we prove the lemma.
\end{proof}

\begin{theorem}\label{theorem:mon_grid_tiling_hardness}
	For every $b > 0$, there exists $\eps > 0$ such that for every $C > 0$ there
    is no algorithm that solves every instance $(x,y,N,\mathcal{S})$ of
    $\mongtiling$ in time $C \cdot 2^{\eps \cdot x \cdot y} \cdot N^{b}$, unless ETH fails.
\end{theorem}

\begin{proof}
	We will prove the claim by giving a reduction from $\ngtiling$.

	Suppose there exists $b > 0$ such that for all $\eps > 0$ there is an algorithm
	that solves every instance $(x,y,N,\mathcal{S})$ of $\mongtiling$
	in time $2^{\eps \cdot x \cdot y} \cdot N^{b}$.
	For all $\eps > 0$, we will describe an algorithm that solves every instance 
	$(x,y,N,\mathcal{S})$ of $\ngtiling$ in time $2^{\eps \cdot (x \cdot y + x^{2} \cdot
	\log(N))} \cdot N^{b}$, which contradicts ETH by \cref{theorem:grid_tiling_hardness}.

	\paragraph{Construction of the instance.} Let $\mathcal{I} = (x,y,N,\mathcal{S})$
	be an instance of $\ngtiling$ where $\mathcal{S} = \{S_{i,j}\}_{(i,j) \in [x] \times [y]}$.
	Without loss of generality, we can assume
	that $N$ is a power of two.
	Define $x' \coloneqq 2 \cdot x$,
	$y' \coloneqq y + 2 \cdot x \cdot \log(N)$ and $N' = N$.
	Finally, define the interval $M \coloneqq [x \cdot \log(N) + 1, x \cdot \log(N) + y]$.

	Intuitively, we copy the original sets in $\mathcal{S}$ and add further rows
	and columns to make use of the monotonicity property and ensure that
	the elements from each $S'_{i,j}$ satisfy the properties in the definition
	of $\ngtiling$.

	For $i \in [x']$ and $j \in M$, we let
	\begin{equation*}
		S'_{i,j} = \begin{cases}
			S_{i', \;j - x \cdot \log(N) } &\text{if } i = 2 \cdot i' - 1\\
			\tilde{S}_{i', \; j - x \cdot \log(N) } &\text{if } i = 2 \cdot i' 			
		\end{cases}
	\end{equation*}
	where $\tilde{S}_{i,j} = \{(x, \overline{s}) \mid (x,s) \in S_{i,j}\}$.

	That is, we double the number of rows and add the complement of each set in the original row.
	For $1 \leq e \leq x$, define the intervals
	\begin{align*}
		L(e) &\coloneqq [(e-1) \cdot \log(N) + 1, e \cdot \log(N)]\\
		R(e) &\coloneqq [(e - 1 + x) \cdot \log(N) + y + 1, (e + x) \cdot \log(N) + y]
	\end{align*}
	and
	\begin{align*}
		\mathcal{L} &\coloneqq \bigcup_{1 \leq e \leq x} L(e)\\
		\mathcal{R} &\coloneqq \bigcup_{1 \leq e \leq x} R(e).
	\end{align*}

	Observe that
	\begin{equation*}
		[y'] = \mathcal{L}  \cup M \cup \mathcal{R}.
	\end{equation*}
	Then, for $i \in [x']$ and $j \in [y'] \setminus M = (\mathcal{L} \cup \mathcal{R})$ we let
	\begin{equation*}
		S'_{i,j} = \begin{cases}
			A_{j \modulo \log(N)} &\text{if } i = 2 \cdot i' - 1 \text{ and } j \in L(i') \\
			A_{(j - y) \modulo \log(N)} &\text{if } i = 2 \cdot i' - 1 \text{ and } j \in R(i') \\
			
			B_{j \modulo \log(N)} &\text{if } i = 2 \cdot i' \text{ and } j \in L(i') \\			
			B_{(j - y) \modulo \log(N)} &\text{if } i = 2 \cdot i' \text{ and } j \in R(i') \\			
			\{0,1\} \times [N] &\text{otherwise}.
		\end{cases}
	\end{equation*}

	The new instance of $\mongtiling$ is $\mathcal{I}' = (x', y', N', \mathcal{S}')$
	where $\mathcal{S}' = \{S'_{i,j}\}_{(i,j) \in [x'] \times [y']}$.

	\begin{figure}[htpb]
		\centering
		\includegraphics[width=0.8\textwidth]{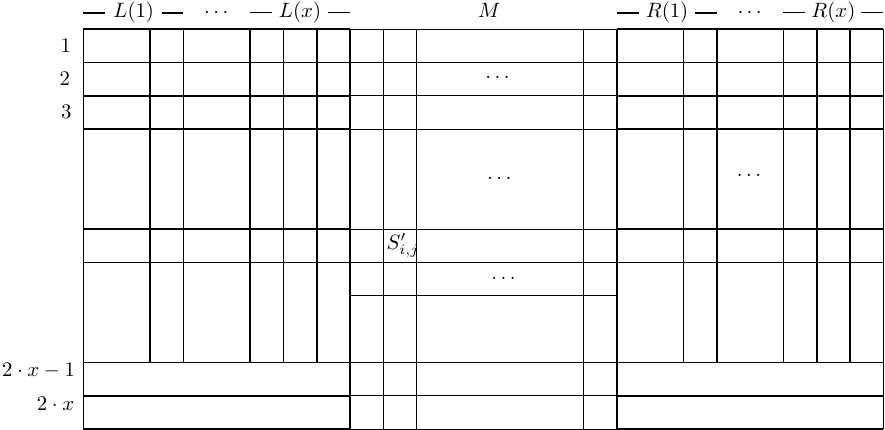}
		\caption{The instance $\mathcal{I}'$ of \mongtiling.}
		\label{fig:inst_mongtiling}
	\end{figure}

	\paragraph{Equivalence of the instances.} Suppose $\mathcal{I}$ is a yes instance,
	i.e. there exists a set $T = \{s_{i,j}\}_{(i,j) \in [x] \times [y]}$ consistent with $\mathcal{I}$.
	Since $T$ is consistent with $\mathcal{I}$, there exists $\{x_j\}_{j \in [y]}$ and $\{a_i\}_{i \in [x]}$
	such that $s_{i,j} = (c_j, a_i)$.

	We construct a solution for $\mathcal{I}'$ as follows.
	For $i \in [x']$ and $j \in M$, we let
	\begin{align*}
		s'_{i,j} = \begin{cases}
			(c_{j - x \cdot \log(N)}, a_{i'}) &\text{if } i = 2 \cdot i' - 1 \\
			(c_{j - x \cdot \log(N)}, \overline{a_{i'}}) &\text{if } i = 2 \cdot i' 
		\end{cases}
	\end{align*}

	For the remaining values of $j \in [y'] \setminus M = (\mathcal{L} \cup \mathcal{R})$, we first define
	\begin{equation*}
		p(j) \coloneqq \begin{cases}
			i' \text{ such that } j \in L(i') &\text{if } j \in \bigcup_{1 \leq e \leq x} L(e)\\
			i' \text{ such that } j \in R(i') &\text{if } j \in \bigcup_{1 \leq e \leq x} R(e). 			
		\end{cases}
	\end{equation*}

	Note that $1 \leq p(j) \leq x$ for all $j \in (\mathcal{L} \cup \mathcal{R})$. Then we let
	\begin{equation*}
		s'_{i,j} = \begin{cases}
			\Bigl(a_{p(j)}[j \modulo \log(N)], \;a_{i'} \Bigr) &\text{if } i = 2 \cdot i' - 1 \text{ and } j \in \mathcal{L}\\ 
			\Bigl(a_{p(j)}[(j - y) \modulo \log(N)],\; a_{i'} \Bigr) &\text{if } i = 2 \cdot i' - 1 \text{ and } j \in \mathcal{R}\\ 			
			\Bigl(a_{p(j)}[j \modulo \log(N)],\; \overline{a_{i'}} \Bigr) &\text{if } i = 2 \cdot i' \text{ and } j \in \mathcal{L} \\ 
			\Bigl(a_{p(j)}[(j - y) \modulo \log(N)],\; \overline{a_{i'}} \Bigr) &\text{if } i = 2 \cdot i' \text{ and } j \in \mathcal{R} 			
			\end{cases}
	\end{equation*}

	For each $(i,j) \in [x'] \times [y']$, let us show that $s'_{i,j}
	\in S'_{i,j}$. If $j \in M$ and $i = 2 \cdot i' - 1$, then $S'_{i,j} = S_{i',j - x \cdot
	\log(N)}$ and by our construction $s'_{i,j} = (c_{j - x \cdot \log(N)},
	a_{i'}) \in S_{i', j - x \cdot \log(N)} = S'_{i,j}$. On the other hand,
	if $i = 2 \cdot i'$, then $S'_{i,j} = \tilde{S}_{i', j - x \cdot
	\log(N)}$ and again by our construction $s'_{i,j} = (c_{j - x \cdot
	\log(N)}, \overline{a_{i'}}) \in \tilde{S}_{i', j - x \cdot \log(N)} = S'_{i,j}$
	since $(c_{j - x \cdot
	\log(N)}, a_{i'}) \in S_{i', j - x \cdot \log(N)}$.

	Now let $i \in [x']$ such that $i = 2 \cdot i' - 1$ and $j \in [y'] \setminus M$.
	(The case of $i = 2 \cdot i '$ is very similar and is left to the reader
	to verify).
	There are three cases, $j \in L(i')$, $j \in R(i')$ or $j \in (\mathcal{L} \cup \mathcal{R}) \setminus (L(i') \cup R(i'))$. If $j \in L(i')$, it holds that $p(j) = i'$ and
	\begin{equation*}
		s'_{i,j} = \Bigl(a_{p(j)}[j \modulo \log(N)], \;a_{i'} \Bigr) = \Bigl(a_{i'}[j \modulo \log(N)], \;a_{i'} \Bigr) \in A_{j \modulo \log(N)} \in S'_{i,j}.
	\end{equation*}
	Similarly, if $j \in R(i')$, then $p(j) = i'$ and we have
	\begin{equation*}
		s'_{i,j} = \Bigl(a_{p(j)}[(j - y) \modulo \log(N)],\; a_{i'} \Bigr) = \Bigl(a_{i'}[(j - y) \modulo \log(N)],\; a_{i'} \Bigr) \in A_{(j - y) \modulo \log(N)} = S'_{i,j}.
	\end{equation*}
	Finally, observe that if $j \in (\mathcal{L} \cup \mathcal{R}) \setminus (L(i') \cup R(i'))$, then $S'_{i,j} = \{0,1\} \times [N]$, and it trivially holds that $s'_{i,j} \in S'_{i,j}$.

	 We have $s'_{i,j} = (a_{p(j)}[j'], a_{i'} )$ where $j' \coloneqq j \modulo \log(N)$.
	If $j \in L(i')$, then $S'_{i,j} = A_{j \modulo \log(N)} = A_{j'}$ and $p(j) = i'$.
	By definition $s'_{i,j} = (a_{p(j)}[j'], a_{i'} )= (a_{i'}[j'], a_{i'} ) \in A_{j'} = S'_{i,j}$.
	The same argument also holds
	when $j \in R(i')$. Finally, observe that if $j \in \biggl([y'] \setminus \Bigl( M \cup L(i') \cup R(i') \Bigr) \biggr)$,
	then $S'_{i,j} = \{0,1\} \times [N]$ and it follows that $s'_{i,j} \in S'_{i,j}$.

	Let us now prove that $\{s'_{i,j}\}_{(i,j) \in [x'] \times [y']}$ is consistent
	with $\mathcal{I}'$.
	Observe that the second coordinate of $s'_{i,j}$
	only depends on whether $i$ is even or not, hence the tuples in $\{s'_{i,j}\}_{j \in [y']}$ agree in the
	second coordinate for each $i \in [x']$.
	Similarly, for any fixed  $j \in [y']$, the tuples in $\{s'_{i,j}\}_{i \in [x']}$
	agree in the first coordinate, which is equal to $c_{j - x \cdot \log(N)}$ or $a_{p(j)[j \modulo \log(N)]}$.
	Therefore there is a set consistent with $\mathcal{I}'$ and hence $\mathcal{I}'$ is a yes instance.

	Now suppose that $\mathcal{I'}$ is a yes instance,
	i.e. there exists a set $T = \{s_{i,j}\}_{(i,j) \in [x'] \times [y']}$ consistent with $\mathcal{I}'$. Let $(c_{i,j}, a_{i,j}) \coloneqq s_{i,j}$. Observe that since $T$ is consistent with $\mathcal{I}'$, for fixed $i \in [x']$ it holds that
	\begin{equation}\label{eq:a_i_ineq}
		a_{i,1} \geq a_{i,2} \geq \ldots \geq a_{i, y'} = a_{i, y  +2 \cdot x \cdot \log(N)}.
	\end{equation}
	Similarly, for fixed $j \in [y']$, we have
	\begin{equation}\label{eq:c_eq}
		c_{1,j} = c_{2,j} = \ldots = c_{x', j} = c_{2 \cdot x, j}.
	\end{equation}
	For $1 \leq i \leq x$, define
	\begin{align*}
		l(i) &\coloneqq (i - 1) \cdot \log(N) + 1\\
		r(i) &\coloneqq (i - 1 + x) \cdot \log(N) + y + 1.
	\end{align*}
	Note that $l(i) \in L(i)$ and $r(i) \in R(i)$. For each $i \in [x]$, it holds that
	$S'_{2 \cdot i - 1, l(i)}  = A_{l(i) \modulo \log(N)} = A_1$ and
	\begin{equation}\label{eq:S_2i-1_eq}
		(S'_{2 \cdot i - 1, l(i)}, S'_{2 \cdot i - 1, l(i) + 1}, \ldots, S'_{2 \cdot i - 1, l(i) + \log(N) - 1}) = (A_1, \ldots, A_{\ell})
	\end{equation}
	where $\ell = \log(N)$.
	Similarly, we have $S'_{2 \cdot i, l(i)} = B_{l(i) \modulo \log(N)} = B_1$ and
	\begin{equation}\label{eq:S_2i_eq}
		(S'_{2 \cdot i , l(i)},\, S'_{2 \cdot i , l(i) + 1},\, \ldots,\, S'_{2 \cdot i , l(i) + \log(N) - 1}) = (B_1, \ldots, B_{\ell}).
	\end{equation}

	By \eqref{eq:a_i_ineq}, \eqref{eq:S_2i-1_eq}, \eqref{eq:S_2i_eq} and \cref{lemma:complement_eq}, it holds that $a_{2 \cdot i - 1, l(i)}$ and $a_{2 \cdot i, l(i)}$ are complements in binary.
	Using the same arguments, we can also show that $a_{2 \cdot i - 1, r(i)}$ and $a_{2 \cdot i , r(i)}$ are complements in binary as well.
	Therefore we have
	\begin{equation}\label{eq:pairs_eq}
		a_{2 \cdot i - 1, l(i)} + a_{2 \cdot i, l(i)} = a_{2 \cdot i - 1, r(i)} + a_{2 \cdot i , r(i)}.
	\end{equation}
	Moreover, by \eqref{eq:a_i_ineq} and the fact that $r(i) \geq l(i)$, we also have
	\begin{align}
		a_{2 \cdot i - 1, l(i)} \geq a_{2 \cdot i - 1, r(i)}\notag\\
		a_{2 \cdot i , l(i)} \geq a_{2 \cdot i , r(i)}\label{eq:a2i_ineq}
	\end{align}
	Therefore, it holds that
	\begin{align*}
		a_{2 \cdot i - 1, l(i)} &\geq a_{2 \cdot i - 1, r(i)}\\
					&= a_{2 \cdot i - 1, l(i)} + a_{2 \cdot i, l(i)} - a_{2 \cdot i , r(i)}\\
					&\geq a_{2 \cdot i - 1, l(i)}
	\end{align*}
	where the equality follows from \eqref{eq:pairs_eq} and the last inequality holds by \eqref{eq:a2i_ineq}. Therefore 
	\begin{equation}\label{eq:horizontal_eq}
		a_{2 \cdot i - 1, l(i)} = a_{2 \cdot i - 1, r(i)}.
	\end{equation}

	In particular, \eqref{eq:a_i_ineq} and \eqref{eq:horizontal_eq} together imply that
	\begin{equation}\label{eq:a_i_final_ineq}
		a_{2 \cdot i - 1,l(i)} = a_{2 \cdot i - 1,l(i) + 1} = \ldots =  a_{2 \cdot i - 1, r(i)}.
	\end{equation}

	Now for each $(i,j) \in [x] \times [y]$ we define $q(i) \coloneqq 2 \cdot i - 1$, $z(j) \coloneqq x \cdot \log(N) + j$ and let
	\begin{equation*}
		s_{i,j} = \Bigl(c_{q(i), z(j)}, \,a_{q(i), z(j)}\Bigr) \in S'_{q(i), z(i)}.
	\end{equation*}

	Note that since $z(j) \in M$ for each $j \in [y]$ and $q(i) \in [x']$ for $i \in [x]$, by definition it holds that
	\begin{equation*}
		S'_{q(i), z(j)} = S_{i,j}.
	\end{equation*}
	Therefore, $s'_{i,j} \in S'_{q(i), z(j)} = S_{i,j}$.

	Let us now show that $Z \coloneqq \{s_{i,j}\}_{(i,j) \in [x] \times [y]}$ is consistent with $I$.
	Let $i \in [x]$ and $j_1, j_2 \in [y]$. The second coordinates of $s_{i, j_1}$ and $s_{i, j_2}$ are $a_{q(i), z(j_1)}$ and $a_{q(i), z(j_2)}$ respectively, which are equal by \eqref{eq:a_i_final_ineq} since $l(i) \leq z(j_1), z(j_2) \leq r(i)$.
	
	Similarly, for $j \in [y]$ and $i_1, i_2 \in [x]$, the first coordinates of $s_{i_1,j}$ and $s_{i_2, j}$ are equal to $c_{q(i_1), z(j)}$ and $c_{q(i_2), z(j)}$ respectively.
	By \eqref{eq:c_eq}, it holds that $c_{q(i_1), z(j)}$ and $c_{q(i_2), z(j)}$
	are also equal to each other since $1 \leq q(i_1), q(i_2) \leq 2 \cdot x$.
	Hence $Z$ is consistent with $\mathcal{I}$ and therefore $\mathcal{I}$ is a yes instance.

	\paragraph{Running Time.} The construction of the instance $\mathcal{I}'$ takes
	time
	\begin{align}
		C \cdot x' \cdot y' \cdot N &= C \cdot 2 \cdot x \cdot (y + 2 \cdot x \cdot \log(N)) \cdot N\notag\\
					     &\leq C \cdot 2^{\frac{\eps}{2} \cdot ( x \cdot y + x^{2} \cdot \log(N))} \cdot N^{b}\label{eq:red_gtil_2_const}
	\end{align}
	for large enough $x,y$ and $N$. Running the hypothetical algorithm takes time
	\begin{align}
		C \cdot 2^{\eps' \cdot x' \cdot y'} \cdot N^{b} &= C \cdot 2^{\eps' \cdot 2 \cdot x \cdot (y + 2 \cdot x \cdot \log(N))} \cdot N^{b}\notag\\
						       &\leq C \cdot 2^{\frac{\eps}{2} \cdot \left( x \cdot y + x^{2} \cdot \log(N) \right) } \cdot N^{b}\label{eq:red_gtil_2_algo}
	\end{align}

	Finally, by \eqref{eq:red_gtil_2_const} and \eqref{eq:red_gtil_2_algo}, the whole algorithm runs in time
	\begin{align*}
		C \cdot 2^{\frac{\eps}{2} \cdot ( x \cdot y + x^{2} \cdot \log(N))}
        \cdot N^{b} + C \cdot 2^{\frac{\eps}{2} \cdot ( x \cdot y + x^{2} \cdot
        \log(N))} \cdot N^{b} &= C \cdot 2 \cdot 2^{\frac{\eps}{2} \cdot ( x \cdot y + x^{2} \cdot \log(N))} \cdot N^{b}\\
																				  &= C \cdot 2^{1 + \frac{\eps}{2} \cdot ( x \cdot y + x^{2} \cdot \log(N))} \cdot N^{b}\\
																				  &= C \cdot 2^{\eps \cdot ( x \cdot y + x^{2} \cdot \log(N))} \cdot N^{b}
	\end{align*}
	for large enough $x,y$ and $N$, which contradicts ETH by \cref{theorem:grid_tiling_hardness}.
\end{proof}

\begin{remark}\label{remark:mongtiling}
    By using parameters in~\cref{remark:grid-tiling} instead
    of~\cref{theorem:grid_tiling_hardness}, we can show that for every $b > 0$,
    $\eps = \delta/(4c)$ and $C > 0$ no $C \cdot (xyN)^{\eps\sqrt{xy}/\log(xy)+b}$ time
    algorithm solves every instance $(x,y,N,\mathcal{S})$ of $\mongtiling$ assuming ETH.
\end{remark}

\section{Lower Bounds for Steiner Tree on Unit Squares/Disks}
In this section, we prove the following theorem:

\begin{theorem}\label{thm:st-lb}
    For every $b > 0$, there exists $\eps > 0$, such that for every $C > 0$ there is no algorithm
    that solves every instance $(k,V,T)$ of \stsquare in time $C \cdot 2^{\eps(k + |T|)}
    (|V|+|T|)^b$, unless ETH fails.
\end{theorem}

To prove~\cref{thm:st-lb}, we reduce from $\mongtiling$ problem with parameters
$(x,y,N,\mathcal{S})$. More specifically, suppose that there exists $b > 0$
such that for all $\varepsilon > 0$ there exists an algorithm that solves every
instance 
$(k,V,T)$ of \stsquare{} in time $2^{\eps(k + |T|)}
(|V|+|T|)^b$. Our reduction from $\mongtiling$ problem will imply
that for each $\varepsilon > 0$ there exists an algorithm
that solves every instance $(x,y,N,\mathcal{S})$ of $\mongtiling$ in time
$2^{\eps \cdot x \cdot y} \cdot N^{b}$, which contradicts ETH by
\cref{theorem:mon_grid_tiling_hardness}.

We will fix $\omega = 11, h = 8 \cdot \omega + 11, \gamma = \frac{0.1}{N}$, $\delta=0.01$.

\subsection{Gadgets}
A unit square $s$ in $\mathbb{R}^{2}$ is represented as a pair $s = (a, b)$, 
where $\left(a, b\right) \in \mathbb{R}^2$ denotes the coordinates of the 
lower-left corner of the square.
We say that $s$ lies at the position $(a,b)$.
The square is assumed to be axis-aligned, meaning its edges are parallel to the coordinate axes. 
We also let $\unitsquare{a,b}$ denote a unit square at the position $(a,b)$.
The right-most and left-most $x$-coordinates of $s$ are given by 
$\xmax{s} \coloneqq a + 1$ and $\xmin{s} \coloneqq a$, respectively.
Similarly,
the top-most and bottom-most $y$-coordinates of $s$ are given by 
$\ymax{s} \coloneqq b + 1$ and $\ymin{s} \coloneqq b$, respectively.

Before we present the reduction, we introduce several gadgets.
A gadget is a pair $\left( \mathcal{G}, \mathcal{D} \right)$
where
\begin{enumerate}
	\item $\mathcal{G}$ is a finite set of axis-aligned unit squares in $\mathbb{R}^{2}$, possibly including terminal squares, and
	\item $\mathcal{D} \subseteq \mathcal{G}$ is called the set of interface squares. A gadget
		may have no interface squares in which case we have $\mathcal{D} = \emptyset$.
\end{enumerate}
A set of squares $\mathcal Z$ is \emph{well-separated} if, whenever it contains
a gadget $\left( \mathcal{G}, \mathcal{D} \right) $, the set $\mathcal{D}$
separates $\mathcal{Z} \setminus\mathcal{G}$ from $\mathcal{G} \setminus \mathcal{D}$
in the touching graph.  Our construction assembles several gadgets to produce
an \stsquare{} instance $\mathcal{Z}$ which is
well-separated. Consequently, if $A\subseteq\mathcal Z$ connects all terminals
of $\mathcal Z$, then $A\cap\mathcal G$ must connect each
terminal inside $\mathcal{G}$ to an interface square in $\mathcal{D}$.


\begin{definition}[Block]\label{definition:block}
	For $N \geq 1$ and $S \subseteq \Bigl( \{0,1\} \times [N] \Bigr)$,
	the gadget $\Block{N, S}$ is a set of axis-aligned unit squares
	such that
	\begin{equation*}
		\Block{N, S} = \Big\{\unitsquare{\gamma \cdot (b-1),\, a \cdot \delta} \,\Big|\, (a,b) \in S\,\Big\}. 
	\end{equation*}
	Let $B$ be a copy of $\Block{N, S}$ with offset $(x,y) \in \mathbb{R}^{2}$.
	Define $\sigma_B \from S \to B$ such that $\sigma_B(a,b)$ is the unique copy in $B$ of the square indexed by $(a,b) \in S$,
	i.e.,
	\begin{equation*}
		\unitsquare{\gamma\cdot \,(b-1)+x,\; a \cdot \delta + y}.
	\end{equation*}
\end{definition}

\begin{figure}[htbp]
  \centering
  \begin{subfigure}[t]{0.40\textwidth}
    \centering
    \includegraphics[page=2,width=\linewidth]{block}
    \caption{The bottom-left corner of squares in $\Block{N, S}$.}
    \label{fig:block-page2}
  \end{subfigure}
  \hfill
  \begin{subfigure}[t]{0.40\textwidth}
    \centering
    \includegraphics[page=1,width=\linewidth]{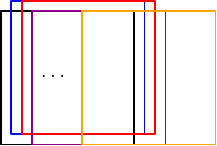}
    \caption{The squares in $\Block{N, S}$ corresponding to the points on the left.}
    \label{fig:block-page1}
  \end{subfigure}

  \caption{Two representations of the gadget $\Block{N, S}$ for some $N \geq 1$ and $S \subseteq \{0,1\} \times [N]$. }
  \label{fig:block-pages}
\end{figure}

\begin{definition}[Wire Gadget]\label{definition:wire_gadget}
	Let $N \geq 1$ and $S \subseteq \Bigl( \{0,1\} \times [N] \Bigr)$ be nonempty.
	For each $1 \leq i \leq \omega$, define
	\begin{equation*}
		B_i \coloneqq \begin{cases}
			\Block{N, S} &\text{if } i = 1\\
			\Block{N, \left( \{0\} \times [N]  \right)} \text{ with offset } \Bigl(i-1, \frac{\delta}{2}\Bigr) &\text{if } 2 \leq i \leq \omega - 1\\
			\Block{N, \overline{S}} \text{ with offset } (\omega - 1, 0) &\text{if } i = \omega,
		\end{cases}
	\end{equation*}
	where $\overline{S} = \bigl\{  \bigl(1 -a,  b\bigr) \,\big|\, \bigl(a,b\bigr) \in S \bigr\} $.
	Then, $\WireGadget{N, S}$ is the collection of squares
	contained in $B_1, \ldots, B_{\omega}$.
	Moreover, the interface vertices of $\WireGadget{N, S}$ is the set $B_1 \cup B_{\omega}$.
\end{definition}
\begin{figure}[htbp]
    \centering
    \includegraphics{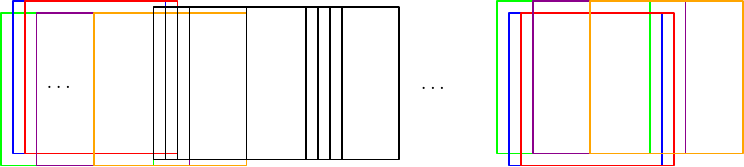}
    \caption{A representation of the gadget $\WireGadget{N, S}$ for some $N \geq 1$ and $S \subseteq \{0,1\} \times [N]$. The colored squares represent the first and last blocks, while the black squares represent the second block $B_2$.}
    \label{fig:wire_gadget}
\end{figure}
In \cref{definition:wire_gadget}, the squares in blocks
$B_2,\dots,B_{\omega-1}$ have a vertical offset $\tfrac{\delta}{2}$.
This offset guarantees that these squares intersect only those in $B_1$ and
$B_{\omega}$.
Now let $W$ be a copy of $\WireGadget{N, S}$ with offset $(x,y) \in \mathbb{R}^{2}$.

Define
\begin{equation}\label{eq:tau-definition}
	\tau_{W} \from S \to W,\qquad
	(a,b)\longmapsto \Biggl(\,\bigcup_{i = 1}^{\omega - 1} \sigma_{B_i}\left( a,b \right) \Biggr) \cup \Big\{\sigma_{B_\omega}(1-a, b)\Big\},
\end{equation}
i.e., for $(a,b) \in S$, $\tau_{W}(a,b)$ is a contiguous chain of $\omega$ unit squares indexed
by $(a,b)$ in $B_1$, $(0,b)$ in $B_2, \ldots, B_{\omega - 1}$ and $(1-a, b)$ in $B_{\omega}$. 
Next, we prove a structural lemma about the gadget $\WireGadget{N, S}$.

\begin{lemma}[Wire Gadget]\label{lemma:wire-gadget}
	Let $N \geq 1$, $S \subseteq \Bigl( \{0,1\} \times [N] \Bigr)$ be nonempty
	and let $G$ be the touching graph of $\WireGadget{N, S}$.
	Moreover, let $H$ be a subgraph of $G$ such that $H$ connects a square
	from $B_1$ to a square from $B_{\omega}$. Then the following holds:
	\begin{enumerate}
		\item $\abs{H} \geq \omega$,
		\item if $\abs{H} = \omega$, then $H$ contains exactly one
			square from each $B_i$ for $1 \leq i \leq \omega$.
	\end{enumerate}
\end{lemma}

\begin{proof}
	To see that $H \geq \omega$, let us consider the smallest horizontal
	distance $H$ should cover. To that end, consider the square in $B_1$ with the maximum
	horizontal offset and the square in $B_{\omega}$ with the minimum
	horizontal offset. Observe that this length is at least
	\begin{equation}\label{eq:wire_gadget_eq}
		\omega - \left( N-1 \right) \cdot \gamma > \omega - 1
	\end{equation}
	since $\gamma \cdot N < 1$ by definition.
	However, if $\abs{H} \leq \omega - 1$, then the maximum horizontal
	distance $H$ could cover would be $\omega - 1$, which contradicts
	\eqref{eq:wire_gadget_eq}.

	Now suppose that $\abs{H} = \omega$. In this case, if one of the blocks
	$B_1, \ldots, B_{\omega}$ contains at least two squares,
	then there exists $1 \leq i \leq \omega$ such that 
	$H$ contains no squares from $B_i$. However, then $H$ cannot connect
	the blocks $B_1, \ldots, B_{\omega}$ which leads to a contradiction.
\end{proof}

\begin{definition}[Crossing Gadget]\label{definition:crossing_gadget}
	Let $\omega$ be a constant such that $\omega \equiv 3 \mod 8$.
	Define the set of squares
	\begin{align*}
		\Omega_1 &\coloneqq \bigcup_{i = 1}^{\frac{h-3}{2}} \unitsquare{\frac{\gamma}{2}, i},  &&\Omega_2 \coloneqq \bigcup_{i = \frac{h+1}{2}}^{h-2} \unitsquare{\frac{\gamma}{2}, i}\\
		\Omega_3 &\coloneqq \bigcup_{i = 1}^{\frac{h-3}{2}} \unitsquare{\omega - 1 - \frac{\gamma}{2}, i}, &&\Omega_4 \coloneqq \bigcup_{i = \frac{h+1}{2}}^{h-2} \unitsquare{\omega - 1 - \frac{\gamma}{2}, i}.
	\end{align*}
	Then, we define the intervals $I_{\text{down}} \coloneqq [\frac{h-3}{2}]$, $I_{\text{up}} \coloneqq \{x + \frac{h-1}{2} \mid x \in I_{\text{down}}\}$, and for $i \in [\omega - 2]$ define:
	\begin{align*}
		\beta_{\text{down}}(i) \coloneqq \begin{cases}
			\{1\} &\text{if } i \equiv 1 \mod 4\\
			I_{\text{down}} &\text{if } i \equiv 2 \mod 4\\
			\{\frac{h-3}{2}\} &\text{if } i \equiv 3 \mod 4\\
			I_{\text{down}} &\text{if } i \equiv 4 \mod 4
		\end{cases}
	\end{align*}
	and
	\begin{align*}
		\beta_{\text{up}}(i) \coloneqq \begin{cases}
			\{h-2\} &\text{if } i \equiv 1 \mod 4\\
			I_{\text{up}} &\text{if } i \equiv 2 \mod 4\\
			\{\frac{h+1}{2}\} &\text{if } i \equiv 3 \mod 4\\
			I_{\text{up}} &\text{if } i \equiv 4 \mod 4.
		\end{cases}
	\end{align*}

	Moreover, we define the following set of squares
	\begin{align*}
		\Omega_5 &\coloneqq \bigcup_{i = 1}^{\frac{\omega - 3}{2}} \bigcup_{j \in B_{\text{down}(i)}} \unitsquare{i, j + \delta}, &&\Omega_6 \coloneqq \bigcup_{i = \frac{\omega + 1}{2}}^{\omega - 2} \bigcup_{j \in B_{\text{down}(i)}} \unitsquare{i, j + \delta}\\
		\Omega_7 &\coloneqq \bigcup_{i = 1}^{\frac{\omega - 3}{2}} \bigcup_{j \in B_{\text{up}(i)}} \unitsquare{i, j - \delta}, &&\Omega_8 \coloneqq \bigcup_{i = \frac{\omega + 1}{2}}^{\omega - 2} \bigcup_{j \in B_{\text{up}(i)}} \unitsquare{i, j - \delta}
	\end{align*}
	together with four terminal squares
	\begin{align*}
		T_N &\coloneqq \unitsquare{\frac{\omega - 1}{2}, h-2 - \delta},\\
		T_S &\coloneqq \unitsquare{\frac{\omega - 1}{2}, 1 + \delta},\\
		T_W &\coloneqq \unitsquare{\frac{\gamma}{2},\frac{h-1}{2}},\\
		T_E &\coloneqq \unitsquare{\omega - 1 - \frac{\gamma}{2}, \frac{h-1}{2}},
	\end{align*}
	and four \emph{interface squares}
	\begin{align*}
		u_{SW} &\coloneqq\unitsquare{\frac{\gamma}{2},0},\\
		u_{SE} &\coloneqq\unitsquare{\omega- 1 - \frac{\gamma}{2},0},\\
		u_{NW} &\coloneqq\unitsquare{\frac{\gamma}{2},h-1},\\
		u_{NE} &\coloneqq\unitsquare{\omega - 1 - \frac{\gamma}{2}, h - 1}.
	\end{align*}
	$\CrossingGadget$ is a set of unit squares consisting of
	\begin{equation*}
		 \Biggl(\bigcup_{i \in [8]} \Omega_i\Biggr) \cup \Big\{T_N,T_S,T_W,T_E,u_{SW},u_{SE},u_{NW},u_{NE}\Big\}.
	\end{equation*}                                            
\end{definition}

Observe that we have
\begin{equation*}
	\abs{\Omega_1} = \abs{\Omega_2} = \abs{\Omega_3} = \abs{\Omega_4} = \frac{h-3}{2}
\end{equation*}
and
\begin{equation*}
	\abs{\Omega_5} = \abs{\Omega_6} = \abs{\Omega_7} = \abs{\Omega_8} = \frac{\omega - 3}{2} \cdot \frac{1}{4} \cdot (h-1) = \frac{\omega - 3}{8} \cdot (h-1).
\end{equation*}

Let $C$ be a copy of $\CrossingGadget$ with offset $(x,y) \in \mathbb{R}^{2}$.
Define $\Delta_C^{1},\Delta_C^{2} \subseteq C$ to be the set of unit squares in $C$
corresponding to
\begin{equation*}
	 \Omega_1 \cup \Omega_5 \cup \Omega_4 \cup \Omega_8 \cup \{u_{SW}, u_{NE}\} 
\end{equation*}
and
\begin{equation*}
	 \Omega_2 \cup \Omega_7 \cup \Omega_3 \cup \Omega_6 \cup \{u_{NW}, u_{SE}\},
\end{equation*}
respectively, where we have
\begin{equation*}
	\abs{\Delta_C^{1}} = \abs{\Delta_C^{1}} = 2 \cdot \biggl(\frac{\omega - 3}{8} \cdot (h-1) + \frac{h-3}{2} + 1\biggr) = \frac{\omega + 1}{4} \cdot (h-1).
\end{equation*}

\begin{lemma}[Crossing Gadget]\label{lemma:crossing-gadget}
    Let $G$ be the touching graph of $\CrossingGadget$ and let $\phi = \{T_N, T_S, T_W, T_E\} $ be the set of
    terminals in $G$ and $U = \{u_{NW}, u_{NE}, u_{SW}, u_{SE}\}$ be the interface vertices.
    Moreover, let $H$ be a subgraph of $G$ such that each terminal in $\phi$
    is connected to an interface vertex in $U$. Then, the following holds:
    \begin{enumerate}
    	\item $\abs{V(H) \setminus \phi} \geq \frac{\omega +1}{4} \cdot (h-1)$,\label{item:crossing_gadget_vertex_lb}
	\item If $\abs{V(H) \setminus \phi} = \frac{\omega +1}{4} \cdot (h-1)$, then $H$ has
		two connected components, each containing two terminals, and
		it holds that
		\begin{equation*}
			\left( V(H) \setminus \phi \right) \in \{\Delta^{1}_{C_{i,j}}, \Delta^{2}_{C_{i,j}}\}.
		\end{equation*} \label{item:minimum_vertices_crossing}
	\item If $H$ connects one of the interface vertices in $\{u_{NW}, u_{NE}\}$ to one of the interface vertices in $\{u_{SW}, u_{SE}\}$, then it holds that $\abs{V(H) \setminus \phi} \geq \frac{\omega +1}{4} \cdot (h-1) + 4 \cdot \omega$.
		In this case we say $H$ connects top to bottom.\label{item:crossing_gadget_top_bottom}		
	\item If $H$ connects all the terminal vertices to an interface vertex in $\{u_{SW},u_{SE}\}$ (or $\{u_{NW}, u_{NE}\}$),
		then it holds that $\abs{V(H) \setminus \phi} \geq \frac{\omega +1}{4} \cdot (h-1) + 4 \cdot \omega$.\label{item:all_terminals_to_one_side}
    \end{enumerate}
\end{lemma}

\begin{proof}
	To establish \cref{item:crossing_gadget_vertex_lb},
	observe that each of the terminal squares $T_S$ and $T_N$
	needs $\frac{\omega - 3}{8} \cdot (h-1)$ additional squares to reach an adjacent column ($\Omega_1 \cup \Omega_2$ or $\Omega_3 \cup \Omega_4$).
	Similarly, $T_E$ and $T_W$ require at least $\frac{h - 3}{2}$
	vertices each to be connected to an interface square. 
	Let us call these basic squares.
	Since at least two further vertices (possibly interface vertices) are needed to keep every terminal in $H$ connected
	to an interface square, we
	conclude that
	\begin{equation*}
		\abs{V(H) \setminus \phi} \geq 2 \cdot \frac{\omega - 3}{8} \cdot (h-1) + 2 \cdot \frac{h - 3}{2} + 2 = \frac{\omega +1}{4} \cdot (h-1).
	\end{equation*}
	
	For \cref{item:minimum_vertices_crossing}, assume
	$\abs{V(H)\setminus\phi}=\frac{\omega+1}{4} \cdot (h-1)$. Then
	$V(H)\setminus\phi$ contains exactly two non-basic squares; otherwise
	$H$ could not connect each terminal square to an interface square.
	Consequently, every terminal reaches an interface square only when $H$
	contains either $\{u_{NW},u_{SE}\}$ or $\{u_{SW},u_{NE}\}$, with each
	terminal linked to the nearer interface square. These two
	configurations are precisely $\Delta^{1}_{C_{i,j}}$ and
	$\Delta^{2}_{C_{i,j}}$.
	
	For \cref{item:crossing_gadget_top_bottom}, suppose $H$ joins an
	interface vertex in $\{u_{NW},u_{NE}\}$ with one in
	$\{u_{SW},u_{SE}\}$. Traversing this vertical gap already uses
	$2\cdot\frac{h-3}{2}+2=h-1$ squares and simultaneously connects one of
	the terminals $T_E$ or $T_W$ to an interface vertex. The other three
	terminals need an additional
	$2\cdot\frac{\omega-3}{8}(h-1)+\frac{h-3}{2}$ squares in total.
	All in all, $H$
	contains at least
	\begin{equation*}
		\biggl(2 \cdot \frac{\omega - 3}{8} \cdot (h-1) + \frac{h-3}{2} + h - 1\biggr) =  \biggl(\frac{\omega +1}{4} \cdot (h-1) + \frac{h-3}{2}\biggr) \geq \frac{\omega +1}{4} \cdot (h-1) + 4 \cdot \omega
	\end{equation*}
	squares.

	For \cref{item:all_terminals_to_one_side}, assume without loss of
	generality that every terminal is connected to an interface vertex in
	$\{u_{NW},u_{NE}\}$. Observe that $H$ needs to contain $\frac{\omega -
	3}{8} \cdot (h-1) + 2 \cdot \frac{h-3}{2}$ vertices to connect $T_S$ to
	an interface vertex. Moreover, at least $\frac{\omega - 3}{8}$ vertices
    are needed to connect $T_N$ to an interface vertex. Similarly, at least
	$\frac{h-3}{2}$ vertices are needed to connect $T_E$ (or $T_W$,
	depending on which one is not connected by the vertices mentioned so
	far) to an interface vertex. All in all, $H$ contains at least
	\begin{align*}
		\frac{\omega - 3}{4}\cdot (h-1) + 2 \cdot \frac{h-3}{2} +  \frac{h-3}{2} &\geq \frac{\omega - 3}{4}\cdot (h-1) + (h-1) +  4 \cdot \omega\\
											 &= \frac{\omega + 1}{4}\cdot (h-1)  +  4 \cdot \omega
	\end{align*}
	non-terminal vertices.
\end{proof}

\begin{definition}[Top Gadget]\label{definition:top_gadget}
	The gadget $\TopGadget$ consists of
	\begin{itemize}
		\item the unit squares $\mathcal{U}_1 \coloneqq \bigcup_{i = 1}^{\frac{\omega - 3}{2}} \unitsquare{i,1}$ and
			$\mathcal{U}_2 \coloneqq \bigcup_{i = \frac{\omega + 1}{2}}^{\omega - 2} \unitsquare{i,1}$
		\item the interface squares $u \coloneqq \unitsquare{\frac{\gamma}{2},0}$ and $v \coloneqq \unitsquare{\omega-1 - \frac{\gamma}{2}, 0}$, and
		\item the terminal square $x \coloneqq \unitsquare{\frac{\omega - 1}{2}, 1}$.
	\end{itemize}
	Let $T$ be a copy of $\TopGadget$ with offset $(x,y) \in \mathbb{R}^{2}$.
	Define $\kappa_T^{1}, \kappa_T^{2} \subseteq T$ to be the set of unit squares corresponding
	to $\Bigl(\mathcal{U}_1 \cup \{u\} \Bigr)$ and $\Bigl(\mathcal{U}_2 \cup \{v\} \Bigr)$, respectively,
	where we have $\abs{\kappa^{1}_{T}} = \abs{\kappa^{2}_{T}} = \frac{\omega - 1}{2}$.
\end{definition}

\begin{figure}[htbp]
\centering
\includegraphics{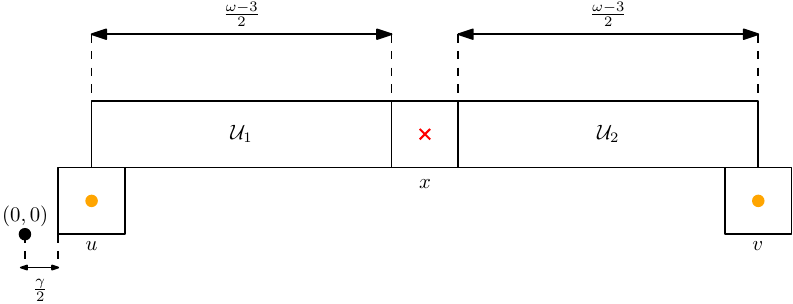}
\caption{Top gadget defined in \cref{definition:top_gadget}. The red cross
  denotes the terminal square, whereas the orange disks denote the interface squares.}
\label{fig:top_gadget}
\end{figure}

\begin{lemma}[Top Gadget]\label{lemma:top_gadget}
	Let $G$ be the intersection graph of $\TopGadget$ and let $H$
	be a subgraph of $G$ such that $x \in V(H)$ and $x$ is connected to one
	of the interface vertices in $V(H)$. Then $H \setminus x$ has at least
	$\frac{\omega - 1}{2}$ squares. Moreover, if $H \setminus x$ has
	exactly $\frac{\omega - 1}{2}$ squares, then it holds that
	\begin{equation*}
		\Bigl(H \setminus x \Bigr) \in \{\kappa^{1}_T, \kappa^{2}_T\}. 
	\end{equation*}
\end{lemma}

\begin{proof}
	Because $\TopGadget$ has only two interface squares, $x$ must be
	connected to either $u$ or $v$ in $H$. Thus $H$ contains all the
	squares in
	\begin{equation*}
		\{\unitsquare{i, 1}\}_{i \in [\frac{\omega - 3}{2}]} \cup \{u\} \text{ or } \{\unitsquare{i, 1}\}_{i \in [\frac{\omega +1}{2}, \omega - 2]} \cup \{v\},
	\end{equation*}
	where each set consists of $\frac{\omega - 1}{2}$ squares.
	Consequently, when $\abs{V(H)}=\frac{\omega - 1}{2}$, $H$ consists
	precisely of $u$ (or $v$) together with the contiguous squares lying
	between $x$ and that interface square.
\end{proof}

The bottom gadget is the mirror image of the top gadget from~\cref{lemma:top_gadget}: it is identical except
that the interface squares now sit in the \emph{top} corners of the bounding
box (i.e.\ it is $\TopGadget$ reflected across the $x$-axis).

\begin{definition}[Bottom Gadget]\label{definition:bottom_gadget}
	The gadget $\BottomGadget$ consists of
	\begin{itemize}
		\item the unit squares $\mathcal{V}_1 \coloneqq \bigcup_{i = 1}^{\frac{\omega - 3}{2}} \unitsquare{i,0}$ and
			$\mathcal{V}_2 \coloneqq \bigcup_{i = \frac{\omega + 1}{2}}^{\omega - 2} \unitsquare{i,0}$,
		\item the interface squares $u\coloneqq\unitsquare{\frac{\gamma}{2},1}$ and $v\coloneqq\unitsquare{\omega-1 - \frac{\gamma}{2},1}$, and
		\item the terminal square $x\coloneqq\unitsquare{\frac{\omega-1}{2},0}$.
	\end{itemize}
	Let $D$ be a copy of $\TopGadget$ with offset $(x,y) \in
	\mathbb{R}^{2}$. Define $\kappa_D^{1}, \kappa_D^{2} \subseteq D$ to be
	the set of unit squares corresponding to $\Bigl(\mathcal{V}_1 \cup
	\{u\} \Bigr)$ and $\Bigl(\mathcal{V}_2 \cup \{v\} \Bigr)$,
	respectively, where we have $\abs{\kappa^{1}_{D}} =
	\abs{\kappa^{2}_{D}} = \frac{\omega - 1}{2}$.
\end{definition}

\begin{lemma}[Bottom Gadget]\label{lemma:bottom_gadget}
	Let $G$ be the intersection graph of $\BottomGadget$ and let $H$
	be a subgraph of $G$ such that $x \in V(H)$ and $x$ is connected to one
	of the interface vertices in $V(H)$. Then $H \setminus x$ has at least
	$\frac{\omega - 1}{2}$ squares. Moreover, if $H \setminus x$ has
	exactly $\frac{\omega - 1}{2}$ squares, then it holds that
	\begin{equation*}
		\Bigl(H \setminus x \Bigr) \in \{\kappa^{1}_D, \kappa^{2}_D\}. 
	\end{equation*}
\end{lemma}

The proof of \cref{lemma:top_gadget} is nearly identical to that of \cref{lemma:top_gadget}
and is omitted.

\begin{definition}[Stem Gadget]\label{definition:stem_gadget}
    For each $x \geq 1$, the gadget
    $\StemGadget{x}$ consists of the following squares:
    \begin{equation*}
	    \unitsquare{0,i} \text{ for } i \in \Bigl\{0,\, \ldots,\, (x-1) \cdot (h+1)\Bigr\}
    \end{equation*}
    together with
    \begin{equation*}
	    \unitsquare{1, (i-1) \cdot (h + 1 + \delta)} \text{ for } i \in [x].
    \end{equation*}
    Moreover, each square of $\StemGadget{y}$ is a terminal square.
\end{definition}
Since the unit squares in $\StemGadget{y}$ are contiguous, the touching graph
is connected for every $y \geq 1$. We now turn to the construction of the
\stsquare instance using the gadgets defined above.

\subsection{Construction}
Let $\mathcal{I} = (x,y,N,\mathcal{S})$ be a $\mongtiling$ instance with non-empty sets
\begin{equation*}
  \mathcal S=\{S_{i,j}\}_{(i,j)\in[x]\times[y]},\qquad
  S_{i,j}\subseteq\{0,1\}\times[N].
\end{equation*}
Without loss of generality, we assume that $x,y,N$ are even integers.
The reduction creates an \stsquare instance $\mathcal{I}'$ with $\mathcal O(x \cdot y \cdot N)$ squares and keeps both the solution size~$k$
and the terminal set~$T$ within $O\left(x \cdot y\right)$.
\paragraph{Gadget layout.}
We instantiate the gadget copies listed in Table \ref{tab:placements},
together with the single terminal squares $R_1, \ldots, R_x$.

\begin{table}[h]
\centering
\begin{tabular}{|l|l|l|}
\hline
Gadget & Copies & Offset\\
\hline
$\StemGadget{y}$ & Single copy denoted $M$ & $(\gamma \cdot ( N-1),\,2+\frac{\delta}{2})$\\[2pt]
$\WireGadget{N,S_{i,j}}$ &
  $W_{i,j}$ for $(i,j)\in[x]\times[y]$ &
  $\bigl(2+(j-1)\cdot \omega,\,2+(i-1)(h+1+\delta)\bigr)$\\[2pt]
$\CrossingGadget$ &
  $C_{i,j}$ for $(i,j)\in[x-1]\times[y]$ &
  $\bigl(2+(j-1)\cdot \omega,\,3+\delta+(i-1)(h+1+\delta)\bigr)$\\[2pt]
$\BottomGadget$ &
  $D_j$ for $j\in[y]$ &
  $\bigl(2+(j-1)\cdot \omega,\,0\bigr)$\\[2pt]
$\TopGadget$ &
  $T_j$ for $j\in[y]$ &
  $\bigl(2+(j-1)\cdot \omega,\,3+\delta+(x-1)(h+1+\delta)\bigr)$\\[2pt]
  $\unitsquare{0,0}$ &
  $R_i$ for $i\in[x]$ &
  $\bigl(2+ y\cdot \omega,\,2+\frac{\delta}{2}+(i-1)(h+1+\delta)\bigr)$\\

\hline
\end{tabular}
\caption{All copies of gadgets used in the construction of $\mathcal{I}'$ and their corresponding offsets.}
\label{tab:placements}
\end{table}

\paragraph{Setting up parameters.}
The gadget counts are
\[
\begin{aligned}
	N_{\text{wire}}   &= x\cdot y,        &\text{(Wire gadgets)}\\
N_{\text{cross}}  &= x \cdot(y-1),    &\text{(Crossing Gadgets)}\\
N_{\text{border}} &= 2 \cdot y,        &\text{(Top and Bottom Gadgets)}\\
\end{aligned}
\]
plus one stem gadget and $x$ terminal squares, namely $R_1, \ldots, R_x$.
Let
\begin{equation*}
	N_{\text{gadget}} \coloneqq N_{\text{wire}} + N_{\text{cross}} + N_{\text{border}}.
\end{equation*}
We set the solution size (terminals excluded) to be
\begin{equation}
  k \coloneqq
  \omega \cdot N_{\text{wire}}
  + \frac{\omega + 1}{4} \cdot (h-1) \cdot N_{\text{cross}}
  + \frac{\omega-1}{2} \cdot N_{\text{border}}.
\end{equation}
Recall that $\omega$ and $h$ are constants.

Let $T$ and $V$ be the sets of all terminal and non-terminal squares
introduced so far, respectively.
Since every gadget contributes only a constant number of terminals, the size of the stem gadget is
\begin{equation*}
	(x-1) \cdot (h+1) + x + 1,
\end{equation*}
and we have introduced
additional $x$ terminals $R_1, \ldots, R_x$, it holds that
\begin{equation}\label{eq:T_x_y_bound}
	\abs{T} \leq \biggl(c_1 \cdot N_{\text{gadget}} + (x-1) \cdot (h+1) + 2 \cdot x + 1\biggr)  \leq c_2 \cdot x \cdot y,
\end{equation}
for some $c_1,c_2 \geq 1$.
By similar arguments we have that
\begin{equation}\label{eq:k_x_y_bound}
	k \leq \abs{V} \leq c_3 \cdot x \cdot y,
\end{equation}
for some $c_3 \geq 1$.

This concludes the construction of the $\stsquare$ instance $\mathcal{I}' = (k,V,T)$.
Now, we focus on the
correctness.

\subsection{Correctness}
In this section we prove the correctness of our reduction.
In other words, we show that the two instances $\mathcal{I}$ and $\mathcal{I}'$ are equivalent.

\begin{lemma}\label{lem:equivalence-part-1}
    If $\mathcal{I} = (x,y,N,\mathcal{S})$ is a $\yes$-instance, then
    $\mathcal{I}'$ is a $\yes$-instance as well,
    i.e.,
    there exists $S \subseteq V$ with $\abs{S} \leq k$ such that the intersection graph
    of $S \cup T$ is connected.
\end{lemma}

\begin{proof}
	Let $\{s_{i,j}\}_{(i,j)\in[x]\times[y]}$ be a solution to $\mathcal I$
	with each $s_{i,j} = (a_{i,j}, b_{i,j})\in S_{i,j}$. Next, we initiate the construction by
	setting $S$ to be the empty set, i.e. $S\coloneqq\emptyset$.

	\subparagraph{Wire Gadgets.}
	For every $(i,j) \in [x] \times [y]$,  we add to $S$ the
    set of squares $\tau_{W_{i,j}} (a_{i,j}, b_{i,j})$ (where $\tau_{W_{i,j}}$
    is defined in~\eqref{eq:tau-definition}). Observe that in
	each case we add to $S$ exactly $\omega$ many squares.

	\subparagraph{Crossing Gadgets.}
	For each $(i,j)\in[x-1]\times[y]$, if $a_{_i,j} = 0$,
	then we add to $S$ the set of unit squares $\Delta^{2}_{C_{i,j}}$,
	otherwise we add to $S$ the set of unit squares $\Delta^{1}_{C_{i,j}}$.
	Note that in each case, we add to $S$ exactly $\frac{\omega + 1}{4} \cdot (h-1)$ squares.

    	\subparagraph{Top and Bottom Gadgets.}
	Let $j \in [y]$ and consider $D_j$. If $a_{1,j} = 0$, we add to $S$ the
	set of unit squares $\kappa^{1}_{T_j}$, and if $a_{1,j} = 1$, then we
	add to $S$ the set of unit squares $\kappa^{2}_{T_j}$.
	Now consider $T_j$. Analogously, if $a_{x,j} = a_{1,j} = 0$, then we
	add to $S$ the set $\kappa^{2}_{T_j}$, and otherwise, we add to $S$ the
	set $\kappa^{1}_{T_j}$.
	Observe that we add to $S$ exactly $\frac{\omega - 1}{2}$ squares.

    \paragraph{Connectedness.}

    Note, that the size of the selected set $S$ is exactly $k$.
    Let $H$ be the touching graph of $S \cup T$.
    It remains to show that $H$ is connected.
    Observe that for each $i \in [x]$, the squares
    \begin{equation*}
	    \Biggl( \,\bigcup_{j \in [y]} \tau_{W_{i,j}}\left( a_{i,j}, b_{i,j} \right)  \Biggr)
    \end{equation*}
    forms a path in $H$. We call the corresponding set of vertices in $H$ the
    $i$\nth row in $H$.

    For each $(i,j)\in[x-1]\times[y]$, $\Delta^{1}_{C_{i,j}}$ (or $\Delta^{2}_{C_{i,j}}$) consists
    of two connected components, each of which is touching the $i$\nth row
    and $(i+1)$\nth row.
    Moreover, for each $j \in [y]$, $\kappa^{1}_{T_j}$ (or $\kappa^{2}_{T_j}$) touches the
    $x$\nth row, and similarly, $\kappa^{1}_{D_j}$ (or $\kappa^{2}_{D_j}$) touches the first row.

    Finally, observe that each row touches $M$ and we have $M \subseteq T$.
    Therefore, it holds that $H$ is connected.
\end{proof}

Now we continue with the other direction of the proof,
namely, if $\mathcal{I}'$ is a yes-instance, then $\mathcal{I}$ is as well.

\begin{lemma}\label{lem:equivalence-part-2}
    If $\mathcal{I}'$ is a yes instance, then $\mathcal{I}$ is also a
    yes-instance.
\end{lemma}

\begin{proof}
	Let us assume that $\mathcal{I}'$ is a yes instance,
	i.e. there exists a solution $X \subseteq V$ of cardinality $k$
	such that $X \cup T$ is connected.
	In the following, we will prove that (without loss of generality)
	further structure on $X$ can be assumed.
	Recall the notion of connecting top to bottom from \cref{lemma:crossing-gadget}.
	Similarly, given a solution $X$ and a wire gadget $W$,
	we say that $W$ is $\lrconn{X}$ if $X$ connects the leftmost block of $W$ to the rightmost block.

	\begin{claim}\label{claim:X_W_lr_conn}
		Let $1 \leq j \leq y$. Then, there exists a solution $X^j$ such that
		for each $1 \leq j' \leq j$, $W_{i,j'}$ is $\lrconn{X}$.
	\end{claim}

	\begin{claimproof}
		We prove the claim by induction on $j$.

		\subparagraph{Base Case ($j = 1$).}
		Let $1 \leq i \leq x$ such that $W_{i,1}$ is not $\lrconn{X}$
		and consider $W_{i,2}$.
		Either at least one terminal is connected to $W_{i,2}$,
		or $X \cap C_{i,2}$ connects the terminal squares of $C_{i,2}$
		to the gadget $W_{i+1, 2}$.
		In the latter case, we can consider $C_{i,2}$ and exchange
		$X \cap C_{i,2}$ with $\Delta^{1}_{C_{i,2}}$ (or $\Delta^{2}_{C_{i,2}}$),
		together with at most $\omega$ squares from each $W_{i,1}, W_{i,2}$ and $W_{i+1,2}$
		to obtain a solution $X'$ where $W_{i,1}, W_{i,2}$ and $W_{i+1,2}$ are
		all $\lrconn{X'}$.
		Observe that by \cref{lemma:crossing-gadget} 
		$\abs{X'} \leq \abs{X}$, and the connectivity properties are preserved.
		Moreover, $W_{i,1}$ is $\lrconn{X'}$.

		In the former case, when at least one terminal is connected to $W_{i,2}$,
		consider a path $P$ from that terminal square to the stem gadget.
		Let $C$ be the first crossing gadget on $P$ such that
		$C$ connects top to bottom.
		In this case, let $F_1, F_2$ be the wire gadgets adjacent to $C$ and replace
		$X \cap C$ with $\Delta^{1}_C$ (or $\Delta^{2}_C$) together with
		at most $\omega$ squares from each $F_1, F_2, W_{i,1}$ and $W_{i,2}$ to
		obtain $X'$, such that $F_1, F_2, W_{i,1}$ and $W_{i,2}$
		are all $\lrconn{X}$.
		Observe that, by \cref{lemma:crossing-gadget} $\abs{X'} \leq \abs{X}$ 
		and connectivity properties are preserved since $W_{i,1}$ and $W_{i,2}$ are connected to the 
		stem gadget.
		Moreover, $W_{i,1}$ is $\lrconn{X'}$.

		Let $X^{1}$ be the solution obtained by going over all $W_{i,1}$
		which is not $\lrconn{X}$. In the end, $X^{1}$ satisfies the properties
		in the claim for $j = 1$.

		\subparagraph{Induction Step ($2 \leq j \leq x-1$).}
		Suppose the induction hypothesis holds for $j-1$.
		Let $1 \leq i < x$ (the $i = x$ case is handled similarly)
		such that $W_{i,j}$ is not $\lrconn{X}$. Then, consider $W_{i, j+1}$.
		Either at least one terminal is connected to $W_{i,j+1}$,
		or $X \cap C_{i,j+1}$ connects the terminal squares of $C_{i,j+1}$
		to the gadget $W_{i+1, j+1}$.
		In the latter case, we can consider $C_{i,j+1}$ and exchange
		$X \cap C_{i,j+1}$ with $\Delta^{1}_{C_{i,j+1}}$ (or $\Delta^{2}_{C_{i,j+1}}$),
		together with at most $\omega$ squares from each $W_{i,j}, W_{i,j+1}$ and $W_{i+1,j+1}$
		to obtain a solution $X'$ where $W_{i,j}, W_{i,j+1}$ and $W_{i+1,j+1}$ are
		all $\lrconn{X'}$.
		Observe that by \cref{lemma:crossing-gadget} 
		$\abs{X'} \leq \abs{X}$, and the connectivity properties are preserved.
		Moreover, $W_{i,1}$ is $\lrconn{X'}$.

		In the former case, when at least one terminal is connected to $W_{i,j+1}$,
		consider a path $P$ from that terminal square to $M$.
		Let $C$ be the first crossing gadget on $P$ such that
		$C$ connects top to bottom.
		In this case, let $F_1, F_2$ be the wire gadgets adjacent to $C$ and replace
		$X \cap C$ with $\Delta^{1}_C$ (or $\Delta^{2}_C$) together with
		at most $\omega$ squares from each $F_1, F_2, W_{i,j}$ and $W_{i,j+1}$ to
		obtain $X'$, such that $F_1, F_2, W_{i,j}$ and $W_{i,j+1}$
		are all $\lrconn{X}$.
		Observe that, by \cref{lemma:crossing-gadget} $\abs{X'} \leq \abs{X}$ 
		and connectivity properties are preserved since $W_{i,j}$ and $W_{i,j+1}$ are connected to the 
		stem gadget by the induction hypothesis.
		Moreover, $W_{i,j}$ is $\lrconn{X'}$.

		Let $X^{j}$ be the solution obtained by going over all $W_{i,j}$
		which is not $\lrconn{X}$. In the end, $X^{j}$ satisfies the properties
		in the claim for $j$.

		\subparagraph{Induction Step ($j = y$).}
		Suppose the induction hypothesis holds for $y-1$.
		Then, let $1 \leq i < x$ (the case$i = x$ holds similarly)
		such that $W_{i,y}$ is not $\lrconn{X}$.
		In that case, for $R_i$ to be connected to $M$,
		$C_{i,y}$ (or $C_{i-1, y}$ depending on $y$, in which case similar arguments hold)
		should connect top to bottom.
		We can replace $X \cap C_{i,y}$ with $\Delta^{1}_{C_{i,y}}$ (or $\Delta^{1}_{C_{i,y}}$)
		together with at most $\omega$ vertices from $W_{i,y}$ and $W_{i+1,y}$ to
		obtain $X'$ such that $W_{i,y}$ and $W_{i+1,y}$ are $\lrconn{X'}$.
		Observe that by the induction hypothesis and the fact that
		$W_{i,y}$ are $\lrconn{X'}$, it holds that the connection properties
		are satisfied.
		Let $X^{y}$ be the solution obtained by going over all $W_{i,y}$
		which is not $\lrconn{X}$. In the end, $X^{y}$ satisfies the properties
		in the claim for $j = y$.
		
		Therefore, the claim holds for all values of $j \in [y]$ by induction.
	\end{claimproof}

	In the following, let $X$ denote the solution $X^{j}$ obtained from \cref{claim:X_W_lr_conn}.
	Observe that since $W_{i,j}$ is $\lrconn{X}$, then $\abs{X \cap W_{i,j}} \geq \omega$
	for each $(i,j) \in [x] \times [y]$ by \cref{lemma:wire-gadget}.
	By \cref{lemma:crossing-gadget,lemma:top_gadget,lemma:bottom_gadget},
	it holds that $\abs{X \cap W_{i,j}} = \omega$ for each $(i,j) \in [x] \times [y]$.
	Because otherwise we would have $\abs{X} > k$ which is a contradiction.
	Furthermore, by \cref{lemma:wire-gadget},
	$X$ contains exactly one square from each block in $W_{i,j}$.
	Similarly, for each $C_{i,j}$, $L_{i,j}$ $D_j$ and $T_j$,
	$X$ contains the minimum number of
	squares from those gadgets.
	In particular, for each $i \in [x]$ and $j \in [y]$, it holds that $\abs{X \cap X_{i,j} = \frac{\omega +1}{4} \cdot (h-1)}$.
	Then, \cref{item:minimum_vertices_crossing} in \cref{lemma:crossing-gadget} implies that
	\begin{equation}\label{eq:Ical_prime_gadget}
		\left( X \cap C_{i,j} \right) \in \bigg\{\Bigl(\Delta^{1}_{C_{i,j}} \cup \phi_{i,j}\Bigr), \Bigl(\Delta^{2}_{C_{i,j}} \cup \phi_{i,j}\Bigr)\bigg\}.
	\end{equation}

	For each $(i,j) \in [x] \times [y]$, let $B_{i,j}$ be the first block of the gadget $W_{i,j}$.
	Then, let $(a_{i,j},b_{i,j}) \in S_{i,j}$ such that
	$\sigma_{B_{i,j}}(a_{i,j},b_{i,j})$ is the (single) square in the $X \cap W_{i,j}$.
	Recall that $X$ contains exactly one square from each block in $W_{i,j}$ as mentioned above,
	therefore $a_{i,j}$ and $b_{i,j}$ are well-defined.

	\begin{claim}\label{claim:horizontal_ineq}
		For each $i \in [x]$, it holds that $b_{i,j} \geq b_{i, j+ 1}$
		for $j \in [y-1]$.
	\end{claim}

	\begin{claimproof}
		Let $i \in [x]$ and suppose that $b_{i,j} < b_{i, j+ 1}$
		for some $j \in [y-1]$. Consider the squares
		\begin{equation*}
			A_j \coloneqq \sigma_{B_{i,j}}\left( a_{i,j}, b_{i,j} \right) \text{ and } A_{j+1} \coloneqq \sigma_{B_{i,j+1}}\left( a_{i,j+1}, b_{i,j+1} \right).
		\end{equation*}
		Recall that $\abs{X \cap W_{i,j}} = \omega$, hence $X$ contains $\omega - 1$ squares
		that connect $A_j$ and $A_{j+1}$.
		However, the horizontal distance between $A_j$ and $A_{j+1}$ is given by
		\begin{align*}
			\biggl(2 + (j+1) \cdot \omega + \gamma \cdot ( b_{j+1} - 1)  \biggr) - \biggl(2 + j \cdot \omega + \gamma \cdot ( b_j - 1) + 1\biggr) &= \omega - 1 + \gamma \cdot \left( b_{j+1} - b_j \right) \\
																			      &> \omega - 1
		\end{align*}
		where the last step holds because $b_{j+1} > b_j$ by our assumption.
		This distance cannot be covered by $\omega - 1$ squares,
		therefore we have $b_{i,j} \geq b_{i,j+1}$.
	\end{claimproof}

	\begin{claim}\label{claim:vertical_eq}
		For each $j \in [y]$, it holds that $a_{i,j} = a_{i+1, j}$ for
		$i \in [x-1]$.
	\end{claim}

	\begin{claimproof}
		Let $j \in [y]$ such that there exists $i \in [x-1]$ such that
		$a_{i,j} \neq a_{i+1, j}$.
		We consider two cases:
        \subparagraph*{Case 1: $a_{i,j} = 0$ and $a_{i+1, j} = 1$.} Observe that since $a_{i,j} = 0$, it holds that
				\begin{equation*}
					\left( X \cap C_{i,j} \right) \neq \left( \Delta^{1}_{C_{i,j}} \cup \phi_{i,j} \right) 
				\end{equation*}
				because otherwise the interface vertex $u_{SW}$ of $C_{i,j}$
				would not be connected to a vertex outside of $C_{i,j}$,
				which is a contradiction.

				Similarly, we cannot have
				\begin{equation*}
					\left( X \cap C_{i,j} \right) \neq \left( \Delta^{2}_{C_{i,j}} \cup \phi_{i,j} \right),
				\end{equation*}
				because otherwise the interface vertex $u_{NW}$ in $C_{i,j}$
				would not be connected to a vertex outside of $C_{i,j}$.

				All in all, this leads to a contradiction by \cref{eq:Ical_prime_gadget}.
                \subparagraph{Case 2: $a_{i,j} = 0$ and $a_{i+1, j} = 1$.} In this case, suppose that
				\begin{equation}\label{eq:vertical_eq_X_choice}
					\left( X \cap C_{i,j} \right) = \left( \Delta^{1}_{C_{i,j}} \cup \phi_{i,j} \right).
				\end{equation}
				The remaining case of $\left( X \cap C_{i,j} \right) = \left( \Delta^{1}_{C_{i,j}} \cup \phi_{i,j} \right)$
				follows from similar arguments.

				Let $B_{i,j}$ denote the last block in $W_{i+1, j}$.
				Observe that by \cref{eq:vertical_eq_X_choice}, the (single)
				square in $X \cap B_{i,j}$ should touch the interface vertex $u_{NE}$ in $C_{i,j}$.
				However, then, none of the interface squares $u_{SE}$ and $u_{SW}$ in $C_{i+1,j}$ can be connected to
				a vertex outside of $C_{i+1,j}$, which leads to a contradiction.
				(if $i+1 = x$, then the same holds for the interface squares $u$ and $v$ of $T_j)$).
		
		In both cases, we arrive at a contradiction, therefore the claim is true.
	\end{claimproof}
	
	Finally, by \cref{claim:horizontal_ineq,claim:vertical_eq},
	\begin{equation*}
		\Big\{\bigl( a_{i,j}, b_{i,j} \bigr) \Big\}_{(i,j) \in [x] \times [y]}
	\end{equation*}
	is a solution for the $\mongtiling$ instance $\mathcal{I}$.
	Therefore, $\mathcal{I}$ is a $\yes$-instance.
\end{proof}

Next, we combine everything together, and prove \cref{thm:st-lb}.
\begin{proof}[Proof of~\cref{thm:st-lb}]
	The correctness of the reduction follows from \cref{lem:equivalence-part-1,lem:equivalence-part-2}.
	Observe that we have $k + T \leq c \cdot x \cdot y$
	for some $c \geq 1$
	by \eqref{eq:T_x_y_bound} and \eqref{eq:k_x_y_bound}.
	Moreover, it also holds that $\abs{V} \leq c' \cdot N \cdot x \cdot y$
	by the construction of the instance $\mathcal{I}'$.
	Constructing the instance $\mathcal{I}'$ takes time polynomial in
	$x,y$ and $N$. Running the hypothetical algorithm takes time
	\begin{align*}
		C \cdot 2^{\varepsilon \cdot (k + \abs{T})} \cdot \left( \abs{V} +
        \abs{T} \right)^{b} &\leq C \cdot 2^{c \cdot \varepsilon \cdot x \cdot y} \cdot x^{b} \cdot y ^{b} \cdot N ^{b} \\
											       &= C \cdot 2^{\varepsilon' \cdot x \cdot y}\cdot N^{b},
	\end{align*}
	which contradicts ETH by \cref{theorem:mon_grid_tiling_hardness}.
\end{proof}

If we use the bound from~\cref{remark:mongtiling} on \mongtiling\ we get the
following corollary of the construction.

\begin{corollary}
    For every $b > 0$, there exists $\eps > 0$, such that for every $C > 0$ there is no algorithm
    that solves every instance $(k,V,T)$ of \stsquare in time
    $C \cdot (|V|+|T|)^{\eps\sqrt{k}/\log{k} + b}$, unless ETH fails.
\end{corollary}

\begin{remark}
    In the above construction, we can replace unit squares with any shape $S
    \subseteq [0,1] \times [0,1]$ that contains $[0,1] \times
    [1/2-\delta,1/2+\delta] \cup [1/2-\delta,1/2+\delta] \times [0,1]$ cross. In
    particular, by setting $\delta$ sufficiently small 
    \cref{thm:st-lb} is also true for unit disks.
\end{remark}

\section{Lower Bound for Almost Squares (Proof of~\cref{th:almostsquarelb})}%
\label{sec:rectangles-lb-proof}
    \newcommand{\SpecialSC}{\textsc{Special-3SC}\xspace}

This result is corollary of the construction of Chan and
Grant~\cite{chan2014exact} who showed APX-hardness of this problem. We include
this construction for completeness. As a starting point, Chan and Grant use
the Vertex Cover problem in the 3-regular graphs.

\begin{theorem}[cf.~\cite{amiri2020fine}]\label{thm:3vc-lb}
    Assuming ETH, there is no $2^{o(n)}$ time algorithm that finds minimum
    Vertex Cover of $3$-regular graphs with $n$ vertices.
\end{theorem}

Next, Chan and Grant~\cite{chan2014exact} define the restricted version of unweighted set cover.

\begin{definition}
    In a \SpecialSC problem we are given a universe $U = A \cup W \cup X \cup Y
    \cup Z$ with $A = \{a_1,\ldots,a_n\}$, $W = \{w_1,\ldots,w_m\}$, $X =
    \{x_1,\ldots,x_m\}$, $Y = \{y_1, \ldots ,y_m\}$ and $Z = \{z_1,\ldots,z_m\}$
    such that $2n = 3m$. Moreover, we are given a family $\mathcal{S}$ of $5m$ subsets
    of $U$ that satisfy the following conditions:
    \begin{itemize}
        \item For every $t \in [n]$ the element $a_t$ is in exactly two sets of
            $\mathcal{S}$, and 
        \item For every $t \in [m]$ there exist integers $1 \le i < j < k \le n$
            such that $\mathcal{S}$ contains the sets $\{a_i,w_t\}, \{w_t,x_t\},
            \{a_j, x_t, y_t\}, \{y_t,z_t\}$ and $\{a_k,z_t\}$.
    \end{itemize}
    The task in the \SpecialSC problem is to find a minimum set cover of the
    universe $U$ with sets in $\mathcal{S}$.
\end{definition}

\begin{lemma}\label{lem:specialsc-lb}
    Assuming ETH, there is no $2^{o(n)}$ time algorithm that solves \SpecialSC.
\end{lemma}
\begin{proof}
    Given an instance of Vertex Cover in $3$-regular graph $G$ with edges
    $\{e_1,\ldots,e_m\}$ and vertices $\{v_1,\ldots,v_n\}$, Chan and
    Grant~\cite{chan2014exact} define an equivalent instance of \SpecialSC as 
    $\{a_i,w_t\}$, $\{w_t,x_t\}$, $\{a_j,x_t,y_t\}$, $\{y_t,z_t\}$ and
    $\{a_k,z_t\}$ for each $(t,i,j,k)$ such that $v_t$ is incident to edges
    $e_i,e_j$ and $e_k$ with $i < j < k$. This concludes the construction of the set
    $\mathcal{S}$.

    Assume $S$ is the solution to \SpecialSC, we take $v_t$ to be the vertex in
    the minimum vertex cover if and only if at least one of $\{a_i,w_t\}$,
    $\{a_j,x_t,y_t\}$, or $\{a_k,z_t\}$ is taken in $S$. Notice that this is a
    feasible solution to the Vertex Cover, as an edge $e_i$ is covered in $G$
    iff $a_i$ is covered by $S$. For the other direction, assume that $v_t$ is a
    vertex in the minimum vertex cover, then to construct a solution $S$ of
    \SpecialSC we take the sets $\{a_i,w_t\}$, $\{a_j,x_t,y_t\}$, $\{a_k,z_t\}$
    and when $v_t$ is not in the vertex cover we take sets $\{w_t,x_t\}$ and
    $\{y_t,z_t\}$. The cardinalities of the solutions are preserved as the sizes
    of these choices differ by $1$.

    Therefore, by~\cref{thm:3vc-lb} no $2^{o(n)}$ time algorithm can solve
    \SpecialSC assuming ETH.
\end{proof}

The proof of~\cref{th:almostsquarelb} is a corollary of the following
statement, by scaling and adjusting the number of terminals:
\begin{lemma}
    Assuming ETH, for any $\eps > 0$ there is no $2^{o(n)}$ algorithm for Steiner Tree of
    intersection graph of a given set of axis-aligned rectangles in
    $\mathbb{R}^2$, even when:
    \begin{itemize}
        \item all non-terminal rectangles have lower-left corner in $[-1,-1-\eps] \times
            [-1,-1-\eps]$ and upper-right corner in $[1,1+\eps] \times [1,1+\eps]$ 
        \item all terminals are points. 
    \end{itemize}
    Both the number of terminals and non-terminals is at least $n$.
\end{lemma}
\begin{proof}
    We let $\Delta < \eps/(10n^2)$. Following~\cite{chan2014exact}, we reduce from \SpecialSC and let $B = W
    \cup X\cup Y\cup Z$ and $A = \{a_1,\ldots,a_n\}$ be as in definition of \SpecialSC. We linearly
    order $B = \{w_1,x_1,y_1,z_1,w_2,x_2,\ldots,z_m\}$ so that $w_t,x_t,y_t,z_t$
    are consecutive for every $t \in [m]$.

    To construct terminals, for every $a_i$ with $i \in [n]$ add point $a'_i :=
    (1+i\Delta,1+i\Delta-2)$. Similarly, for every $w_i,x_i,y_i,z_i$ with $i \in
    [m]$ add points $w'_i := ((4i+1) \Delta - 1, (4i+1)\Delta+1)$, 
    $x'_i := ((4i+2) \Delta - 1, (4i+2)\Delta+1)$, 
    $y'_i := ((4i+3) \Delta - 1, (4i+3)\Delta+1)$ and $z'_i := ((4i+4) \Delta - 1, (4i+4)\Delta+1)$.

    For a set $S \in \mathcal{S}$ containing points
    $a_j,x_t,y_t$ we let $r_s$ be the minimum rectangle with lower-left corner
    in $[-1,-1-\eps] \times [-1,-1+\eps]$ and upper-left corner in $[1,1-\eps]
    \times [1,1+\eps]$ that contains points $a_j',x_t',y_t'$. Analogously we
    construct the remaining sets and notice that by minimality $r_s$ contains only
    points of $S$ of $A\cup B$. Finally, notice that all rectangles share a
    point $(0,0)$. Therefore, the solution to \SpecialSC is equivalent to the
    solution to the Steiner tree of its intersection graph. Hence
    by~\cref{lem:specialsc-lb} there is no $2^{o(n)}$ time algorithm for Steiner
    Tree of an intersection graph.
\end{proof}


\phantomsection
\addcontentsline{toc}{section}{References}
\bibliography{main}

\end{document}